\documentclass[
twocolumn
]{aastex63}

\usepackage{multirow}
\usepackage[fleqn]{amsmath}
\usepackage{graphicx}
\usepackage[FIGTOPCAP]{subfigure}
\usepackage{txfonts}
\usepackage{natbib}
\bibpunct{(}{)}{;}{a}{}{,}

\usepackage{color}

\newcommand{\structuring}[1]{}

\newcommand{\vONE}[1]{#1}
\newcommand{\vTWO}[1]{{#1}}

\received{\today}
\revised{}
\accepted{}
\submitjournal{ApJS}

\shorttitle{Makemake + Sedna}
\shortauthors{Kuiper, Yorke, \& Mignone}

\begin{document}

\title{Makemake + Sedna: \\A Continuum Radiation Transport and Photoionization Framework for Astrophysical Newtonian Fluid Dynamics}

\correspondingauthor{Rolf Kuiper}

\author[0000-0003-2309-8963]{Rolf Kuiper}
\email{rolf.kuiper@uni-tuebingen.de}
\affiliation{Institut f\"ur Astronomie und Astrophysik, Universit\"at T\"ubingen, Auf der Morgenstelle 10, D-72076 T\"ubingen, Germany}

\author{Harold W.~Yorke}
\affiliation{SOFIA Science Center, Universities Space Research Association (USRA), M/S 232-12, Moffett Field, CA 94035-1000, USA}

\author[0000-0002-8352-6635]{Andrea Mignone}
\affiliation{Dipartimento di Fisica Generale, Universita di Torino, via Pietro Giuria 1, 10125 Torino, Italy}

\vONE{
\begin{abstract}
Astrophysical fluid flow studies often encompass a wide range of physical processes to account for the complexity of the system under consideration. 
In addition to gravity, a proper treatment of thermodynamic processes via continuum radiation transport and/or photoionization is becoming the state of the art.
We present a major update of our continuum radiation transport module, Makemake, and a newly developed module for photoionization, Sedna, coupled to the magnetohydrodynamics code PLUTO.
These extensions are currently not publicly available; access can be granted on a case-by-case basis.
We explain the theoretical background of the equations solved, elaborate on the numerical layout, and present a comprehensive test suite for radiation--ionization hydrodynamics.
The grid based radiation and ionization modules support static one-dimensional, two-dimensional, and three-dimensional grids in Cartesian, cylindrical, and spherical coordinates.
Each module splits the radiation field into two components, one originating directly from a point source -- solved using a ray-tracing scheme -- and a diffuse component -- solved with a three-dimensional flux-limited diffusion (FLD) solver.
The FLD solver for the continuum radiation transport makes use of either the equilibrium one-temperature approach or the linearization two-temperature approach.
The FLD solver for the photoionization module enables accounting for the temporal evolution of the radiation field from direct recombination of free electrons into hydrogen's ground state as an alternative to on-the-spot approximation. 
A brief overview of completed and ongoing scientific studies is given to explicitly illustrate the multipurpose nature of the numerical framework presented.
\end{abstract}
}

\keywords{Radiative transfer --- Magnetohydrodynamics (MHD)  --- Methods: numerical  --- Stars: formation  --- H II regions}

%
%
\section{Introduction}
\label{sect:introduction}
Code development for astrophysical research can be categorized based on the generality of the implementations:
often, algorithms are implemented to model a specific system or physical behavior unique to that system; examples from our own numerical studies are the subgrid modules for protostellar outflow feedback \citep{2015ApJ...800...86K, 2016ApJ...832...40K} and the stellar evolution solver \citep{2013ApJ...772...61K}.
These specific applications rely on underlying software modules that, by contrast, treat the more general basic equations common to a variety of problems. In astrophysics, these are, e.g., magnetohydrodynamics (MHD), \emph{N}-body, and dust and line radiation transport solver packages.

A variety of such general purpose codes, solving the MHD equations, were developed in the past and are commonly applied in astrophysical studies.
Without claiming completeness, commonly used open-source and Message Passing Interface (MPI)-parallelized MHD software packages include
Zeus \citep[e.g.][]{2000RMxAC...9...66N, 2006ApJS..165..188H, 2015A&A...574A..81R},
PLUTO \citep{2007ApJS..170..228M, 2012ApJS..198....7M},
Flash \citep{2000ApJS..131..273F, 2009arXiv0903.4875D, 2012ApJS..201...27D, 2014ApJ...797....4K},
Ramses \citep[e.g.][]{2002A&A...385..337T, 2006A&A...457..371F, 2011A&A...529A..35C, 2013MNRAS.436.2188R, 2014A&A...563A..11C, 2015A&A...578A..12G, 2015MNRAS.449.4380R, 2016arXiv160508032D, 2017A&A...603A.105D}, 
Nirvana \citep[e.g.][]{2011ascl.soft01006Z, 2013A&A...560A..93G}, and
Enzo \citep[e.g.][]{2004astro.ph..3044O, 2010ascl.soft10072O, 2009AIPC.1171..260N, 2018FrASS...5...34N, 2010ApJS..186..308C, 2011MNRAS.414.3458W, 2014ApJS..211...19B}.
The gravito-MHD equations, accounting for gravity of point sources and/or the self-gravity of the gas, can be implemented into such general purpose software quite easily.
By contrast, implementing continuum radiation transport into an existing MHD code is not straightforward.
One reason is the huge computational effort associated with solving the general radiation transport equation in three spatial dimensions, perhaps even including frequency dependence and scattering.
For this reason, radiation MHD frameworks do not solve the general radiation transport equation but make use of multiple approximations, such as frequency averaging, moment methods, short and long characteristics, thus limiting the general applicability of the code.
Analogous arguments apply for the photoionizing radiation in combination with MHD.

In addition to the breadth of physical processes being modeled, the applicability of a software package is limited by the underlying data structure as well.
Here, we will focus on grid based codes, although a generalization of our radiation transfer modules for smooth particle hydrodynamics would be possible in a hybrid scheme.
The different grid based approaches used currently include
regular static grids or grids composed of multiple regular grids, using Cartesian, cylindrical, and spherical coordinates,
nested or adaptive mesh refinement (AMR) grids, usually done in Cartesian coordinates,
and unstructured grids such as triangulations.
Each of these different approaches has its own advantages and disadvantages.

For astrophysical applications, the grid structure often has to cover a broad dynamical range in e.g.~spatial dimension and/or mass density.
This feature is perhaps easiest to achieve using unstructured grids.
But at the same time, astrophysical applications often require higher-order integration schemes, e.g.~to properly account for shocks.
Higher-order schemes are naturally easier to realize on structured grids.
Nested and AMR grids combine these two features, but currently those implementations are usually done in Cartesian coordinates \citep[but see also][for a curvilinear AMR approach for MHD]{2012ApJS..198....7M}.
One reason for the wide use of Cartesian coordinates  is that the currently available open-source and MPI-parallelized grid libraries are restricted to this type of coordinate system, as is the case for Paramesh \citep{2000CoPhC.126..330M, 2011ascl.soft06009M} and Chombo \citep{Chombo2015, 2012ascl.soft02008A}.
A regular grid in spherical coordinates has increased spatial resolution toward the center, alleviating the necessity for a nested or AMR grid for a large variety of astrophysical systems that require higher resolution near the coordinate origin.

The continuum radiation transport and photoionization solver modules presented here both make use of a hybrid ansatz, which combines a ray-tracing routine along one coordinate axis with a three-dimensional flux-limited diffusion (FLD) solver.
Thus, our numerical framework is clearly tailored toward applications for which either a single source dominates the radiation field (spherical coordinates) or for plane-parallel setups (which can be solved using Cartesian coordinates).
The supported grids and solvers are suitable for a huge variety of astrophysical systems, such as
star formation,
planet formation,
planetary atmospheres,
accretion disks, 
disk photoevaporation,
planet--disk interaction,
common envelope,
late stages of stellar evolution,
planetary nebulae,
H II regions,
black hole accretion, and 
active galactic nucleus (AGN) physics.

The manuscript is organized as follows:
in Sect.~\ref{sect:methods}, we introduce the newly developed continuum radiation transport and photoionization framework;
in Sect.~\ref{sect:tests}, we present a comprehensive test suite of the code; and
in Sect.~\ref{sect:applications}, we give an overview of successfully completed and ongoing research projects utilizing the described numerical tool.

%
%
\structuring{\newpage}
\section{Methods}
\label{sect:methods}
The central workhorse of the numerical framework presented is the open-source MHD code PLUTO (\href{http://plutocode.ph.unito.it}{http://plutocode.ph.unito.it}), as presented in \citet{2007ApJS..170..228M, 2012ApJS..198....7M}.
PLUTO is not only used to solve the equations of motion for the gas (as described below in detail), but we also use the parallel layout and I/O structure for the multiphysics numerical framework presented.
We currently use PLUTO in its version~4.1.

\begin{figure*}[htbp]
\centering
\includegraphics[width=0.99\textwidth]{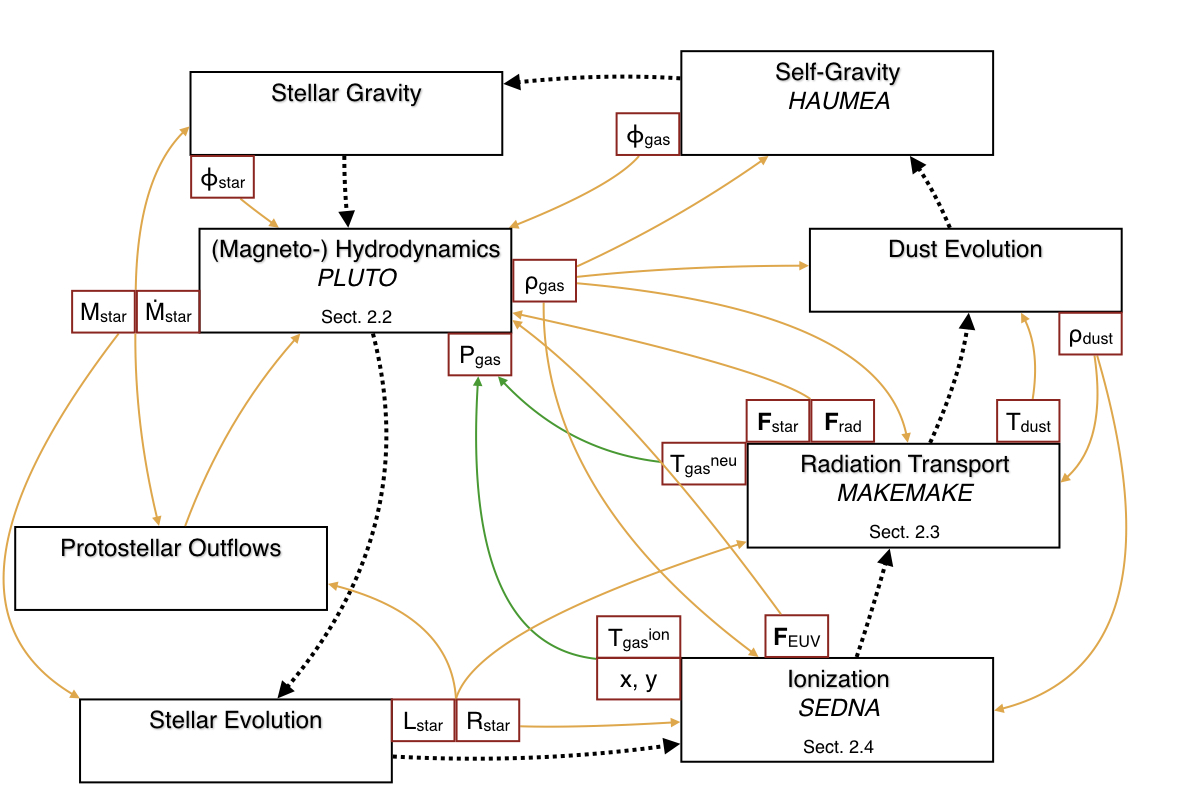}
\caption{
Flowchart of the overall multiphysics numerical framework for astrophysical fluid dynamics.
Black boxes represent a module for a specific physical task.
Red boxes connected to the modules represent their output quantities;
output quantities of a module are only shown if they denote an input for another module.
Black dotted arrows represent the call sequence of the different modules; 
the main loop starts from prefdefined initial conditions by calling the self-gravity module.
Yellow arrows denote input quantities.
Green arrows denote update of a dependent quantity.
}
\label{fig:flowbelt}
\end{figure*}

A brief overview of the multiphysics framework is presented as a flowchart of their dependencies and interlinks in Fig.~\ref{fig:flowbelt}.
Detailed description of the physics and numerics of each of the radiation--ionization modules are given in the following sections.

The modules added to the PLUTO code either
address specific physics of star formation and accretion disks such as 
stellar evolution, 
protostellar outflows, 
dust evaporation and sublimation, 
disk shear-viscosity prescriptions, 
or
involve implicit solution methods on the basic spatial dimensions (while PLUTO solves for the MHD in an explicit fashion) such as 
self-gravity, 
ray-tracing of thermal continuum radiation (Sect.~\ref{sect:irradiation}) and photoionizing radiation (Sect.~\ref{sect:directionization})
FLD of thermal radiation energy density (Sect.~\ref{sect:FLD}) , and 
evolution of diffuse direct recombination EUV photons (Sect.~\ref{sect:recombination}).

Although the developed numerical framework denotes a straightforward extension of our earlier work in the field of
radiation hydrodynamics \citep{2010A&A...511A..81K, 2012A&A...537A.122K, 2013A&A...555A...7K},
stellar feedback in cloud collapse \citep{2010ApJ...722.1556K, 2015ApJ...800...86K, 2016ApJ...832...40K, 2013ApJ...763..104K, 2013ApJ...772...61K},
and disk formation simulations \citep{2011ApJ...732...20K}
all routines and algorithms were newly written from scratch and have undergone major changes in the code design.
The most recent code development, not present in these earlier studies, is the ionizing radiation solver module.
Furthermore, the other physics modules have been updated by, e.g., the inclusion of a so-called two-temperature FLD solver, a model for time-dependent dust evaporation and sublimation, a flared disk model for the shear-viscosity description, and an algebraic multigrid preconditioner for fast, incomplete matrix conversion.
The latter improves the parallel scalability of the implicit solvers for self-gravity, thermal diffusion, and diffuse ionization.

%
%
\subsection{Grids}
\label{sect:methods_grids}
The MHD code PLUTO is capable of treating static grids of logically rectangular coordinates (Cartesian, cylindrical, and spherical) as well as exploiting AMR techniques \citep{2012ApJS..198....7M} via the Chombo library.
The extensions presented herein are restricted to the static grids of rectangular coordinates.
The implicit solvers (self-gravity, thermal diffusion, diffuse ionization) work in all three coordinate systems.
The ray-tracing is only done along the first coordinate axis, i.e.~along the $x$ direction in Cartesian coordinates, and the radial direction in cylindrical and spherical coordinates.
These static grids can be arbitrarily stretched and stacked (but not nested),
i.e.~the cell size in the $n$th coordinate direction can be an arbitrary function of the $n$th coordinate, but is independent of the two other directions.

Two grids that we use regularly are spherical grids in log-radial and cos-polar.
We define the spherical coordinates as the spherical radius $r$, going from $r_\mathrm{min}$ to $r_\mathrm{max}$; the polar angle $\theta$, going in its maximum extent from $\theta_\mathrm{min} = 0$ at the upper polar axis to $\theta_\mathrm{max} = \pi$ at the lower polar axis; and the azimuthal angle $\phi$, going in its maximum extent from $\phi_\mathrm{min} = 0$ to $\phi_\mathrm{max} = 2\pi$.
In spherical coordinates, the resolution of grid cells linearly decreases toward smaller radii in the polar and azimuthal direction.
To achieve the same behavior in the radial direction, the radial resolution can be set to be a linear function of the radius itself, a so-called log-radial grid.
Such a grid will cover very large spatial regions within the computational domain with increased resolution toward the central region. 

Although the log-radial grid has the same extensions in all three coordinate dimensions in the grid cells of the midplane ($\theta = \pi/2$), the size of grid cells in the azimuthal direction decreases proportionally to $\sin \theta$ toward the poles.
Accordingly, the volume of the grid cells decreases proportionally to $\sin \theta$ toward the poles.
In order to achieve grid cells of comparable volume at the same radius but for different polar angles, the size of grid cells in the polar direction can be set to be uniform in $\cos \theta$ space.
Such a grid allows for higher spatial resolution (in the polar direction) toward the midplane of the computational domain, as, e.g., used to study accretion disks.
We have used PLUTO with such a grid for the first time in 3D simulations of the formation of protoplanetary atmospheres, see \citet{2015MNRAS.447.3512O} for further details.

%
%
\subsection{Fluid Dynamics}
As mentioned above, we coupled our continuum radiation transport and photoionization framework (as well as the self-gravity, stellar evolution, dust evolution, and protostellar outflows module) with the open-source MHD code PLUTO \citep{2010ascl.soft10045M}.
PLUTO is a grid based code, which solves the MHD equations using Godunov-type shock-capturing schemes.
It provides a variety of different solvers, interpolation schemes, and slope limiters.
For details, we refer the interested reader to the original technical reports by \citet{2007ApJS..170..228M, 2012ApJS..198....7M}.

In general, PLUTO allows the user to add ``external'' acceleration or potentials to the momentum and energy equation.
We utilize this interface to introduce the additional accelerations from absorption and reemission of radiative fluxes in the continuum and EUV regime respectively
$\vec{a}_\mathrm{ext}^\mathrm{~tot} = \vec{a}_\mathrm{ext}^\mathrm{~rad} + \vec{a}_\mathrm{ext}^\mathrm{~ion}$.

%
%
\structuring{\clearpage}
\subsection{Radiation Transport}
\label{sect:radiationtransport}
In this section, we describe the physics and numerics of the updated radiation transport module named Makemake.
\begin{figure*}[htbp]
\centering
\includegraphics[width=0.99\textwidth]{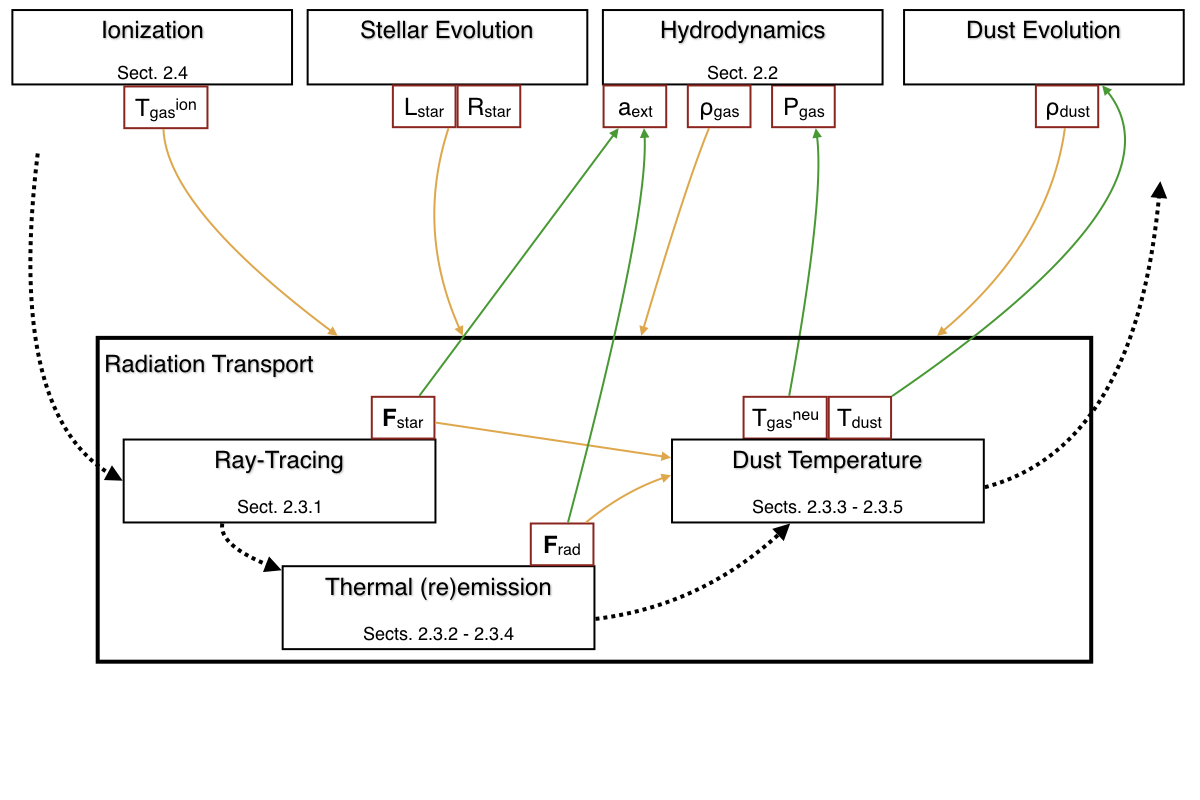}
\caption{
Flowchart of the continuum radiation transport module Makemake.
The legend is the same as in Fig.~\ref{fig:flowbelt}.
}
\label{fig:flowMakemake}
\end{figure*}
A predecessor was introduced in \citet{2010A&A...511A..81K} as the generalization of the hybrid radiation transfer module of \citet{1997A&A...327..317R} written in cylindrical coordinates.
We describe the radiation transport methods implemented, their derivation, and how they are implemented in our hybrid approach, whereby the radiation field is split into multiple components and each component is handled by a different appropriate solver method.
Determining the appropriate solver method means finding a good balance between physical accuracy and computational speed.
Starting from the general radiation transport equation given below, 
we describe the different approximations applied in each solver method,
discuss their applicability to different components of the total radiation field, 
and describe the numerical algorithms used to solve the final equations.

Consider the basic time-dependent radiation transport equation
\vTWO{
\citep[see e.g.][]{1984frh..book.....M}
}
\begin{equation}
\label{eq:radiationtransportequation}
\left(
\frac{1}{c} ~ \partial_t
+
\vec{\Omega} \cdot \vec{\nabla}
+
\chi_\mathrm{ext}
\right)
~ I_\mathrm{rad}
=
\vTWO{
\chi_\mathrm{ext}
~
S
}
\end{equation}
with
the radiation intensity $I_\mathrm{rad}$,
the direction $\vec{\Omega}$ of the radiative flux,
\vTWO{
and the source function $S$.
}
The extinction coefficient $\chi_\mathrm{ext}$ along the radiative direction comprises the coefficient for absorption and scattering $\chi_\mathrm{ext} = \chi_\mathrm{abs} + \chi_\mathrm{scat}$.

\subsubsection{Irradiation}
\label{sect:irradiation}
The irradiation routine handles the transport of a user-defined luminosity and spectrum along the first coordinate via ray-tracing.
Although in principle this routine can be used to compute ray-tracing along the $x$ coordinate in Cartesian geometry, e.g., to model a part of an atmosphere in locally plane-parallel approximation, we will focus the description of the routine on a central source at the origin of a spherical coordinate system.
The ray-tracing computes the absorption of the source function of spectral photons along each ray direction,
the handling of sources from thermal (re)emission along the ray is shifted to the FLD solver described below.
Hence, in the case of ray-tracing, Equation \eqref{eq:radiationtransportequation} reduces to the form without source terms:
\begin{equation}
\frac{1}{c} ~ \partial_t ~ I_\mathrm{rad}
+
\vec{\Omega} \cdot \vec{\nabla} ~ I_\mathrm{rad}
+
\chi_\mathrm{ext} ~ I_\mathrm{rad}
=
0
\end{equation}

The radiation transport for $I_\mathrm{rad}$ is computed for the same time step as the simultaneously running MHD.
If the photon travel time up to its first absorption or scattering is short compared to the time step of the hydrodynamics, we can ignore the time derivative of the radiation intensity on the left-hand side:
\begin{equation}
\vec{\Omega} \cdot \vec{\nabla} ~ I_\mathrm{rad}
+
\chi_\mathrm{ext} ~ I_\mathrm{rad}
=
0
\end{equation}

If we solve the remaining equation for a source at the origin of a spherical coordinate system, 
\vTWO{
}
the resulting differential equation
\vTWO{
\begin{equation}
\partial_r ~ I_\mathrm{rad}
=
-
\chi_\mathrm{ext} ~ I_\mathrm{rad} 
\end{equation}
}
has the solution
\vTWO{
\begin{equation}
\label{eq:radintenssol}
I_\mathrm{rad}
=
I_\mathrm{rad}(r_\mathrm{min})
~
\exp(-\tau)
\end{equation}
}
with the source function $I_\mathrm{rad}(r_\mathrm{min})$ at the minimum integration radius $r_\mathrm{min}$ and the optical depth along the ray direction 
\begin{equation}
\label{eq:opticaldepth}
\tau(r) = \int_{r_\mathrm{min}}^r ~ \chi_\mathrm{ext} ~ \mbox{d}r.
\end{equation}
The radiation intensity and optical depth are computed for each ray direction.
For frequency-dependent ray-tracing, the radiation intensity and the optical depth are computed for each frequency bin.
\vTWO{
Equation \eqref{eq:radintenssol} states that in the absence of extinction (and emission) the radiation intensity is conserved along a ray.
The corresponding solution for the irradiation flux emitted by an isotropic emitter at the origin of the spherical coordinate system is given by
\begin{equation}
\label{eq:radfluxsol}
F_\mathrm{irr}(r)
=
F_\mathrm{irr}(r_\mathrm{min}) 
~ \exp(-\tau)
~ \left( \frac{r_\mathrm{min}}{r} \right)^2
\end{equation}
}
The term $(r_\mathrm{min} / r)^2$ corresponds to the geometrical attenuation along the radially outgoing ray \vTWO{direction};
for the \vTWO{analogous} ray-tracing 
\vTWO{
of a plane-parallel flux
}
along the $x$-
\vTWO{
direction
}
of a Cartesian coordinate system, this term becomes unity.

\subsubsection{Flux-limited-diffusion Approximation}
\label{sect:FLD}
In the following derivation, we write the equations in the gray (non-frequency-dependent) approximation without loss of generality.
Besides the issue of a gray versus multifrequency approach, the FLD equation is the result of a sequence of approximations, which we will outline step by step.
As a first step, we integrate Equation \eqref{eq:radiationtransportequation} over all solid angles, \vTWO{neglect scattering,} and use the definitions for
radiation energy density $E_\mathrm{rad}$ and radiation energy flux density $\vec{F}_\mathrm{rad}$ 
\begin{eqnarray}
E_\mathrm{rad} &=& \frac{1}{c} ~ \int_{4\pi}  ~ I_\mathrm{rad} ~ \mbox{d}\Omega \\
\vec{F}_\mathrm{rad} &=& \int_{4\pi}  ~ I_\mathrm{rad} ~ \vec{\Omega} ~ \mbox{d}\Omega
\end{eqnarray}
to obtain
\begin{equation}
\label{eq:RT_conservation}
\partial_t ~ E_\mathrm{rad}
+
\vec{\nabla} \cdot \vec{F}_\mathrm{rad}
=
\chi_\mathrm{abs} ~ 
\vTWO{
\left(4 \pi ~ B_\mathrm{rad} -  c ~ E_\mathrm{rad} \right)
}
\end{equation}
Within a given volume, the change in radiation energy density per time (leftmost term) is either the result of a flux over the volume boundary (second term on left-hand side) or caused by the source and sink terms within the volume (right-hand side terms).
The right-hand side of Equation \eqref{eq:RT_conservation} describes the cooling and heating balance of the local emission and the local radiation field.
In the case of $B_\mathrm{rad} > E_\mathrm{rad}$, the emission yields a decrease of the local temperature of the medium (cooling) and an increase of the local radiation field.
In the case of $B_\mathrm{rad} < E_\mathrm{rad}$, the local radiation field decreases with time, and the energy is deposited as an increase in the local temperature of the medium (heating).

Equation \eqref{eq:RT_conservation} relates the zeroth moment of the radiation field $E_\mathrm{rad}$ to the first moment $\vec{F}_\mathrm{rad}$. One can obtain the next higher moment equation by multiplying 
Equation \eqref{eq:radiationtransportequation} by 
$\vec{\Omega}$ and integrating over all solid angles. 
For simplicity, we assume the time-independent case and obtain
\begin{equation}
\label{eq:2ndMoment}
c \vec{\nabla} \cdot \vec{P}_\mathrm{rad}
+
\chi_\mathrm{abs} ~  \vec{F}_\mathrm{rad}
= 0
\end{equation}
where we have introduced the radiation pressure tensor defined by 
\begin{equation}
\vec{P}_\mathrm{rad} = \frac{1}{c} ~ \int_{4\pi} ~ I_\mathrm{rad} ~ \vec{\Omega} \vTWO{\otimes} \vec{\Omega} ~ \mbox{d}\Omega ~ .
\end{equation}
By defining the \vTWO{dimensionless} radiation diffusion tensor by $\vec{D}_\mathrm{rad} = \vec{P}_\mathrm{rad} / E_\mathrm{rad} $, Equation \eqref{eq:2ndMoment} becomes
\vTWO{
\begin{equation}
\label{eq:2ndMomentDiffusion}
\vec{F}_\mathrm{rad} = - \frac{c}{\chi_\mathrm{abs}} ~ \vec{\nabla} \cdot \left( \vec{D}_\mathrm{rad} ~ E_\mathrm{rad} \right).
\end{equation}
}
Up to now, we have not made any approximations, other than time independence.  
For the so-called FLD approximation, the radiation diffusion tensor is approximated by a scalar diffusion coefficient $D_\mathrm{rad}$
\begin{equation}
\label{eq:diffusioncoefficient}
D_\mathrm{rad} = \frac{\lambda ~ c}{\chi_\mathrm{R}}
\end{equation}
with the flux limiter function $\lambda$
\vTWO{
and the Rosseland mean absorption coefficient $\chi_\mathrm{R}$. 
For convenience, the speed of light and the absorption coefficient are here included in the definition of the scalar diffusion coefficient, hence, in contrast to the dimensionless diffusion tensor defined above, the scalar diffusion coefficient has a unit of $\mbox{cm}^2 \mbox{ s}^{-1}$. 
}
With this \vTWO{FLD} approximation, Equation \eqref{eq:2ndMomentDiffusion} becomes
\begin{equation}
\label{eq:fluxlimiteddiffusionapproximation}
\vec{F}_\mathrm{rad} = -D_\mathrm{rad} ~ \vec{\nabla} ~ E_\mathrm{rad} ~ .
\end{equation}
In the classical diffusion limit, the isotropic diffusion coefficient is given as $D_\mathrm{rad} = c / (3 \chi_\mathrm{R})$ (i.e. $\lambda = 1/3$).
\vTWO{
}
Applying this flux limiter to optically thin radiative flows (where the diffusion limit is a priori not satisfied) leads to unphysical infinite flow velocities, because the diffusion coefficient approaches infinity in the case of vanishing absorption ($\chi_\mathrm{abs} \rightarrow 0$).
This deficiency can be circumvented by allowing the flux limiter function to vary.
The choice of the flux limiter function is relatively free and can be adapted to special cases but should fulfill the limiting values $\lambda \rightarrow 1/3$ in the optically thick limit (for $\tau \rightarrow \infty$) and $\lambda \rightarrow \chi_\mathrm{\vTWO{R}} ~ E_\mathrm{rad} / |\vec{\nabla} ~ E_\mathrm{rad}|$ in the optically thin limit (for $\tau \rightarrow 0$).
The physical reason for the latter is that the velocity $\vec{v} = - D_\mathrm{rad} ~ \vec{\nabla} ~ E_\mathrm{rad} / E_\mathrm{rad}$ of the radiative flux is limited to the speed of light.

Finally, inserting the FLD approximation \eqref{eq:fluxlimiteddiffusionapproximation} into the conservation Equation \eqref{eq:RT_conservation} yields the time evolution of the radiation energy density as
\begin{equation}
\label{eq:RT_fluxlimitediffusion}
\partial_t ~ E_\mathrm{rad}
-
\vec{\nabla} \cdot \left( D_\mathrm{rad} ~ \vec{\nabla} ~ E_\mathrm{rad} \right)
=
\chi_\mathrm{abs} ~ 
\vTWO{
\left(4 \pi ~ B_\mathrm{rad} -  c ~ E_\mathrm{rad} \right).
}
\end{equation}
The FLD equation \eqref{eq:RT_fluxlimitediffusion} involves two unknown quantities, namely 
the radiation energy density $E_\mathrm{rad}$ and
the local temperature of the medium 
$B_\mathrm{rad} = 
\vTWO{
\frac{1}{\pi} ~ \sigma_\mathrm{SB}
}
~ T^4$,
\vTWO{
with the Stefan--Boltzmann constant $\sigma_\mathrm{SB}$
}
which are coupled to each other via heating and cooling processes.

In principle, \vONE{Equation }\eqref{eq:RT_fluxlimitediffusion} can be solved without further approximations by simultaneously solving for the temporal evolution of the local internal energy 
$E_\mathrm{int} =  c_\mathrm{V} ~ \rho_\mathrm{gas} ~ T$ 
with the specific heat capacity of the medium $c_\mathrm{V} = R / (\mu (\gamma -1))$ with the universal gas constant $R$, the molar mass $\mu$, and the adiabatic index $\gamma$.
Its temporal evolution due to the thermodynamics of cooling and heating is given by 
\begin{equation}
\label{eq:RT_internalenergy}
\partial_t ~ E_\mathrm{int} 
= 
- \chi_\mathrm{abs} ~ 
\vTWO{
\left(4 \pi ~ B_\mathrm{rad} -  c ~ E_\mathrm{rad} \right).
}
\end{equation}
This solution method is also known as the two-temperature approach, because the evolution of both the radiation and the internal energy is determined.
The drawback of this approach is the numerical cost, especially in the case of a stiff system of equations, which can easily \vONE{follow from} the nonlinear dependence of the absorption coefficient $\chi_\mathrm{abs}$ on the local temperature.
The internal energy can also change due to nonthermodynamic processes such as advection.
\vTWO{
}
These terms are not included in the equation above; hence, these are solved for in the MHD module. 

The system of coupled equations \eqref{eq:RT_fluxlimitediffusion} and \eqref{eq:RT_internalenergy} can be further reduced to a single evolution equation of the radiation energy density only.
Below in sections \ref{sect:equilibriumtemperatureapproach} and \ref{sect:linearizationapproach}, we consider two different approximations, namely
the equilibrium temperature approach (also called one-temperature approach) and
the linearization approach (also called two-temperature linearization approach).

Due to the extreme simplification of the general radiation transport equation \eqref{eq:radiationtransportequation} by the FLD equation \eqref{eq:RT_fluxlimitediffusion}, the applicability of the FLD approximation remains rather narrow.
The method was first intended to solve for the radiation transport in one-dimensional problems.
Moreover, first applications focused on the interior of stars with the medium mostly in the optically thick regime with only one transition from optically thick to thin at the stellar surface.
In such media, the approximations applied are (reasonably) valid, and the method allows for a very efficient way of solving the radiation transport equation.
However, in a multidimensional problem with multiple transitions from optically thick to thin regions and vice versa, defining a usable flux limiter that accurately mimics the relation between radiative energy density and radiation pressure is difficult.
Moreover, the FLD approximation implicitly assumes that the flow direction is given by the gradient of the radiation energy density, which in general cause\vONE{s} unphysical behavior of the radiative flux. 
For example, optically thick obstacles illuminated by a single source are unable to cast sharp shadows when the FLD approximation is invoked; such a situation is e.g.~given by a protostar surrounded by its circumstellar disk \citep[see, e.g.,][]{2013A&A...555A...7K}.

\vONE{Further, we have the gray approximation, assuming} that the absorption coefficient is a pure function of the local radiation temperature. 
In general, however, the absorption coefficient depends on the energy of the absorbed photons, which can originate from any location.
As an example, regions directly irradiated by stellar sources would absorb these photons on average according to the Planck mean opacity with respect to the photospheric stellar temperature; in the gray FLD approximation, these stellar photons are absorbed according to the mean opacity with respect to the local temperature. 
To alleviate this problem, \citet{2002ApJ...569..846Y} used a frequency-dependent FLD solver; the herein presented code development is based on an FLD approach in gray approximation (plus ray-tracing along the first coordinate axis).

\subsubsection{Equilibrium Temperature Approach}
\label{sect:equilibriumtemperatureapproach}
The equilibrium temperature approach has been implemented in the previous version of the star formation framework as presented in \citet{2010A&A...511A..81K}.
As discussed above, both the equilibrium temperature approach and the linearization approach are meant to simplify the system of coupled equations for FLD,\eqref{eq:RT_fluxlimitediffusion}, and internal energy, \eqref{eq:RT_internalenergy}, to a single equation.
In the equilibrium temperature approach, this reduction is achieved via the assumption that the local radiation field is in equilibrium with the temperature of the medium.
Hence, radiation energy and internal energy are related to the same temperature.
Adding up the combined Equations \eqref{eq:RT_fluxlimitediffusion} and \eqref{eq:RT_internalenergy}, 
\begin{eqnarray}
\partial_t ~ E_\mathrm{int} &=& 
- \chi_\mathrm{abs} ~ 
\vTWO{
\left(4 \pi ~ B_\mathrm{rad} -  c ~ E_\mathrm{rad} \right)
}
\\
\partial_t ~ E_\mathrm{rad}
-
\vec{\nabla} \cdot \left( D_\mathrm{rad} ~ \vec{\nabla} ~ E_\mathrm{rad} \right)
&=&
+ \chi_\mathrm{abs} ~ 
\vTWO{
\left(4 \pi ~ B_\mathrm{rad} -  c ~ E_\mathrm{rad} \right)
}
\end{eqnarray}
leads to a single equation including both energies,
\begin{equation}
\partial_t ~ (E_\mathrm{rad} +  E_\mathrm{int})
-
\vec{\nabla} \cdot \left( D_\mathrm{rad} ~ \vec{\nabla} ~ E_\mathrm{rad} \right)
=
0.
\end{equation}
But due to the fact that the internal energy $E_\mathrm{int} = c_\mathrm{V} ~ \rho_\mathrm{gas} ~ T$ and the radiation energy $E_\mathrm{rad} = a ~ T^4$ is assumed to refer to the same temperature, their time derivatives can be expressed as
$\partial_t E_\mathrm{int} = c_\mathrm{V} ~ \rho_\mathrm{gas} / \left(4 ~ a ~ T^3 \right)  ~ \partial_t E_\mathrm{rad}$
and the equation above reduces to a modified diffusion equation,
\begin{equation}
\label{eq:RT_FLDequilibrium}
\partial_t ~ E_\mathrm{rad}
-
f_\mathrm{c} ~ \vec{\nabla} \cdot \left( D_\mathrm{rad} ~ \vec{\nabla} ~ E_\mathrm{rad} \right)
=
0
\end{equation}
with the energy ratio of
\begin{equation}
f_\mathrm{c} = \left( \frac{c_\mathrm{V} ~ \rho_\mathrm{gas}}{4 ~ a ~ T^3} + 1 \right)^{-1}.
\end{equation}
This diffusion equation allows one to directly solve for the radiation field $E_\mathrm{rad}$.
In the case of a single radiation field $E_\mathrm{rad}$, the temperature of the local medium is then determined by the equilibrium condition
\begin{equation}
\label{eq:RT_temperatureupdateequilibrium}
\chi_\mathrm{abs} ~ a ~ T^4 = \chi_\mathrm{abs} ~ E_\mathrm{rad}
\end{equation}
This equation is valid only if the FLD approximation is used to determine the total radiation field, i.e. not for the hybrid scheme discussed below.
The factor $\chi_\mathrm{abs}$ of course cancels out in the equation above, but we show the equation in this form to allow direct comparison to the hybrid equilibrium equation.

\subsubsection{Linearization Approach}
\label{sect:linearizationapproach}
In the linearization approach, the radiation field and the temperature of the medium are allowed to evolve as two different properties with no equilibrium assumption a priori.
But instead of solving the two Eqs.~\eqref{eq:RT_fluxlimitediffusion} and \eqref{eq:RT_internalenergy} simultaneously, linearizing on the right-hand side of Equation \eqref{eq:RT_fluxlimitediffusion} the implicit dependence on the temperature of the medium,
\begin{equation}
\label{eq:linearization}
\left( T(t+\Delta t) \right)^4 \approx 4 \left( T(t) \right)^3 \times T(t+\Delta t) - 3 \left( T(t) \right)^4,
\end{equation}
reduces the fourth-power dependence to only a linear dependence on the new temperature $T(t+\Delta t)$ at the current time step $\Delta t$.
The concept of linearization of the $T^4$ dependence is used regularly in radiative transfer literature and was pursued in the pioneering work by \citet{1968ApJ...151..311A} on stellar atmospheres (their Equation (7)); 
in the context of radiation hydrodynamics, this linearization approach to decouple the two evolution equations was — to the best of our knowledge — first presented in \citet{2011A&A...529A..35C}.

Utilizing this linearization, we can solve for the radiation energy density according to Equation \eqref{eq:RT_fluxlimitediffusion} and afterwards solve for the new temperature via
\begin{equation}
\label{eq:RT_temperatureupdatelinearization}
\frac{T(t+\Delta t)}{T(t)} = 
\frac{
E_\mathrm{int}(t) 
+ 
\chi_\mathrm{abs} ~ \Delta t ~ \left(
\vTWO{
12 \pi
}
~ B_\mathrm{rad}(t) + \vTWO{c} ~ E_\mathrm{rad}(t+\Delta t) \right)
}{
E_\mathrm{int}(t) 
+ 
\chi_\mathrm{abs} ~ \Delta t ~ 
\vTWO{
16 \pi
} 
~ B_\mathrm{rad}(t)
}.
\end{equation}
\vTWO{
This equation includes on the right-hand-side an implicit dependence of the absorption coefficient on the new temperature.
Accordingly, this equation is solved via an iterative Newton--Raphson update.
}
As in the equilibrium approach, this equation is valid only if \vTWO{the} FLD approximation is used to determine the total radiation field.
Changes due to hybrid radiation transport schemes are discussed below.

The drawback of the linearization approach is that the time step $\Delta t$ has to be small enough to assure that the linearization \eqref{eq:linearization} is valid, i.e.~changes in temperature have to be small within a single time step.
Hence, the temperature difference has to be monitored and limited throughout the course of simulations.
For example, in the presence of strong shocks or the direct irradiation of previously shadowed regions, the solution method becomes CPU expensive.

\subsubsection{Hybrid Schemes}
\label{sect:hybridschemes}
In our hybrid radiation transport scheme, as presented in \citet{2010A&A...511A..81K}, we split the total radiation field into an irradiation source and thermal (re)emission.
Such a splitting of the total radiation field into multiple components allows us to use different radiation transport solvers for the different components.
In this way, it is possible to adapt the methods (and their approximations) closely to the properties of the individual radiation field components and through that achieve a good balance between accuracy of the solution and speed of the solver.
The irradiation source, here denoted as 
\vTWO{
$\vec{F}_\mathrm{irr}$,
}
is solved via ray-tracing along the first coordinate axis; see Section \ref{sect:irradiation}.
This component can either be solved in the gray approximation or for multiple frequency bins.
The thermal (re)emission is solved in the gray FLD approximation; see Section \ref{sect:FLD}. 

In the case of a hybrid radiation transport scheme, the absorbed irradiated flux enters the equation of internal energy, previously Equation \eqref{eq:RT_internalenergy}, as a source term:
\begin{equation}
\label{eq:RT_hybridinternalenergy}
\partial_t ~ E_\mathrm{int} 
= 
- \chi_\mathrm{abs} ~ 
\vTWO{
\left(4 \pi ~ B_\mathrm{rad} -  c ~ E_\mathrm{rad} \right)
}
- 
\vec{\nabla} \cdot \vec{F}_\mathrm{irr}.
\end{equation}
Hence, in the equilibrium approach, the modified diffusion Equation \eqref{eq:RT_FLDequilibrium} is given as 
\begin{equation}
\label{eq:RT_hybridFLDequilibrium}
\partial_t ~ E_\mathrm{rad}
-
f_\mathrm{c} ~ \vec{\nabla} \cdot \left( D ~ \vec{\nabla} ~ E_\mathrm{rad} \right)
=
- f_\mathrm{c} ~ \vec{\nabla} \cdot \vec{F}_\mathrm{irr}
\end{equation}
and the equilibrium condition \eqref{eq:RT_temperatureupdateequilibrium} for the temperature update becomes
\begin{equation}
\label{eq:RT_hybridtemperatureupdateequilibrium}
\chi_\mathrm{abs} ~ a ~ T_\mathrm{dust}^4 = \chi_\mathrm{abs} ~ E_\mathrm{rad} + \sum_\nu \chi_\mathrm{\nu} ~ \frac{F_\mathrm{irr,\nu}}{c} \, .
\end{equation}

In the linearization approach, the temperature update, formerly Equation \eqref{eq:RT_temperatureupdatelinearization}, becomes
\begin{equation}
\label{eq:RT_hybridtemperatureupdatelinearization}
\frac{
T(t+\Delta t)}{T(t)} = \frac{E_\mathrm{int} + \xi ~ \left( 3 ~ 
\vTWO{
a (T(t))^4
} 
+ E_\mathrm{rad}(t+\Delta t) \right) 
- 
\Delta t ~ \vec{\nabla} \cdot \vec{F}_\mathrm{irr}
}{
E_\mathrm{int}(t) + 4 ~ \xi ~ 
\vTWO{
a (T(t))^4
}
}
\end{equation}
with the dimensionless abbreviation $\xi = \chi_\mathrm{abs} ~ c ~ \Delta t$.
\vTWO{
Analogous to Equation \eqref{eq:RT_temperatureupdatelinearization}, the equation above is solved via an iterative Newton--Raphson update to take into account the temperature-dependence of the absorption coefficient.
}
The derivation and basic tests of such a hybrid scheme -- splitting the total radiation field into irradiation sources and thermal emission --  was given in \citet{2010A&A...511A..81K}.
In the current code description, we have augmented this algorithm including the linearization approach by \cite{2011A&A...529A..35C}.

In the introductory remark about the validation of the FLD approximation, we pointed out two caveats, namely problems in anisotropic multi-dimensions, where FLD is not capable of modeling shadows, and the underestimation of radiative forces in directly irradiated regions when using the gray FLD approximation.
The hybrid scheme was intentionally introduced in the past to overcome these two caveats: 
the shadow due to the stellar irradiation of the inner disk rim can be very accurately reproduced using the frequency-dependent hybrid scheme in contrast to the FLD approximation alone as shown in \citet{2013A&A...555A...7K}.
An accurate computation of the radiative force within directly irradiated regions turned out to be a crucial necessity to properly model the evolution of low-density bipolar cavities (see \citet{2012A&A...537A.122K} for details and the simulation outcomes in a direct numerical comparison of the hybrid and FLD-only scheme).

In general, the hybrid scheme is independent of the choice of the coordinate system or the geometry of the computational grid.
Recently, this hybrid scheme was also implemented on an AMR grid in Cartesian coordinates \citep{2014ApJ...797....4K}.
Furthermore, the splitting of the total radiation field into multiple components and the use of different solver methods per component can also be applied using other radiation transport methods such as Monte Carlo or M1.

\subsubsection{Radiative Forces}
\paragraph{Physics}
The coupling of radiation transport and MHD includes the radiative force, which acts on the absorbing and emitting medium due to momentum conservation.
The resulting acceleration from the direct irradiative flux can be computed as
\begin{equation}
\label{eq:raytracingacceleration}
\vec{a}_\mathrm{irr} = - \frac{\vec{\nabla} \cdot \vec{F}_\mathrm{irr}}{c ~ \rho_\mathrm{gas}} \vec{\Omega} \; .
\end{equation}

\paragraph{Numerics}
In the literature, radiative acceleration is often specified as $\vec{a}= \kappa ~ \vec{F} / c$ according to \citet{1984frh..book.....M} with the opacity $\kappa = \chi_\mathrm{ext} / \rho_\mathrm{gas}$.
Although this equation is identical to the equation above, its direct numerical implementation has to be handled with care due to spatial discretization.
For a detailed proof of the equality of the two expressions, we refer the interested reader to \citet{2010A&A...511A..81K}.

No problems would occur if optically thick regions are resolved by as many grid cells that all individual grid cells have a local optical depth less than unity.
But problems can arise in the case where individual grid cells in the computational domain are optically thick.
Physically, the maximum acceleration of a specific volume of gas is given by the case where this volume absorbs the full incoming radiative flux.
This upper limit is only accounted for correctly in the discretized version of the equation $\vec{a} = \kappa ~ \vec{F} / c$, if the radiative flux acting on a specific volume or grid cell of gas is computed as the integrated mean value over the volume.

As a thought experiment, we can think of a volume of gas or a grid cell, with an opacity corresponding to an optical depth of $\tau = 10^3$.
The radiative acceleration of this volume is given by the upper limit where all photons are absorbed within the cell and does not change if the opacity and optical depth would become a factor of 10 larger; the number of absorbed photons within the volume is still the same.
Hence, to eliminate the proportionality on opacity in $\vec{a} = \kappa ~ \vec{F} / c$, one has to take into account the fact that the mean radiative flux within the volume scales in the optically thick limit with the inverse of the opacity.
In other words, the full acceleration as given by $\kappa ~ \vec{F} / c$ is acting only on a fraction $1/\tau$ of the optically thick volume, and the remnant part of the volume feels no acceleration at all.

In contrast to this (careful) handling, the discretized version of Equation \eqref{eq:raytracingacceleration} automatically includes the upper limit by computing the difference of incoming and outgoing fluxes as given by the gradient of radiative flux.
Note that the attenuation of the direct irradiated flux given by $\vec{\nabla} \cdot \vec{F}_\mathrm{irr}$ becomes part of the source of the diffuse radiation field as discussed in Sect.~\ref{sect:hybridschemes}.

The total radiative acceleration $\vec{a}_\mathrm{ext}^{rad}$ must also include the contribution of the diffuse radiation field. To calculate this contribution, we apply $\vec{a}_\mathrm{diff} = \kappa ~ \vec{F}_\mathrm{rad} / c$ together with our definition of $\kappa$ to Eqs.~\eqref{eq:diffusioncoefficient} and \eqref{eq:fluxlimiteddiffusionapproximation}:
\begin{equation}
\label{eq:radiativeacceleration}
\vec{a}_\mathrm{ext}^{rad} = 
- \frac{\vec{\nabla} \cdot \vec{F}_\mathrm{irr}}{c ~ \rho_\mathrm{gas}} \vec{\Omega} 
- \frac{\lambda}{\rho_\mathrm{gas}} \vec{\nabla} E_\mathrm{rad} \; .
\end{equation}

%
%
\structuring{\clearpage}
\subsection{Photoionization}
\label{sect:ionization}
\begin{figure*}[htbp]
\centering
\includegraphics[width=0.99\textwidth]{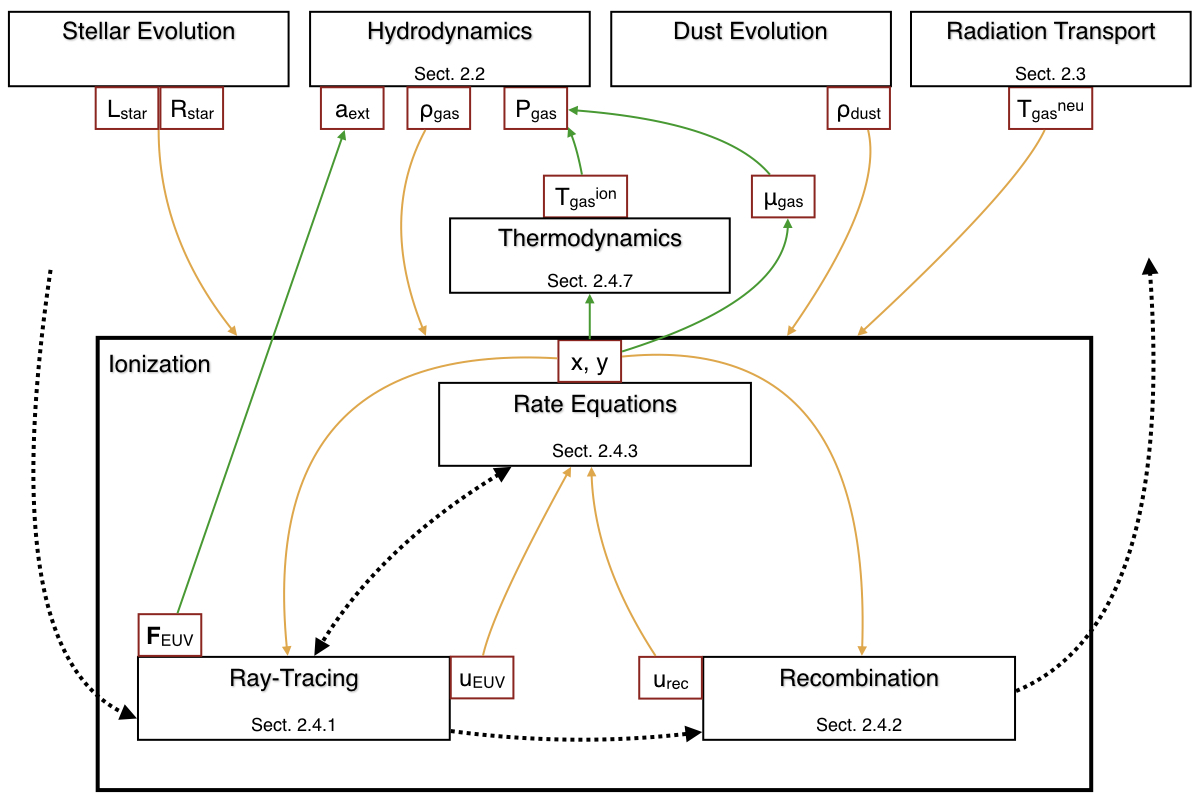}
\caption{
Flowchart of the photoionization module Sedna.
The legend is the same as in Fig.~\ref{fig:flowbelt}.
}
\label{fig:flowsedna}
\end{figure*}

As a completely new ingredient, a hydrogen photoioni\-za\-tion module, Sedna (Figure \ref{fig:flowsedna}), is included in this multiphysics framework.
If the ionization module is switched on, the spectrum of the direct radiation field from the central source is divided into two spectral regimes, one with a photon energy lower than 13.6~eV and one with a higher photon energy.
The part of the spectrum with higher energy is -- in addition to the continuum absorption of the radiation transport step described above -- handled by the ionization module.
Furthermore, we do not only ray-trace the high-frequency part of the central object's emission spectrum, but also compute for the diffuse EUV radiation field created in (partly) ionized regions due to direct recombination of free electrons into the hydrogen's ground state.
The physical description and the basic numerical implementation of both of these components follow the description by \citet{1996A&A...315..555Y} and \cite{1997A&A...327..317R} but makes use of modern state-of-the-art numerical solvers and algorithms.

In addition to photon momentum absorption, ionization couples to the hydrodynamics through the gas pressure due to its effect on both the temperature $T_\mathrm{gas}$ and molecular mass $\mu_\mathrm{gas}$ of the gas. $\mu_\mathrm{gas}$ depends on the degree of ionization, 
\begin{equation}
x = \frac{n_\mathrm{H^+}}{n_\mathrm{H^0} + n_\mathrm{H^+}}
\end{equation}
and neutral fraction, 
\begin{equation}
y = \frac{n_\mathrm{H^0}}{n_\mathrm{H^0} + n_\mathrm{H^+}} = 1 - x, 
\end{equation}
which are determined in the rate equation solver of the ionization module.
$n_\mathrm{H^0}$ and $n_\mathrm{H^+}$ denote the neutral and ionized hydrogen number densities. For simplicity, we do not include molecular hydrogen. A more rigorous treatment would include the formation and destruction of molecular hydrogen via Lyman--Werner-band photons and its associated contribution to $\mu_\mathrm{gas}$.

Details of the rate equation solver for ionization/recombination are given in Sect.~\ref{sect:rateequations}.
Solving the rate equations requires knowledge of the local ionizing radiation flux.
The determination of those is described in Sections \ref{sect:directionization} and \ref{sect:recombination} for the ray-tracing and diffuse flux, respectively.
Medium properties appropriate for present-day star formation regions are presented in Sect.~\ref{sect:materialproperties}.

\subsubsection{Ray-tracing of Ionizing Radiation}
\label{sect:directionization}
Ray-tracing of the direct radiative EUV flux with photon energy $h \nu \ge 13.6 \mbox{ eV}$ of a specified spectrum is performed along the first coordinate axis of the computational domain.
In principle, this routine can be used in Cartesian coordinates to model, e.g., plane-parallel atmospheres, the routine works also for any user-defined irradiation spectrum.
Nonetheless, we focus in the following description on a grid in spherical coordinates and ray-tracing of a given radiation field denoted as $F_\mathrm{EUV}$.
Analogous to the ray-tracing 
\vTWO{
solution
Equation \eqref{eq:radfluxsol}
}
described in Sect.~\ref{sect:irradiation}, the radiative EUV flux is given as a function of radius as
\begin{equation}
F_\mathrm{EUV}(r) = F_\mathrm{EUV}(r_\mathrm{min}) ~ \left(\frac{r_\mathrm{min}}{r}\right)^2 ~ \exp(-\tau_\mathrm{tot}),
\end{equation}
where $F_\mathrm{EUV}(r_\mathrm{min})$ denotes the initial flux of ionizing radiation at the minimum radius of the computational domain.
The factor $\left(r_\mathrm{min}/r\right)^2$ denotes the geometrical attenuation of the flux due to the increase in the traversed area for
\vTWO{
an isotropic
}
source at the origin of a grid in spherical coordinates (i.e., this factor depends on the geometry of the 
\vTWO{
source}
).
The factor $\exp(-\tau_\mathrm{tot})$ denotes the total extinction of the radiative flux along the path.
The total optical depth $\tau_\mathrm{tot} = \tau_\mathrm{ion} + \tau_\mathrm{ext}$ is the sum of the optical depth due to ionization of hydrogen and continuum extinction by dust grains and gas.
The optical depth for the hydrogen-ionizing flux is given as
\begin{equation}
\tau_\mathrm{ion}(r) = \int_{r_\mathrm{min}}^r n_\mathrm{H} ~ y ~ \sigma_\mathrm{EUV} ~ \mbox{d}r
\end{equation}
with the total hydrogen number density $n_\mathrm{H} = n_\mathrm{H^0} + n_\mathrm{H^+}$ and
the photon cross section $\sigma_\mathrm{EUV}$ of the medium with respect to ionizing radiation from the ray-tracing source.

The optical depth $\tau_\mathrm{ext}$ due to continuum extinction of the EUV photons is described above in the radiation transport module in Equation \eqref{eq:opticaldepth} and is given as
\begin{equation}
\tau_\mathrm{ext}(r) = \int_{r_\mathrm{min}}^r \chi_\mathrm{ext} ~ \mbox{d}r.
\end{equation}
In practice, the ray-tracing of the EUV radiative flux is not split into radiation and ionization modules, but is done only once while simultaneously computing the heating of dust grains (formally belonging to the radiation transport module) and the ionization of the hydrogen gas (formally belonging to the ionization module).
That is, the two modules share the ray-tracing routine for the EUV part of the specified spectrum.

The optical depth $\tau_\mathrm{ion}$ depends on the ionization fraction $x = 1 - y$ of the medium along the ray path.
The ionization fraction is solved at each location via rate equations, which depend in turn on the local ionizing radiation field.
Hence, the rate equations have to be solved locally while simultaneously solving for the ionizing radiative flux along the ray.
Furthermore, the implicit dependence of the flux on $x$ and vice versa requires a Newton--Raphson iterative update of these quantities in each grid cell during the advance of the ray-tracing.

The EUV flux $F_\mathrm{EUV}$ of the ray-tracing step enters the rate equation solver in terms of the EUV photon number density $u_\mathrm{EUV}$, which are related to each other via
\begin{equation}
u_\mathrm{EUV} = \frac{1}{\langle h \nu \rangle_\mathrm{EUV}} ~ \frac{F_\mathrm{EUV}}{c}
\end{equation}
with the mean photon energy $\langle h \nu \rangle_\mathrm{EUV}$ of the ray-traced spectrum.

\subsubsection{Diffuse Ionizing EUV Flux from Direct Recombination of Free Electrons into the Ground State of Hydrogen Atoms}
\label{sect:recombination}
Within a (partly) ionized region, free electrons will recombine directly into the hydrogen's ground state.
This process is accompanied by spontaneous emission of an EUV photon.
We solve for the further ionization of gas by these photons by following the evolution of the recombination photon number density $u_\mathrm{rec}$ of this diffuse ionizing radiation field.
For an alternative method, the so-called on-the-spot approximation, please see Sect.~\ref{sect:onthespot}.
The temporal evolution of the diffuse EUV radiation field is governed by the conservation equation:
\begin{equation}
\label{eq:recombination}
\partial_t ~ u_\mathrm{rec}
+ 
\frac{\vec{\nabla} \cdot \vec{F}_\mathrm{rec}}{\langle h \nu \rangle_\mathrm{rec}}
=
+
\alpha_1(T_\mathrm{gas}) ~ n_\mathrm{H}^2 ~ x^2
- 
\chi_\mathrm{rec} ~ u_\mathrm{rec} ~ c
\end{equation}
with the flux of the recombination radiative energy density $\vec{F}_\mathrm{rec}$,
the recombination rate $\alpha_1(T_\mathrm{gas})$ of free electrons directly into the ground state of hydrogen atoms, and
the recombination absorption coefficient $\chi_\mathrm{rec}$, which denotes the inverse of the mean free path of the recombination EUV photons and is accordingly determined as
\begin{equation}
\chi_\mathrm{rec} 
= 
n_\mathrm{H} ~ y ~ \sigma_\mathrm{rec} 
+ 
\chi_\mathrm{ext},
\end{equation}
with the continuum extinction coefficient $\chi_\mathrm{ext}$ introduced in the ray-tracing description of the nonionizing radiation transport module (Sect.~\ref{sect:irradiation}).

For the sake of computation, we make use of the FLD approximation for the diffuse recombination EUV radiation field, i.e.~we assume that the flux is proportional to the gradient of the recombination photon number density,
\begin{equation}
\vec{F}_\mathrm{rec} = - \langle h \nu \rangle_\mathrm{rec} ~ D_\mathrm{rec} ~ \vec{\nabla} u_\mathrm{rec},
\end{equation}
with the diffusion coefficient 
\begin{equation}
D_\mathrm{rec} = \frac{\lambda_\mathrm{rec} ~ c}{\chi_\mathrm{rec}}.
\end{equation}
The flux limiter $\lambda_\mathrm{rec}$ is set according to \citet{1981ApJ...248..321L} to
\begin{equation}
\lambda_\mathrm{rec} = \frac{2 + R_\mathrm{rec}}{6 + 3 R_\mathrm{rec} + R_\mathrm{rec}^2}
\end{equation}
with
\begin{equation}
R_\mathrm{rec} = \frac{| \vec{\nabla} u_\mathrm{rec} |}{\chi_\mathrm{rec} ~ u_\mathrm{rec}}.
\end{equation}
The recombination photon number density $u_\mathrm{rec}$ that is finally solved enters the rate equations solver, described in the following section, as a source of ionizing photons.

\structuring{\clearpage}
\subsubsection{Rate Equations for Ionization--Neutral Fraction}
\label{sect:rateequations}
The temporal evolutions of the ionization fraction $x$ and neutral fraction $y$ are given by
\begin{eqnarray}
\partial_t ~ (\rho_\mathrm{gas} ~ x) 
&+& 
\vec{\nabla} \cdot \left(\rho_\mathrm{gas} ~ x ~ \vec{u}_\mathrm{gas} \right) 
=\nonumber \\
&+& 
\rho_\mathrm{gas} ~ y ~ \left( \sigma_\mathrm{EUV} ~ u_\mathrm{EUV} + \sigma_\mathrm{rec} ~ u_\mathrm{rec} \right) ~ c  \nonumber \\
&+&
\rho_\mathrm{gas} ~ C(T_\mathrm{gas}) ~ n_\mathrm{H} ~ x ~ y  \nonumber \\
&-& 
\rho_\mathrm{gas} ~ \alpha^{(1)}(T_\mathrm{gas}) ~ n_\mathrm{H} ~ x^2 
\label{eq:ionizationfraction}
\\
\partial_t ~ (\rho_\mathrm{gas} ~ y) 
&+& 
\vec{\nabla} \cdot \left(\rho_\mathrm{gas} ~ y ~ \vec{u}_\mathrm{gas} \right) 
=\nonumber \\
&-& 
\rho_\mathrm{gas} ~ y ~ \left( \sigma_\mathrm{EUV} ~ u_\mathrm{EUV} + \sigma_\mathrm{rec} ~ u_\mathrm{rec} \right) ~ c  \nonumber \\
&-&
\rho_\mathrm{gas} ~ C(T_\mathrm{gas}) ~ n_\mathrm{H} ~ x ~ y  \nonumber \\
&+& 
\rho_\mathrm{gas} ~ \alpha^{(1)}(T_\mathrm{gas}) ~ n_\mathrm{H} ~ x^2
\label{eq:neutralfraction}
\end{eqnarray}
with the recombination rate $\alpha^{(1)}(T_\mathrm{gas})$ of free electrons into any of the states of the hydrogen atoms and 
the collisional excitation coefficient $C(T_\mathrm{gas})$.

These equations describe the change of ionization and neutral fraction with time (first term on the left-hand side) due to
hydrodynamic advection (second term on the left-hand side),
radiative ionization (first term on the right-hand side),
collisional excitation (second term on the right-hand side), and
recombination (third term on the right-hand side).

In principle, one of the two equations above is redundant, due to the fact that the ionization fraction $x$ and neutral fraction $y$ have to sum up to unity, $x + y = 1$.
That is why the rate equations above are identical to each other with switched signs of the source terms on the right-hand side (the source of one quantity is the sink of the other and vice versa).
But the numerical floating point operation to compute, e.g., the neutral fraction as $y = 1 - x$ implicates a severe loss of significance for $x \approx 1$.
Hence, to accurately solve for very small neutral fractions in highly ionized regions as well as for very small ionization fractions in highly neutral regions, we solve both evolution equations given above.
The additional constraint of $x + y = 1$ is actually used as an automatic internal solver check.

For Newtonian gas flows, radiative ionization, collisional excitation, and recombination are \vTWO{commonly} much faster processes than the hydrodynamic advection.
Because all terms on the right-hand side are greater than the advection term by many orders of magnitude, we ignore the advection term in these equations for simplicity.
For the same reason, the remaining terms have to be solved numerically in an implicit fashion, i.e.~the time derivative is discretized via 
$\partial_t ~ x(t) \rightarrow \left( x^{n+1} - x^n \right) / \Delta t$ with $\Delta t = \left( t^{n+1} - t^{n} \right)$
and the ionization and neutral fractions on the right-hand side have to be evaluated at an advanced point in time $t^{n+1}$.

Thus, the discretized representation of equation \eqref{eq:ionizationfraction} leads to a standard quadratic equation of the form
\begin{equation}
A ~ \left( x^{n+1} \right)^2 + B ~ x^{n+1} + C = 0
\end{equation}
with
\begin{eqnarray}
A &=& b + c \\
B &=& 1 + a - b \\
C &=& -x^n - a
\end{eqnarray}
with
\begin{eqnarray}
a &=& c ~ \Delta t ~ (\sigma_\mathrm{EUV} ~ u_\mathrm{EUV} + \sigma_\mathrm{rec} ~ u_\mathrm{rec}) \\
b &=& n_\mathrm{H} ~ \Delta t ~ C(T_\mathrm{gas}) \\
c &=& n_\mathrm{H} ~ \Delta t ~ \alpha_1(T_\mathrm{gas}).
\end{eqnarray}

Analogously, the discretized representation of equation \eqref{eq:neutralfraction} is given as
\begin{equation}
A ~ \left( y^{n+1} \right)^2 + B ~ y^{n+1} + C = 0
\end{equation}
with
\begin{eqnarray}
A &=& b + c \\
B &=& - 1 - a - b - 2c \\
C &=& y^n + c.
\end{eqnarray}

\subsubsection{Diffuse EUV versus On-the-spot Approximation}
\label{sect:onthespot}
In at least partly ionized regions, free electrons will undergo direct recombination into the ground state of hydrogen and by that release an EUV photon, capable of ionizing another hydrogen atom.
We solve for the ionization feature of this diffuse-like EUV radiation field by utilizing the FLD approximation as presented in Sect.~\ref{sect:recombination}.
Another common approach is the so-called on-the-spot approximation.
By that, the released EUV photons are assumed to immediately ionize an atom locally. 
Hence, there is no need to compute the evolution of this radiation field in time.
By definition, this approximation is only valid for regions that are optically thick for EUV photons, but this is rarely the case for regions of high ionization fraction, where the diffuse EUV field has its most important source.
But the on-the-spot approximation also works fine for one-dimensional problems such as the R-type or D-type expansion of a spherically symmetric H II region around a luminous star; see e.g.~\citet{2015MNRAS.453.1324B}. 
The reason for the validity here lies in the fact that the H II region denotes a region of very high ionization fraction, which drops to a nearly neutral ionization fraction on a comparably small length scale.
Hence, the diffuse EUV photons, which are created inside the sphere, will anyway only contribute to the total ionization fraction inside the H II region, and on average, the exact location of their contribution can be disregarded.
Therefore, assuming these diffuse photons ionize the medium locally does not, e.g., change the expansion rate of the global H II region.

We have implemented the on-the-spot approximation into our numerical star formation framework in addition to the time- and space-dependent evolution equations for the diffuse field.
To account for the on-the-spot approximation, the equations presented in the previous sections have to be modified accordingly;
in Equations \eqref{eq:ionizationfraction} and \eqref{eq:neutralfraction}, the recombination rate $\alpha^{(1)}(T_\mathrm{gas})$ into any state of the hydrogen atom is substituted by the recombination rate $\alpha^{(2)}(T_\mathrm{gas})$ into any state of the hydrogen atom besides the ground state: 
$\alpha^{(1)}(T_\mathrm{gas})\rightarrow \alpha^{(2)}(T_\mathrm{gas})$.
Through this, the rate equation solver automatically includes the approximation that direct recombination into the ground state leads to a local ionization event again.
The recombination photon number density $u_\mathrm{rec}$ is not solved for anymore, and in the rate equations solver, it is just replaced by $u_\mathrm{rec} = 0$.

We note that for multidimensional cases, when sharp shadows on the direct EUV field are cast, the more generalized treatment of the diffuse EUV recombination photons should be used to determine the ionizing radiation in the shadowed regions.

\subsubsection{Ionizing Radiation Forces}
The coupling of ionizing radiation transport and MHD includes -- along with changes of the gas components and their thermodynamics -- radiative forces, which act on the absorbing medium due to momentum 
\vTWO{transfer.}
The resulting acceleration from the direct irradiation and diffuse EUV radiation field is analogous to the nonionizing radiation (see Equation \eqref{eq:radiativeacceleration}) given as
\begin{equation}
\label{eq:ionizationacceleration}
\vec{a}_\mathrm{ext}^{ion} = 
- \frac{\vec{\nabla} \cdot \vec{F}_\mathrm{EUV}}{c ~ \rho_\mathrm{gas}} \vec{\Omega} 
- \frac{\lambda_\mathrm{rec}}{\rho_\mathrm{gas}} \vTWO{~ \langle h \nu \rangle_\mathrm{rec} ~} \vec{\nabla} u_\mathrm{rec}.
\end{equation}

\structuring{\clearpage}
\subsubsection{Ionization-related Properties of Gaseous Media and Stellar Photospheres}
\label{sect:materialproperties}
The ray-tracing equations of ionizing photons, diffuse ionizing photons from recombinations, and the rate equations for ionization and neutral fraction involve several material properties, which have to be derived from laboratory experiments or theoretical models of the underlying microphysics.
In this section, we present the relations implemented for these material properties as well as their origin or derivations.

\paragraph{Stellar atmospheres}
In star forming regions, high-mass luminous stars denote a major source of ionizing radiation.
Commonly, stellar evolution is therefore solved simultaneously with radiation (magneto)hydrodynamics.
These solvers return the stellar luminosity $L_\mathrm{star}$, its effective temperature $T_\mathrm{star}$, and radius $R_\mathrm{star}$ at each point in time during the course of the simulation.
If one would approximate the star's spectrum by a blackbody, the number of ionizing photons from the stellar photosphere \vONE{per unit time} would be given as 

\begin{eqnarray}
\label{eq:photonrateBB}
N_\mathrm{BB}(T_\mathrm{star}) 
&=& 4\pi ~ R_\mathrm{star}^2 ~ \int_{\nu_\mathrm{L}}^\infty \frac{\vTWO{\pi} ~ B_\nu(\nu, T_\mathrm{star})}{h \nu} ~ \mbox{d}\nu \\
&=& 4\pi ~ R_\mathrm{star}^2 ~ \int_{\nu_\mathrm{L}}^\infty \frac{\nu^2}{c^2} \frac{\vTWO{2 \pi}}{\exp(h \nu / k_\mathrm{B} T_\mathrm{star}) - 1} ~ \mbox{d}\nu.
\end{eqnarray}
The factor $\vTWO{\pi}$ in front of the Planck spectrum results from the fact that $\vTWO{B(T_\mathrm{star}) =} \int B_\nu ~ \mbox{d}\nu = \sigma_\mathrm{SB} ~ T_\mathrm{star}^4 \vTWO{~/~\pi}$ and we are interested in the flux of radiation energy density 
\vTWO{
$F = \pi ~ B(T_\mathrm{star})$
}
\vONE{
with the Stefan--Boltzmann constant $\sigma_\mathrm{SB}$
\vTWO{
see, e.g., \citet{1984frh..book.....M}.
}remit
}

Actually, a hot star emits fewer EUV photons into its surroundings than given by the blackbody spectrum due to the fact that the stellar atmosphere absorbs a fraction of the generated EUV photons and remits them at lower frequencies (UV-line blanketing).
To appropriately account for this effect, we implemented a stellar atmosphere model based on \citet{1979ApJS...40....1K}.
We compute the number of emitted EUV photons \vONE{per unit time} $N_\mathrm{L}$ from a luminous star as a correction of its blackbody emission,
\begin{equation}
N_\mathrm{L}(T_\mathrm{star}) = f(T_\mathrm{star}) \times N_\mathrm{BB}(T_\mathrm{star})
\end{equation}
and determine the correction function $f(T_\mathrm{star})$ via analytical polynomial fits from the tabulated data of stellar-generated EUV photons given in \citet{1984ApJ...283..165T}; 
the tabulated data are also reprinted in the book by \cite{1988rmgm.book.....K}, Table 3-3 on p.~241.

\begin{figure}[htbp]
\centering
\includegraphics[width=0.49\textwidth]{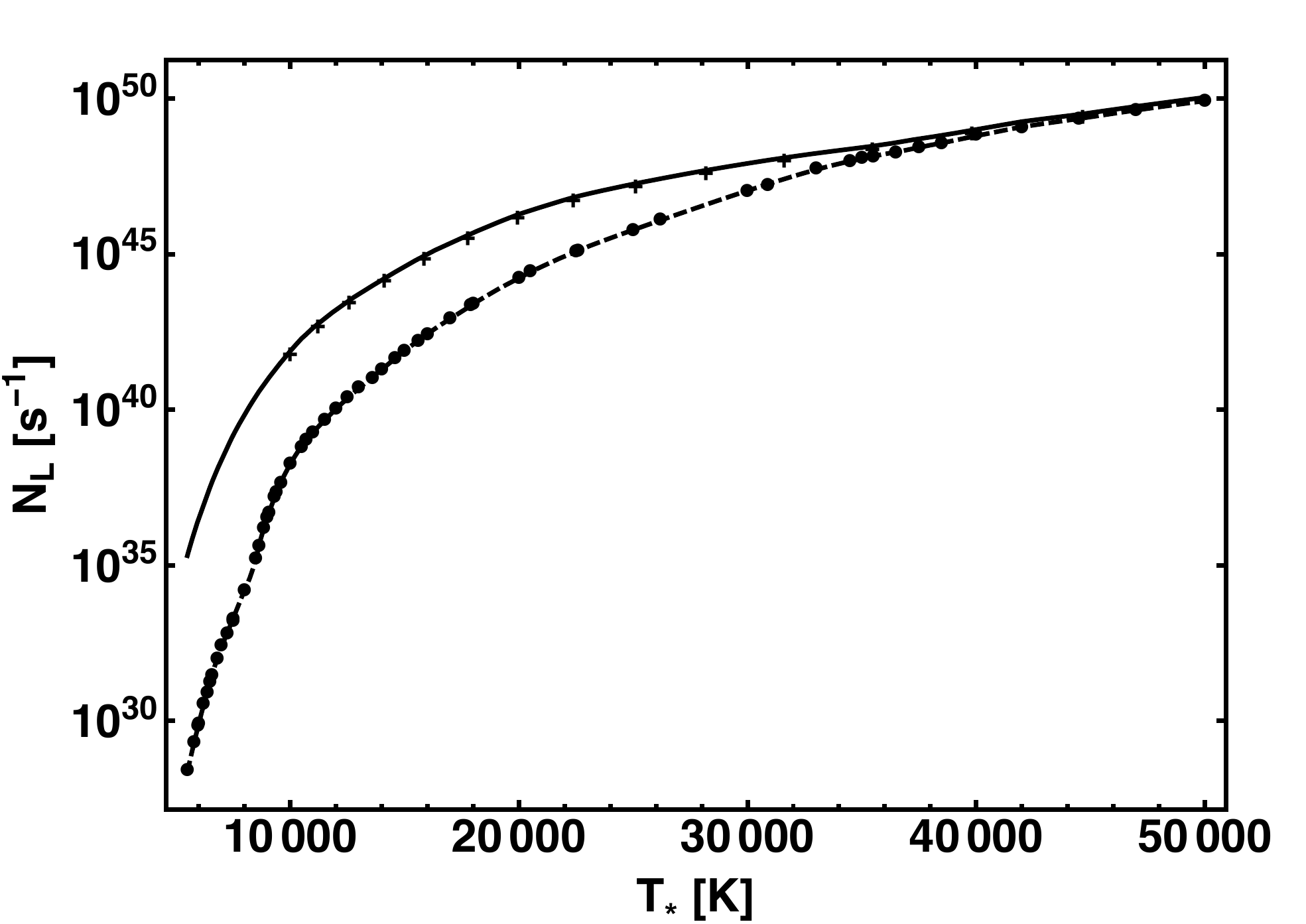}
\caption{
EUV photon emission rate $N_\mathrm{L}$ as function of stellar effective temperature $T_\mathrm{star}$ for 
a blackbody (solid line) and 
a \citet{1979ApJS...40....1K} atmosphere model (dashed line).
Values of the blackbody curve are computed via Equation \eqref{eq:photonrateBB}.
Values for the atmosphere model represent polynomial fits to the tabulated data of \citet{1984ApJ...283..165T}, including their dependence of the stellar radius on temperature.
The original tabulated data is shown as crosses and dots.
}
\label{fig:EUVphotonrate}
\end{figure}

The resulting EUV photon generation rates are shown in comparison to the original data in Fig.~\ref{fig:EUVphotonrate}.
The deviation of the polynomial fits to the tabulated data remains smaller than 10\% for cooler stars $T_\mathrm{star} \approx 10,000 \mbox{ K}$ and decreases toward larger temperature to less than 2\%.
For comparison, such a difference in the number of generated EUV photons would change the extent of a classical Str\"omgren sphere by 0.7\% and 3\%, respectively.

\begin{figure}[htbp]
\centering
\includegraphics[width=0.49\textwidth]{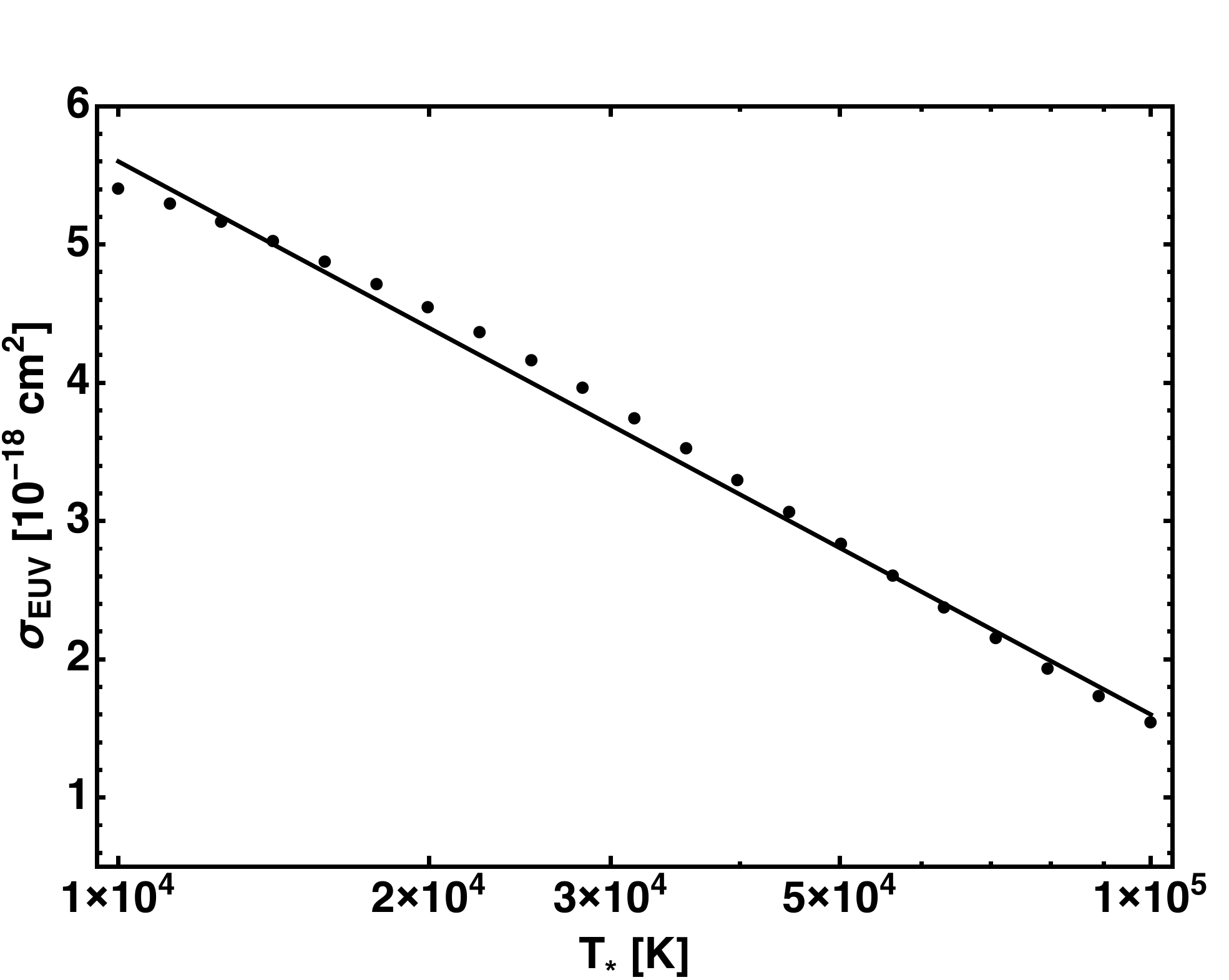}
\caption{
Hydrogen ionization cross section $\sigma_\mathrm{EUV}$ as function of stellar temperature $T_\mathrm{star}$.
Dots represent the original data by Spitzer, while the solid line denotes an analytical fit.
}
\label{fig:sigmastar}
\end{figure}

\paragraph{Stellar irradiation}
For the hydrogen ionization cross section of the gaseous medium with respect to stellar photospheric photons, we use as well an analytical fit function to the original data from \cite{1988rmgm.book.....K}, Table 3-4 on p.~243.
The data are based on solar abundances and a \citet{1979ApJS...40....1K} atmosphere model.
The approximated relation is chosen to be
\begin{equation}
\sigma_\mathrm{EUV} = \left( 21.6 - 4 ~ \log(T_\mathrm{star}/\mbox{K}) \right) \times 10^{-18} \mbox{ cm}^2.
\end{equation}
A comparison of the original data with the approximate analytical relation is presented in Fig.~\ref{fig:sigmastar}.
The analytical relation gives a reasonable fit to the original data with a maximum deviation of 5\% at $T \approx 10^4 \mbox { K}$, 4\% at $T \approx 3 \times 10^4 \mbox { K}$, and less than 2\% for higher temperatures.

\vONE{
In this approach, we treat the EUV frequency range in a single bin.
The implemented ray-tracing scheme supports multiple frequency bins.
If multiple frequency bins should be used within the EUV range, the hydrogen ionization cross section of the gaseous medium should be updated to a frequency-dependent table.
}

\paragraph{Recombination}
A recombination event is given by
\begin{equation}
\mbox{e}^- + \mbox{H}^+ \rightarrow \mbox{H}^0 + h \nu
\end{equation}
For the determination of the recombination rates, we follow the derivation by \citet{1978ppim.book.....S}.
The recombination rate of free electrons into an energy level $n$ is denoted by $\alpha_n$,
the recombination rate into any level $\ge n$ by $\alpha^{(n)}$.
For the rate equation solver of the hydrogen ionization module, we require knowledge of the recombination rates into all possible levels $\alpha^{(1)}$.
For the diffuse EUV recombination field solver, we need the recombination into the ground state only, $\alpha_1$.
And if as an alternative approach the on-the-spot approximation is used, the only required recombination rate is the one into all possible levels except hydrogen's ground state $\alpha^{(2)}$.

By definition,
\begin{equation}
\alpha^{(n)} = \sum_n^\infty \alpha_n.
\end{equation}
We actually use the relation above to compute $\alpha^{(2)}$ as
\begin{equation}
\alpha^{(2)} = \alpha^{(1)} - \alpha_1.
\end{equation}
Hence, only values for $\alpha^{(1)}$ and $\alpha_1$ have to be determined.
Following \citet{1978ppim.book.....S}, we first relate the recombination rates $\alpha^{(n)}$ to the so-called recombination coefficient functions $\Phi_n$, which are dimensionless temperature-dependent functions.
The relationship is defined as
\begin{equation}
\alpha_{(n)}(T_\mathrm{gas}) = 2 ~ A_r ~ \sqrt{\frac{2 ~ k_\mathrm{B} ~ T_\mathrm{gas}}{\pi ~ m_e}} ~ \beta ~ \Phi_n(\beta)
\end{equation}
with the energy ratio $\beta = h ~ \nu_l / (k_\mathrm{B} ~ T_\mathrm{gas})$,
the reference frequency $\nu_l = Z^2 ~c ~ R_\infty$,
the atomic number $Z$ ($Z = 1$ for hydrogen),
the Rydberg constant $R_\infty = \alpha_\mathrm{fs}^2 ~ m_e ~ c / (2 ~ h)$,
the dimensionless fine-structure constant $\alpha_\mathrm{fs}$,
the electron mass $m_e$,
the Boltzmann constant $k_\mathrm{B}$,
the so-called recapture constant $A_r = 2^4 / 3^{3/2} \times h ~ e^2 / (m_e^2 ~ c^3)$, and
the elementary charge $e$.

Evaluating the zoo of physical constants eventually yields
\begin{equation}
\alpha_{(n)}(T_\mathrm{gas}) \approx \xi ~ \frac{Z^2}{\sqrt{T_\mathrm{gas}}} ~  \Phi_n(T_\mathrm{gas})
\end{equation}
with $\xi = 2.065 \times 10^{-11} \mbox{ cm}^3 \mbox{ s}^{-1} \sqrt{\mbox{K}}$.

\begin{figure}[htbp]
\centering
\includegraphics[width=0.465\textwidth]{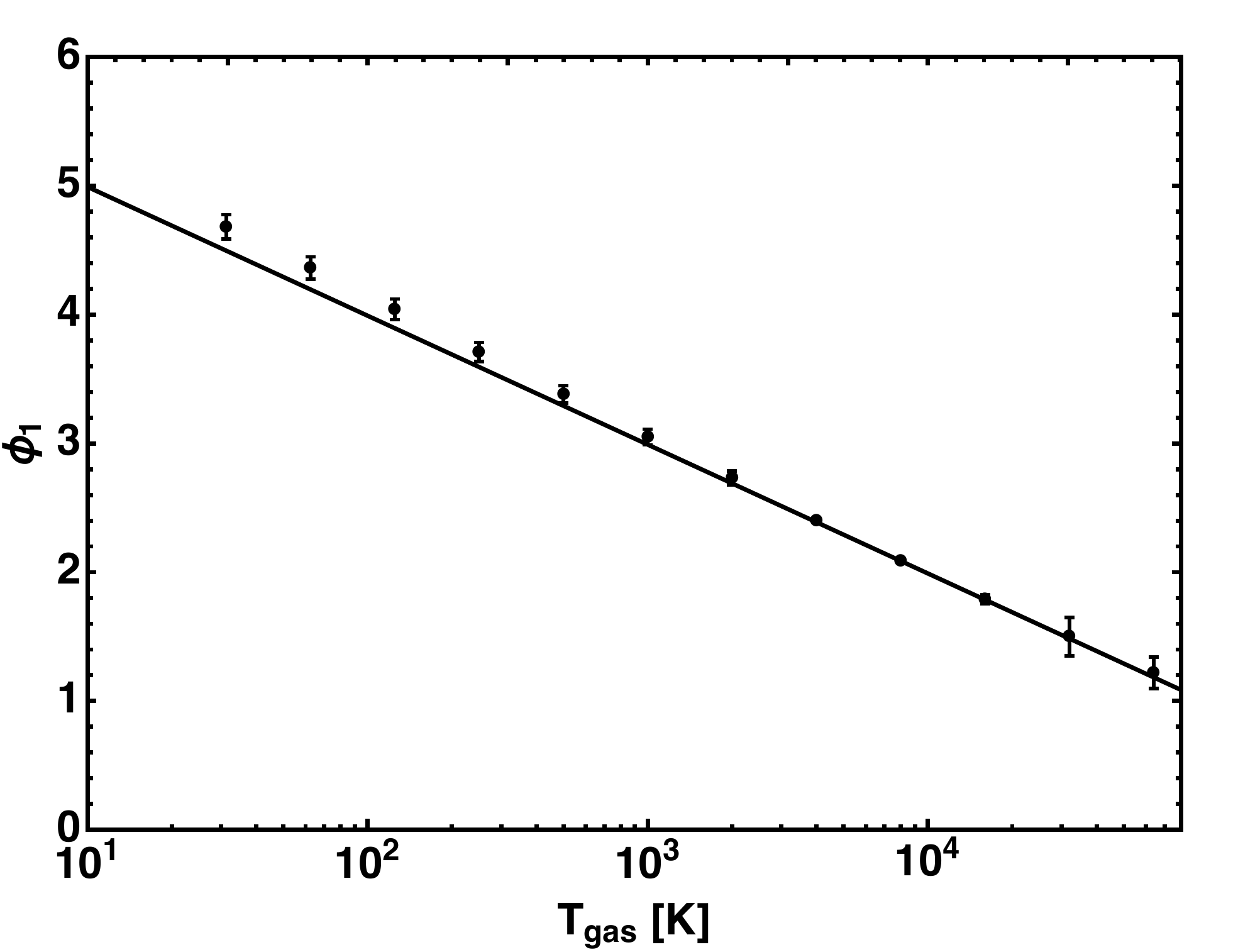}\\
\hspace{-4mm}\includegraphics[width=0.49\textwidth]{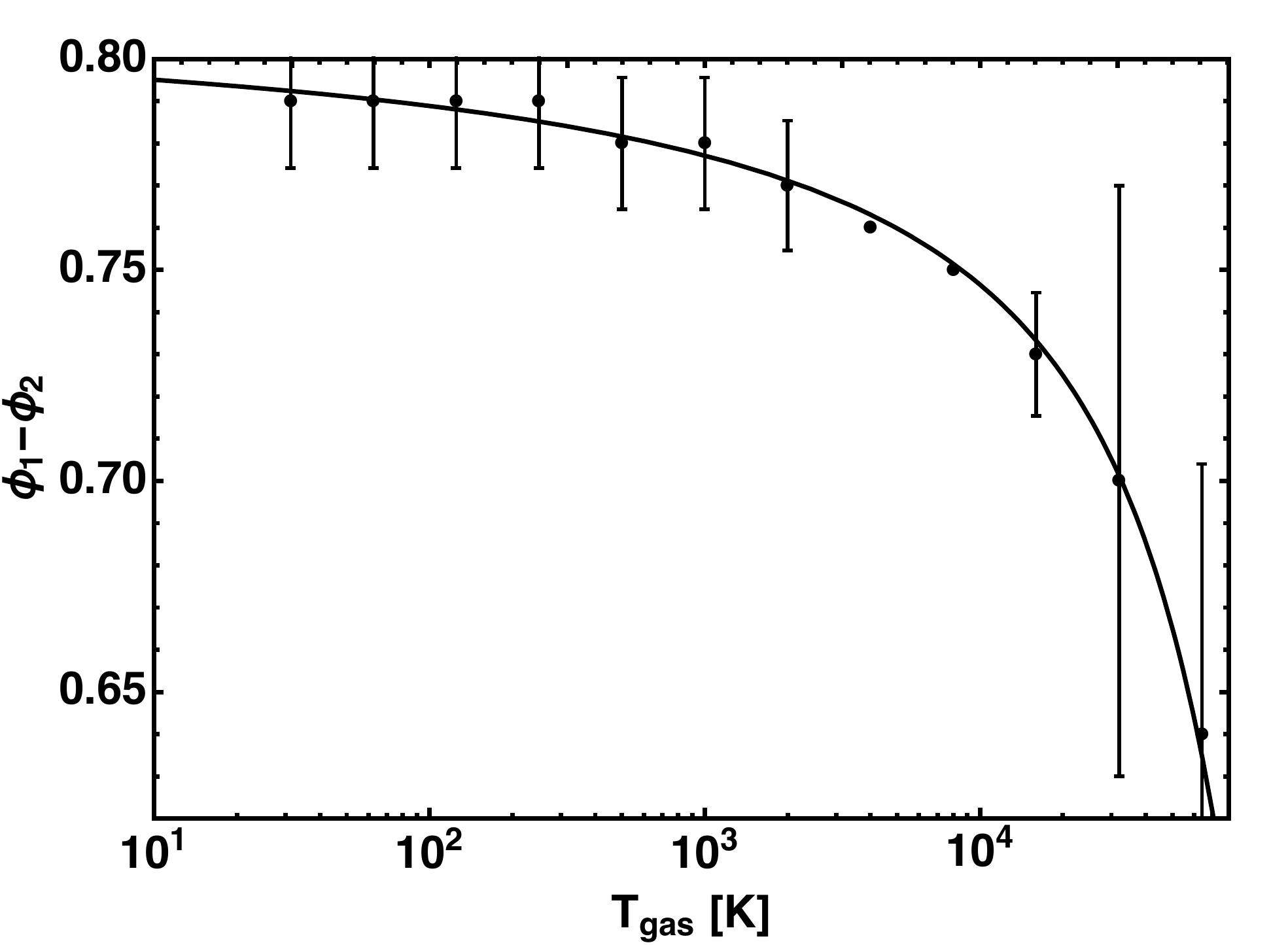}
\caption{
Recombination functions $\Phi_1$ (upper panel) and $(\Phi_1 - \Phi_2)$ (lower panel) as a function of gas temperature $T_\mathrm{gas}$.
Dots represent the original data by \citet{1978ppim.book.....S}, while the solid line denotes our analytical fit.
The error bars correspond to the error estimate given by \citet{1978ppim.book.....S}.
}
\label{fig:recombinationfunctions}
\end{figure}

The two required recombination rates for the ionization module are related to the dimensionless recombination coefficient functions $\Phi_1$ and $\Phi_2$ only:
\begin{eqnarray}
\alpha^{(1)}(T_\mathrm{gas}) &=& \xi ~ \frac{Z^2}{\sqrt{T_\mathrm{gas}}} ~ \Phi_1(T_\mathrm{gas}) \\
\alpha_1(T_\mathrm{gas}) &=& \xi ~ \frac{Z^2}{\sqrt{T_\mathrm{gas}}} ~  \left(\Phi_1(T_\mathrm{gas}) - \Phi_2(T_\mathrm{gas}) \right).
\end{eqnarray}
The functions $\Phi_1$ and $\Phi_2$ are tabulated in \citet{1978ppim.book.....S} as a function of gas temperature.
Here, we provide convenient analytical fit functions to the original tabulated data in form of 
\begin{equation}
\Phi_1 \approx 5.99 - \log(T_\mathrm{gas} / \mbox{K})
\end{equation}
and
\begin{equation}
\Phi_1 - \Phi_2 \approx - 0.815 ~\left(\frac{\arctan(3.1 \log(T_\mathrm{gas} / \mbox{K}) - 16.1)}{\pi} - 0.5 \right).
\end{equation}
The original tabulated data sets as well as the fit functions are presented in Fig.~\ref{fig:recombinationfunctions}.
The analytical expressions resemble the original data values within reasonable accuracy.
The $\Phi_1 - \Phi_2$ data points are within $0.5\%$ of the fitting curve.
The $\Phi_1$ data points are within $2\%$ of the fitting curve for $T \ge 10^3 \mbox{ K}$ and within $4\%$ for lower temperatures.
The error estimate for the original data given by \citet{1978ppim.book.....S} denotes $2\%$ for $T \le 16,000 \mbox{ K}$ and $< 10\%$ for higher temperatures.

\begin{figure}[htbp]
\centering
\includegraphics[width=0.455\textwidth]{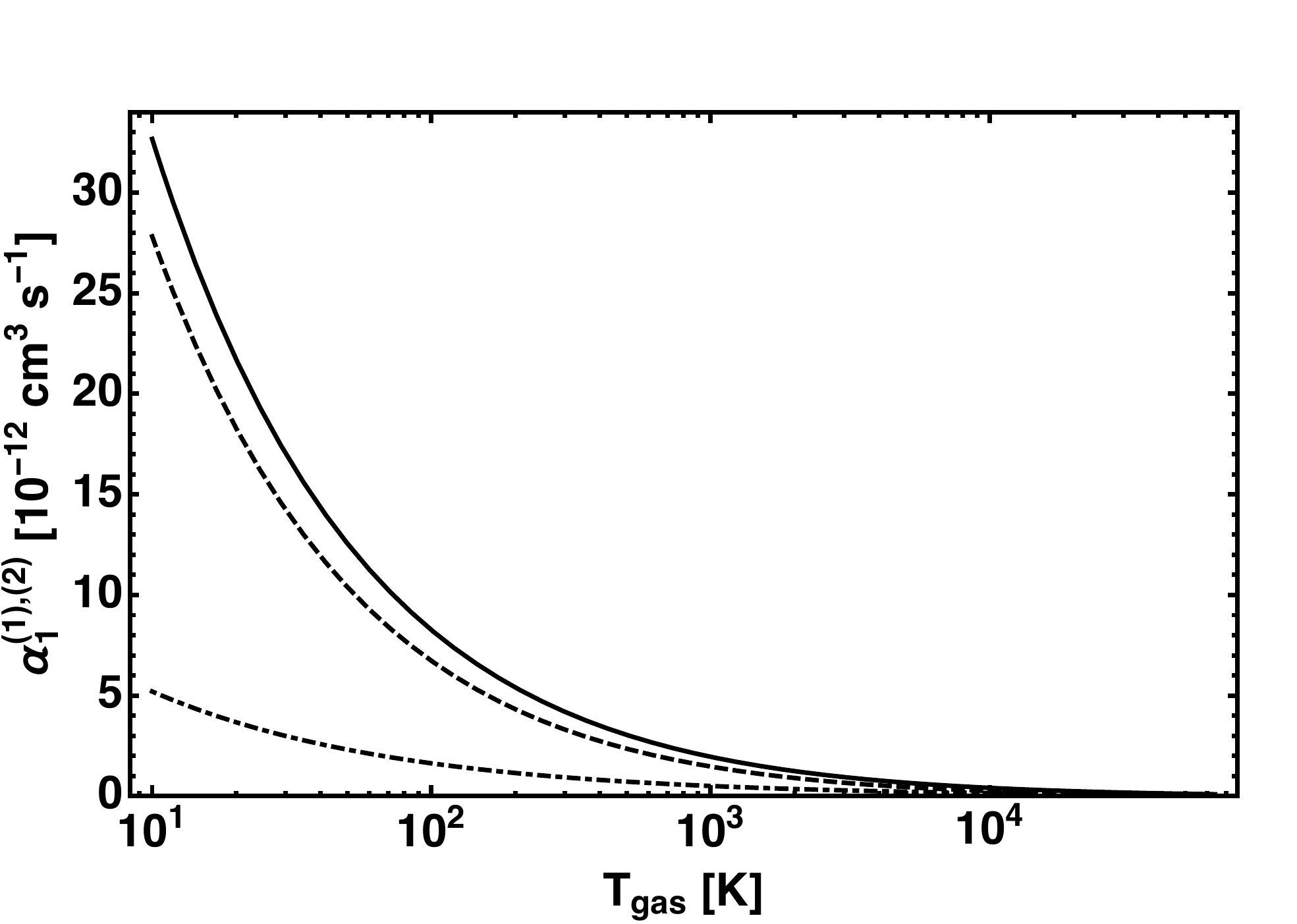} \\
\hspace{-4mm}\includegraphics[width=0.49\textwidth]{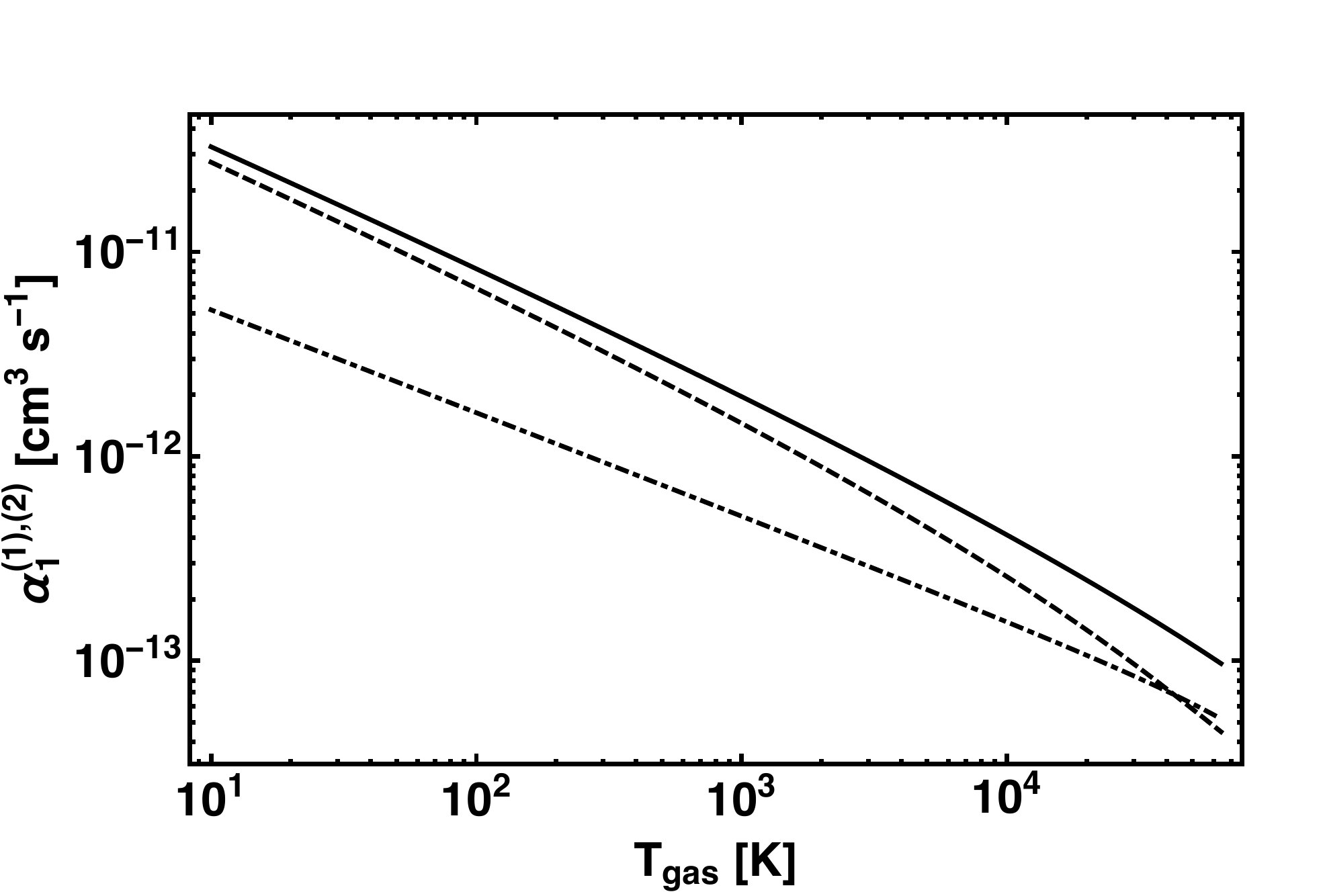}
\caption{
Recombination rates 
$\alpha^{(1)}$ (solid line), 
$\alpha^{(2)}$ (dashed line), and 
$\alpha_1$ (dotted-dashed line) of free electrons into varying states of hydrogen atoms as function of gas temperature.
The upper panel shows a linear scale, the lower panel a log scale on the vertical axis.
See the main text for derivation details.
}
\label{fig:recombinationrates}
\end{figure}

The resulting recombination rates $\alpha_1(T_\mathrm{gas})$ of free electrons directly into the hydrogen ground state,
the recombination rates $\alpha^{(1)}(T_\mathrm{gas})$ of free electrons into any state of the hydrogen atoms, and
the recombination rates $\alpha^{(2)}(T_\mathrm{gas})$ of free electrons into any state of the hydrogen atoms except of the ground state 
are shown in Fig.~\ref{fig:recombinationrates}.

Direct recombination of free electrons into the hydrogen ground state results in the emission of an EUV photon, which is again capable of ionizing neutral hydrogen.
For hydrogen ionization due to these EUV photons created via direct recombination into the hydrogen ground state, the ionization cross section of the gaseous medium reads
\begin{equation}
\sigma_\mathrm{rec} = \sigma_\mathrm{L} \left(\frac{\nu_\mathrm{L}}{\nu}\right)^3
\end{equation}
with the Lyman cross section $\sigma_\mathrm{L} = 6.3 \times 10^{-18} \mbox{ cm}^2$.
The frequency ratio is determined via the Lyman photon energy $h ~ \nu_\mathrm{L} = 13.6 \mbox{ eV}$ and the mean energy of the photons due to recombination $\langle h \nu \rangle_\mathrm{rec}$.
This recombination mean photon energy is a function of gas temperature.
Due to the fact that electrons with low kinetic energy are favored in the recombination process, the mean photon energy has an upper limit of
\begin{equation}
\langle h \nu \rangle_\mathrm{rec} \le h \nu_\mathrm{L}  + \frac{2}{3} ~ k_\mathrm{B} ~ T_\mathrm{gas}.
\end{equation}
If required, temperature-dependent values can be obtained from the tables presented in \citet{1989agna.book.....O}.
For our current applications of the module in present-day high-mass star formation, we estimate the recombination cross section as a constant value based on the tabulated data:
for a typical gas temperature within fully ionized regions of $T_\mathrm{gas} \approx 8000 \mbox{ K}$, 
the mean photon energy is $\langle h \nu \rangle_\mathrm{rec} = 14.2 \mbox{ eV}$, and 
the resulting cross section is $\sigma_\mathrm{rec} = 5.53 \times 10^{-18} \mbox{ cm}^2$.

\begin{figure}[htbp]
\centering
\includegraphics[width=0.49\textwidth]{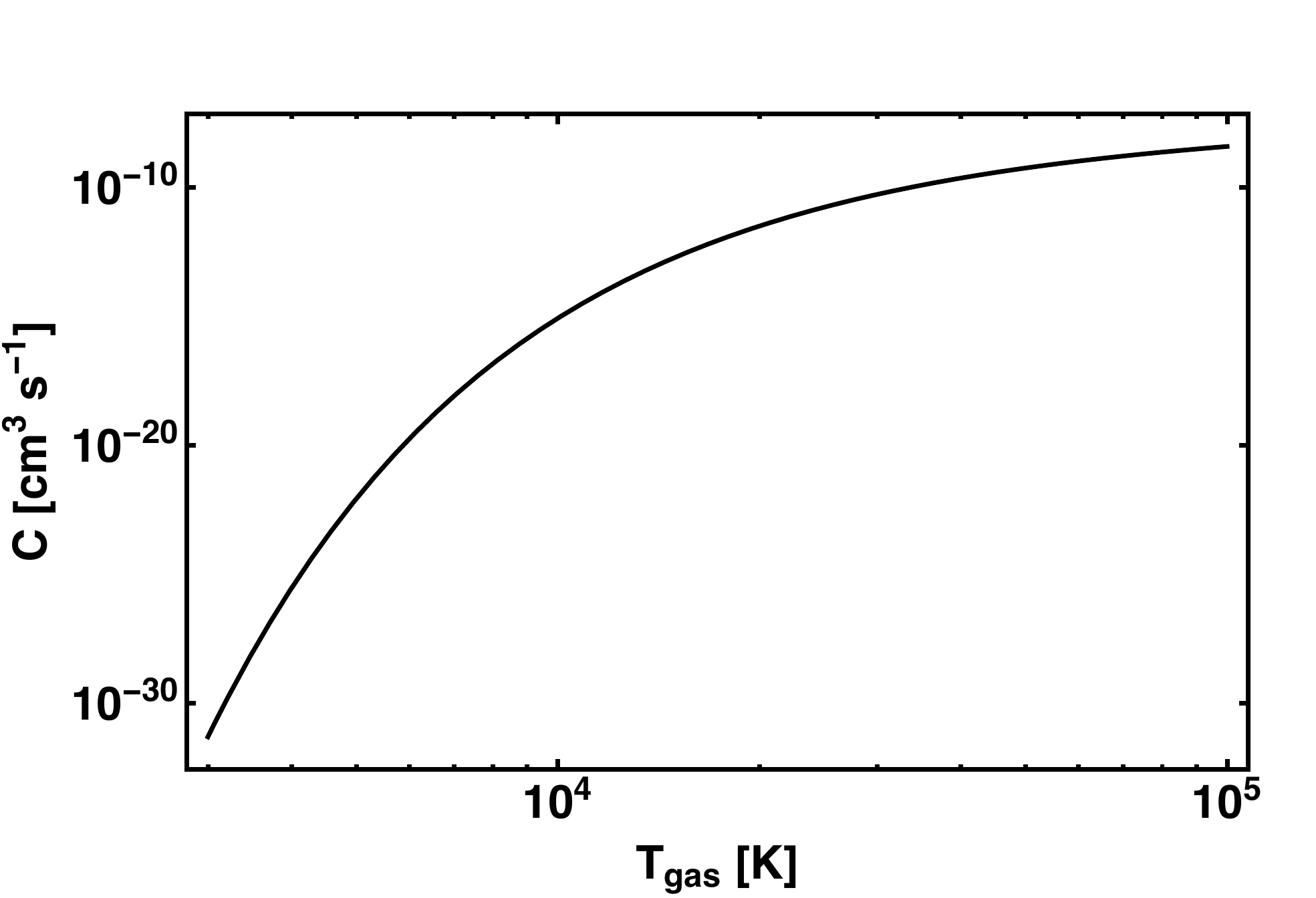}
\caption{
Collisional ionization rates $C$ as function of gas temperature $T_\mathrm{gas}$.
}
\label{fig:collision}
\end{figure}

\paragraph{Collisions}
Within (partly) ionized regions, collisions between free electrons and neutral hydrogen increase the ionization fraction via the reaction
\begin{equation}
\mbox{e}^- + \mbox{H} \rightarrow \mbox{H}^+ + 2 \mbox{e}^-
\end{equation}
The collisional ionization rate is determined via
\begin{equation}
C(T_\mathrm{gas}) = \pi ~ R_\mathrm{B}^2 ~ u_\mathrm{e}(T_\mathrm{gas}) ~ \exp \left(- \frac{h ~ \nu_\mathrm{L} }{k_\mathrm{B} ~ T_\mathrm{gas}} \right)
\end{equation}
with the thermal electron drift velocity of
$u_\mathrm{e}(T_\mathrm{gas}) = \sqrt{3 ~ k_\mathrm{B} ~ T_\mathrm{gas}/m_\mathrm{e}}$.
These rates are shown as function of gas temperature in Fig.~\ref{fig:collision}.

\structuring{\clearpage}
\subsubsection{Thermodynamics}
\label{methods_thermodynamics}
Within ionized regions, the gas temperature is governed by the balance of the heating and cooling mechanisms of the hot gas.
The major heating source within a H II region is given by the excess energy of the ionizing photons; the excess energy is first transformed into kinetic energy of the free electron and afterwards thermalizes due to collisions, more precisely Coulomb interactions between the charged particles.
This process is called the photoelectric heating of gas.

Related to this dominant process,
photoelectric heating of gas can also happen by freeing electrons from dust grains within the gaseous medium.
Further heating processes include
chemical heating on dust grain surfaces,
compressional heating via hydrodynamic shocks,
and
potentially magnetic reconnection.

The most important cooling mechanism is the emission of line radiation.
Collisions between gas particles leading to excitation of one or both of the colliding particles into higher energy levels remove kinetic energy from the gas.
Deexcitation via spontaneous emission will remove energy from the volume of gas,
if the emitted photon is either absorbed by a dust grain or freely escapes from the hot gaseous region.
Hence, the main ``cooling lines'' are a result of appropriate level populations, ionization states, and the frequency-dependent optical depth of the medium.

Gas cooling can also take place via grain--gas collisions, in which thermal energy is transferred from the hot gas to the colder dust grains.\footnote{In neutral regions, it is possible that the dust grains are warmer than the gas and grain--gas collisions would lead to gas heating rather than cooling.}
Further cooling mechanisms include, e.g., free-free emission or Bremsstrahlung,
\vONE{
collisional ionization cooling,
recombination cooling,
}
and hydrodynamic expansion of the gas.

In spite of the complexity of the heating and cooling processes, which ultimately control the gas temperature,
the resulting gas temperature of a fully ionized H II region generally exhibits only small internal variations in comparison to the large temperature difference between ionized and neutral regions.
Hence, for our current simulations, we use a very simplistic and rather convenient way of setting the gas temperature within the ionized regions:
we assume that the equilibrium gas temperature of fully ionized regions $T_\mathrm{gas}^\mathrm{ion}$ is constant.
Within the H II region, the dust temperature is largely decoupled from the gas temperature, because grain heating due to collisions with the hotter electrons and ions is several orders of magnitude smaller than cooling via gray body emission of the dust grains.
Hence, we use the hybrid radiation transport module as described above to determine the dust temperature $T_\mathrm{dust}$ in both ionized and neutral regions.
The gas temperature $T_\mathrm{gas}^\mathrm{neu}$ of the neutral medium, which is shielded from the EUV radiation, is assumed to be in equilibrium with the dust temperature. 
Hence, the temperature is determined by the hybrid radiation transport module as well.
Transition regions between the fully ionized gas and the neutral medium are very confined and only marginally resolved on the numerical grid, if at all.
For simplicity, the gas temperature within these boundary layers is set by a linear regression between $T_\mathrm{gas}^\mathrm{ion}$ and $T_\mathrm{gas}^\mathrm{neu}$ based on the local ionization fraction $x$:
\begin{equation}
T_\mathrm{gas} = y ~ T_\mathrm{gas}^\mathrm{neu} + x ~ T_\mathrm{gas}^\mathrm{ion}.
\end{equation}
If required for an astrophysical problem on hand, this thermodynamics routine can in principle be coupled to a chemical network solver to account for heating and cooling in a more self-consistent way; see, e.g., \citet{2018ApJ...857...57N, 2018ApJ...865...75N} for a more sophisticated chemical--thermodynamical treatment.

%
%
\structuring{\clearpage}
\section{Tests}
\label{sect:tests}
In this section we discuss a variety of tests that have been performed using the numerical framework presented.
We present this as a useful test suite for future code development in the field of astrophysical fluid flows.
Hence, we focus on reproducibility more than modeling a particular astrophysical environment.
In the case of a community's interest in a common code comparison project, including realistic modeling of specific astrophysical conditions, we are happy to join such an effort.

\subsection{Radiation Transport}
We first present tests using the thermal continuum radiation transport solver Makemake without taking into account photoionization or hydrodynamics.

\subsubsection{Optically Thin Irradiation}
The first radiation test describes a dominating source of radiation irradiating an optically thin environment.
The advantage of such a simple initial test is that it allows us to first compare the result of the ray-tracing scheme to an analytical solution as described below.

\paragraph{Physical setup}
The radiating source is chosen to be Sun type, i.e., a point source with solar luminosity and 
\vONE{
a photospheric temperature corresponding to a stellar size of
}
one solar radius.
The optically thin environment is represented by a sphere of radius $r_\mathrm{max} = 1000 \mbox{ au}$ with a uniform negligibly small gas density of $10^{-40} \mbox{ g cm}^{-3}$.
The initial gas temperature can be set to an arbitrarily (low) value.

\paragraph{Numerical configuration}
The problem is solved on a one-dimensional grid in spherical coordinates.
The star is placed at the origin.
The computational domain extends from the innermost radius $r_\mathrm{min} = 1 \mbox{ au}$ to the outermost radius of $r_\mathrm{max} = 1000 \mbox{ au}$.
We use logarithmically increasing radial widths $\Delta r$ toward larger radii with $1000$ grid cells.
At the innermost radial boundary, we use zero gradient in thermal radiation energy density.
At the outermost radial boundary, we use zero gradient in flux of thermal radiation energy density, computed in the optically thin limit.
We compute the equilibrium temperature slope via gray ray-tracing only, gray ray-tracing plus FLD in the equilibrium temperature approach, and gray ray-tracing plus FLD in the linearization temperature approach.

\begin{figure}[htbp]
\centering
\includegraphics[width=0.49\textwidth]{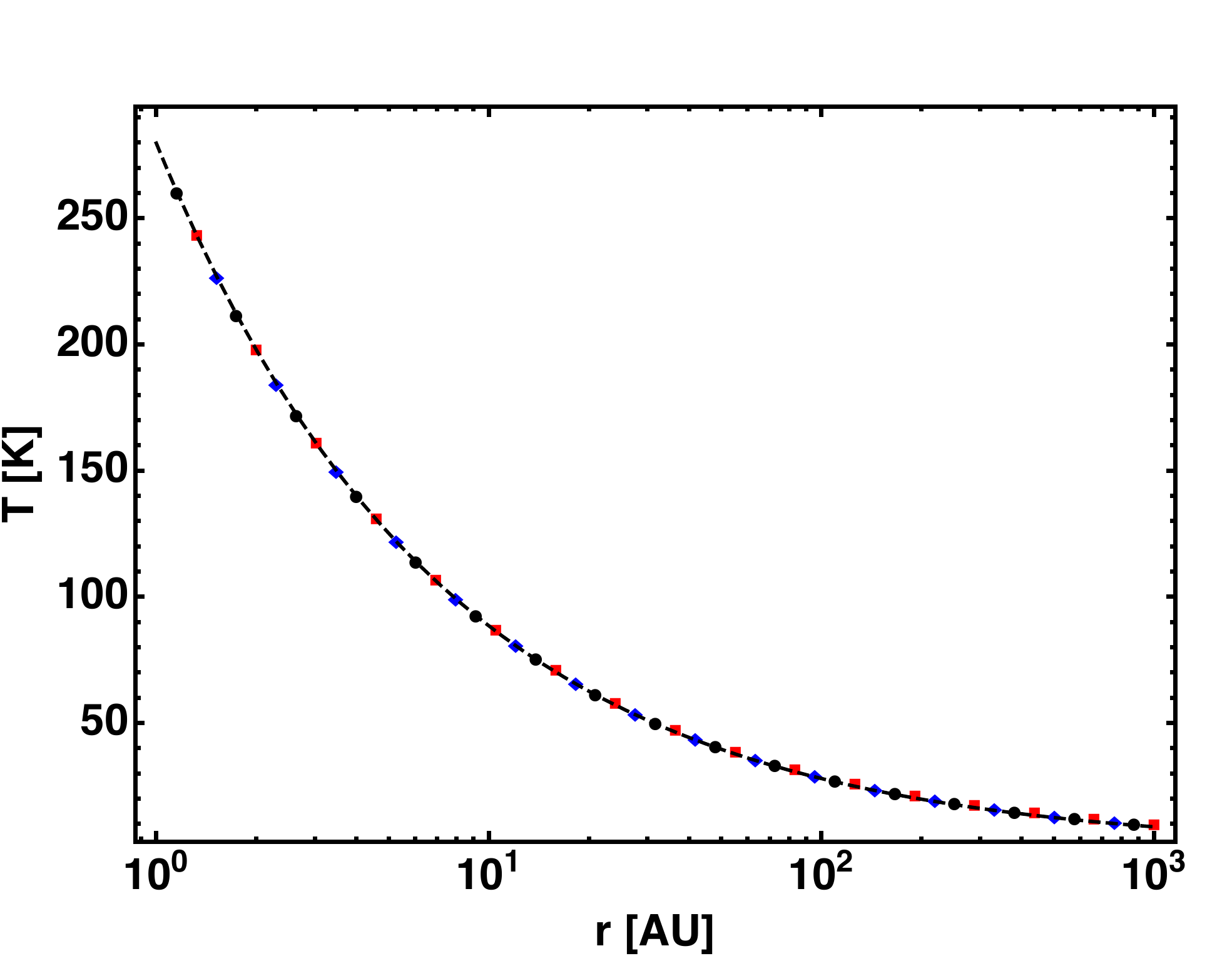}\\
\caption{
Radiation test of optically thin irradiation.
The analytical solution (dashed line) and numerical results (markers) of the temperature slope around a solar-type star. 
Black dots denote ray-tracing only.
Red squares denote ray-tracing plus equilibrium temperature approach FLD.
Blue diamonds denote ray-tracing plus linearization temperature approach FLD.
}
\label{fig:tests_radiation_opticallythin}
\end{figure}

\paragraph{Result}
An analytical solution for directly irradiated regions far away from the radiating source is given by \citet{1968dms..book.....S} as
\begin{equation}
T(r) = T_\mathrm{star} ~ \left(\frac{r}{2~R_\mathrm{star}}\right)^{-0.5}.
\end{equation}
In Fig.~\ref{fig:tests_radiation_opticallythin}, we compare the numerical results of the different solver methods with the analytical estimate.
The three radiation transport methods yield the correct temperature distribution very accurately with relative deviations of the order of $\Delta = 10^{-6}$.
As expected, the different solver methods also yield the same slope due to the fact that the radiation field is fully dominated by the central star, which is solved via the same ray-tracing approach for all three methods.
\vONE{
Due to the extremely low optical depth of the environment in this test case, the effect of the remitted radiation field is negligible.
Hence, this test checks the reliability of the ray-tracing part only.
}

\subsubsection{Irradiated Circumstellar Disk of \citet{2004A&A...417..793P}, the $\tau_\mathrm{550 nm} = 0.1$ Case}
The second and third radiation tests are defined by the most optically thin and the most optically thick test cases of \citet{2004A&A...417..793P}: a two-dimensional benchmark study of continuum radiative transfer for \vONE{circumstellar} disk configurations.
\vONE{We have utilized these setups} in our earlier technical radiation transport studies in \citet{2010A&A...511A..81K} and \citet{2013A&A...555A...7K}.
In contrast to these early studies, the new version of Makemake presented here includes the capability of the two-temperature linearization approach to solve the FLD equation.
We will see in the following, however, that because \vONE{the test} only \vONE{aims for} the final equilibrium temperature for a radiation transport problem without compressional heating, the equilibrium and linearization approaches yield identical results.
The newly implemented two-temperature solving technique becomes important for the radiation hydrodynamics tests presented in the following section.

\vONE{
\paragraph{Physical setup}
The gas and dust mass density distribution describes a flared circumstellar disk setup with
\begin{equation}
\rho_\mathrm{gas}(R, z) = \rho_0 \times \frac{R_\mathrm{d}}{R} \times \exp\left( - \frac{\pi}{4} \left(\frac{z}{z_\mathrm{d}} \left(\frac{R_\mathrm{d}}{R}\right)^{1.125}\right)^2 \right) 
\end{equation}
with the cylindrical radius $R$ and height $z$, $R_\mathrm{d} = 500 \mbox{ au}$, and $z_\mathrm{d} = 125 \mbox{ au}$.
$\rho_0$ is a free parameter that allows us to specify the total optical depth of the disk's midplane in the radial direction.
The radiating source is chosen to be Sun type, i.e., a point source with a photospheric temperature of $5800 \mbox{ K}$ and a stellar luminosity corresponding to a stellar size of one solar radius.
The star is assumed to radiate as a blackbody.
The dust-to-gas mass ratio is set constant to 1\%. 
Scattering is ignored.
The initial gas temperature can be set to an arbitrarily (low) value.

\paragraph{Numerical configuration}
The problem is solved on a two-dimensional grid in spherical coordinates, assuming axial symmetry.
The star is placed at the origin.
The computational domain extends from the innermost radius $r_\mathrm{min} = 1 \mbox{ au}$ to the outermost radius of $r_\mathrm{max} = 1000 \mbox{ au}$.
We use logarithmically increasing radial widths $\Delta r$ toward larger radii with $200$ grid cells.
The polar grid ranges from $0$ to $\pi$ and consists of $64$ grid cells with a uniform coverage in angle.

At the innermost radial boundary, we use zero gradient in thermal radiation energy density.
At the outermost radial boundary, we use zero gradient in flux of thermal radiation energy density, computed in the optically thin limit.
We compute the equilibrium temperature slope via frequency-dependent ray-tracing plus FLD in the equilibrium temperature approach and frequency-dependent ray-tracing plus FLD in the linearization temperature approach; those results are compared to the numerical solution obtained by using the radmc Monte Carlo dust continuum radiation transport code \citep{2011ascl.soft08016D}.
}

\begin{figure*}[htbp]
\centering
\includegraphics[width=0.48\textwidth]{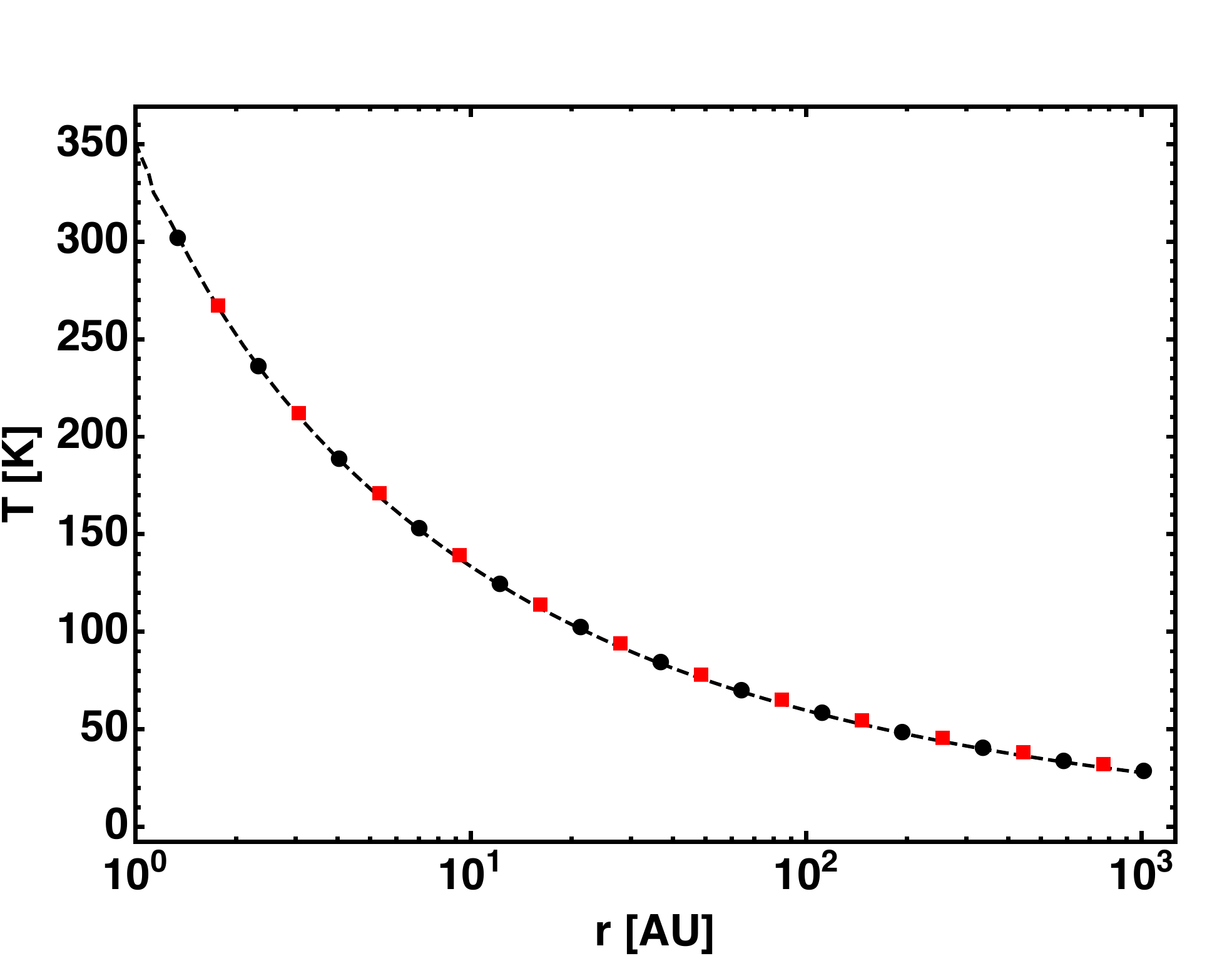}
\includegraphics[width=0.48\textwidth]{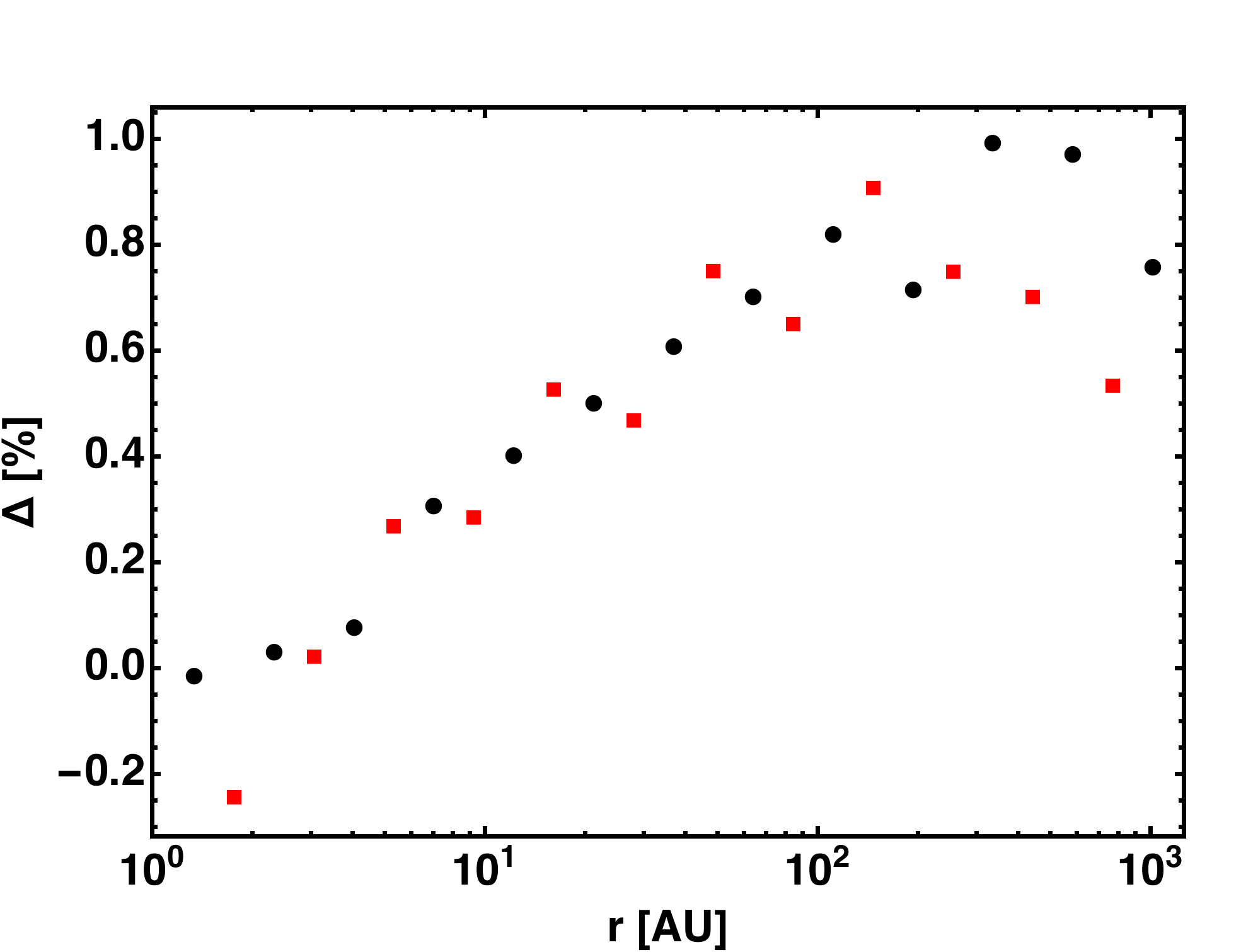}
\caption{
Radiation test of an optically thin disk from \citet{2004A&A...417..793P}.
The dashed line shows the result from the radmc Monte Carlo continuum radiation transport code.
Black dots denote results from ray-tracing plus FLD using the equilibrium temperature approach.
Red squares denote results from ray-tracing plus FLD using the linearization temperature approach.
The left panel shows the resulting temperature distributions through the disk's midplane.
The right panel shows the deviation of the two methods with respect to the radmc Monte Carlo continuum radiation transport code.
}
\label{fig:tests_radiation_Pascucci_opticallythin}
\end{figure*}

\paragraph{Result}
The resulting temperature distributions through the disk's midplane and the deviation in temperature as compared to the radmc Monte Carlo solution \citep{2011ascl.soft08016D} are shown in Fig.~\ref{fig:tests_radiation_Pascucci_opticallythin} for the optically thin case.
The two radiation transport methods reproduce the same temperature distribution with deviations much less than 0.1\%.
As expected, the different solver methods also yield the same distribution due to the fact that this test denotes an equilibrium problem and the radiation field is highly dominated by the central star, which is solved via the same ray-tracing approach.

\subsubsection{Irradiated Circumstellar Disk of \citet{2004A&A...417..793P}, the $\tau_\mathrm{550 nm} = 100$ Case}
\vONE{
\paragraph{Physical setup and numerical configuration}
The physical setup and numerical configuration are identical to the previous test, but the normalization density of the irradiated environment is set to a higher value to obtain an optical depth of $\tau_\mathrm{550 nm} = 100$ through the disk's midplane.
}

\begin{figure*}[htbp]
\centering
\includegraphics[width=0.48\textwidth]{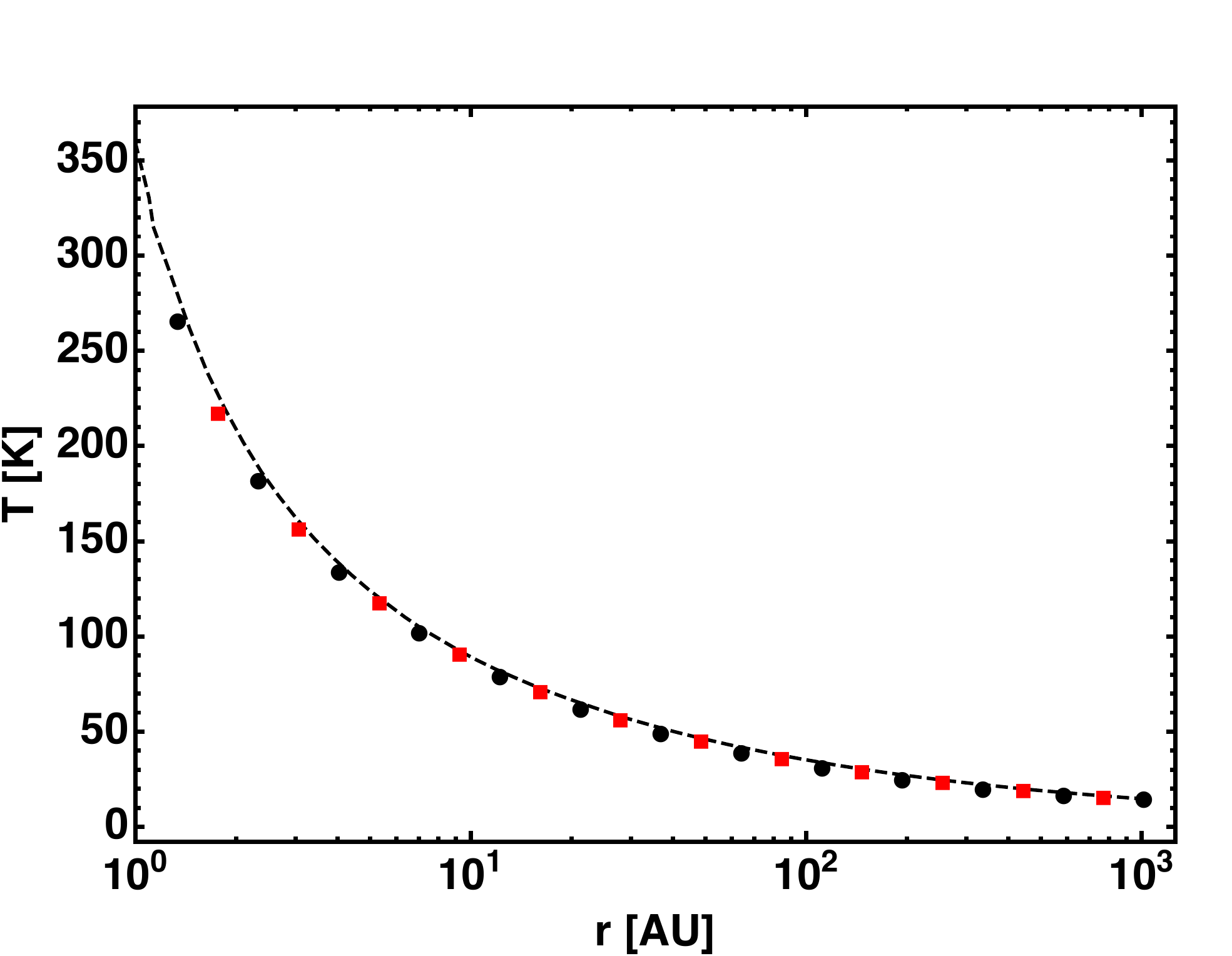}
\includegraphics[width=0.48\textwidth]{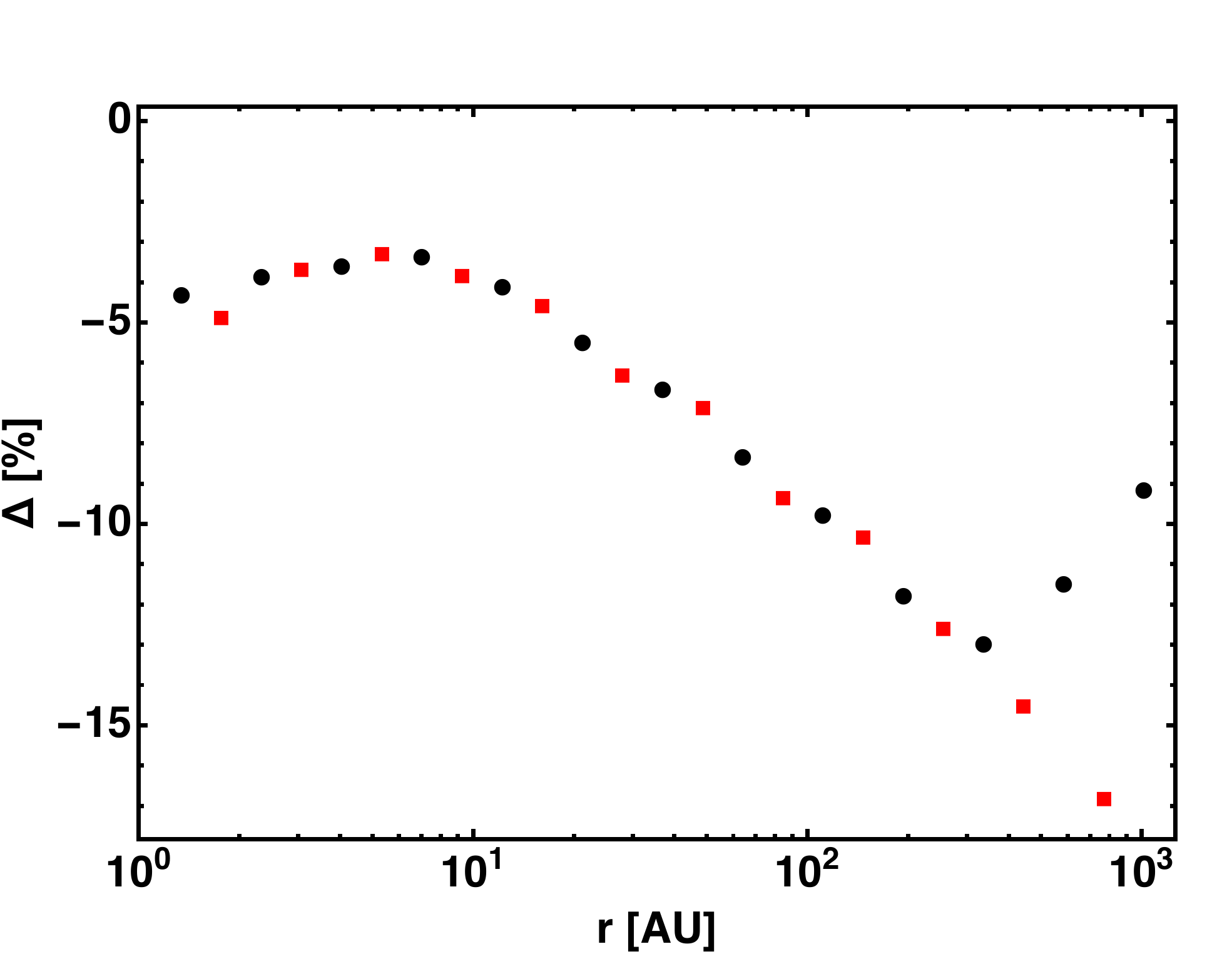}
\caption{
Same as Fig.~\ref{fig:tests_radiation_Pascucci_opticallythin}, but for a 
radiation test of an optically thick disk from \citet{2004A&A...417..793P}.
}
\label{fig:tests_radiation_Pascucci_opticallythick}
\end{figure*}

\begin{figure*}[htbp]
\centering
\includegraphics[width=0.48\textwidth]{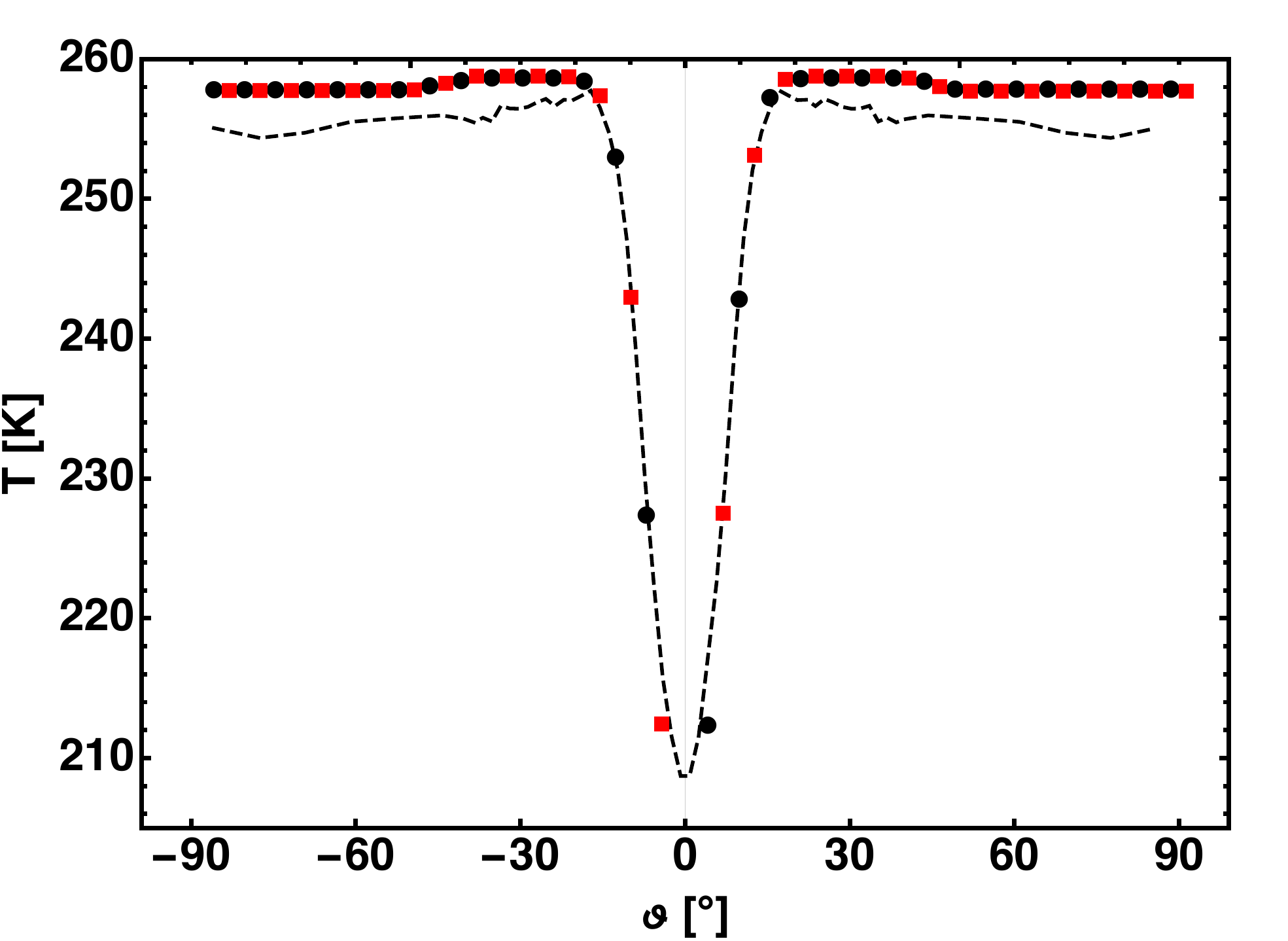}
\includegraphics[width=0.48\textwidth]{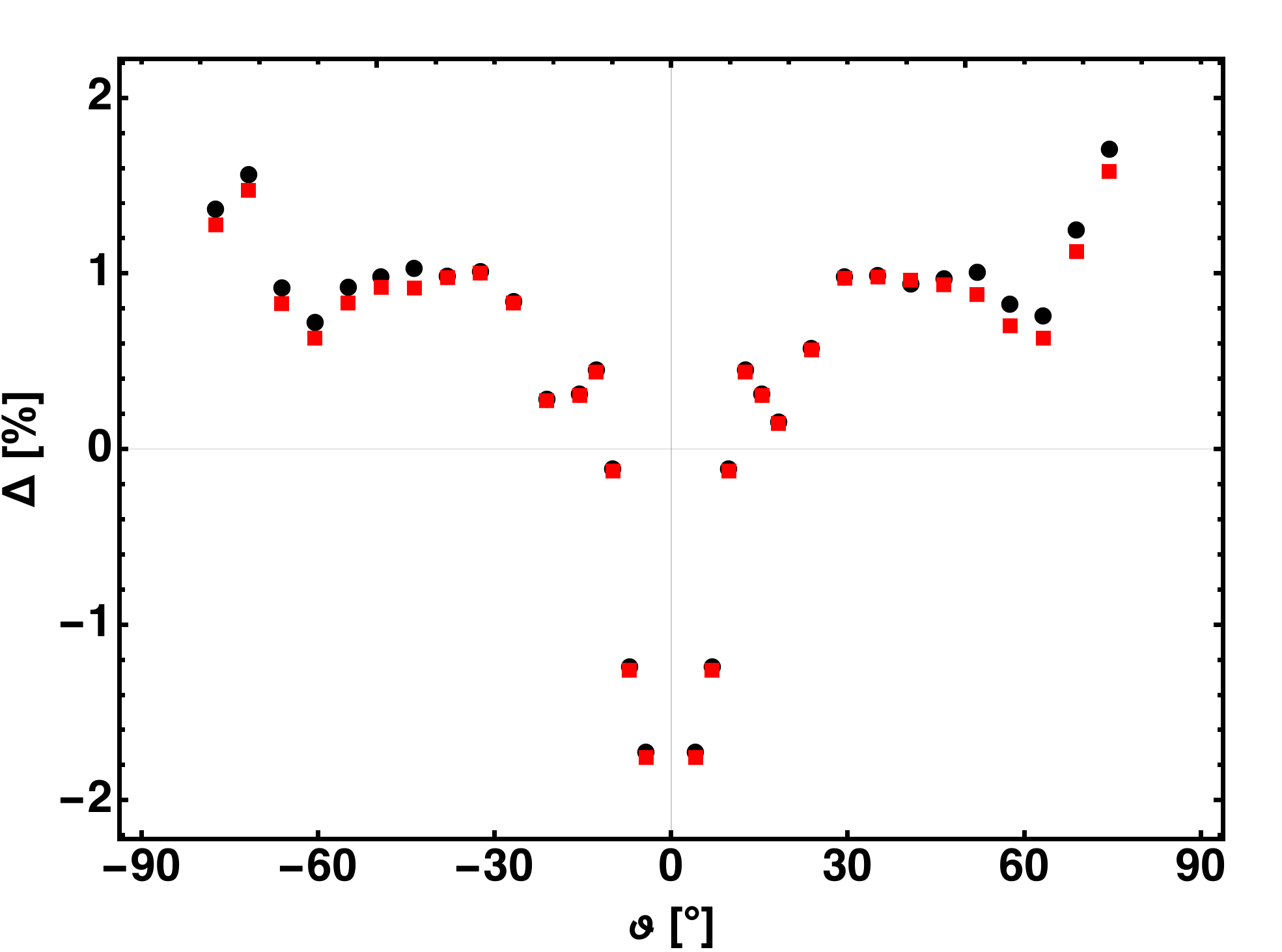}
\caption{
Same as Fig.~\ref{fig:tests_radiation_Pascucci_opticallythick}, but for a cut in the polar direction at $2 \mbox{ au}$ in radius.
}
\label{fig:tests_radiation_Pascucci_opticallythick_polar}
\end{figure*}

\paragraph{Result}
The resulting temperature distributions through the disk's midplane as well as along a polar cut at $2 \mbox{ au}$ and the corresponding temperature deviations from the Monte Carlo solution are presented in Figs.~\ref{fig:tests_radiation_Pascucci_opticallythick} and \ref{fig:tests_radiation_Pascucci_opticallythick_polar}, respectively.
The two radiation transport methods represent the temperature distribution very accurately with deviations from 5\% at the inner disk rim up to 15\% at the outer disk rim; these differences are similar to those found by \citet{2004A&A...417..793P} for the different Monte Carlo and ray-tracing codes participating in the benchmark tests.
The temperature variations in the Monte Carlo solution at higher latitudes, visible in Fig.~\ref{fig:tests_radiation_Pascucci_opticallythick_polar} are actually due to Monte Carlo noise, and the resulting temperature distribution in this optically thin part of the stellar environment should be constant along a radial cut for the higher latitudes, as obtained by the numerical results of the two solvers in use.
As expected, the two different solver methods used here yield the same results due to the fact that this test is an equilibrium problem.

\clearpage

\structuring{\clearpage}
\subsection{Radiation Hydrodynamics}
The tests of the previous section are devoted to radiation transport only.
In this section, we further check the coupling and interaction with the hydrodynamic flow.

\subsubsection{Radiative Shock Tube, Subcritical}

\begin{figure*}[!p]
\begin{center}
\includegraphics[width=0.4\textwidth]{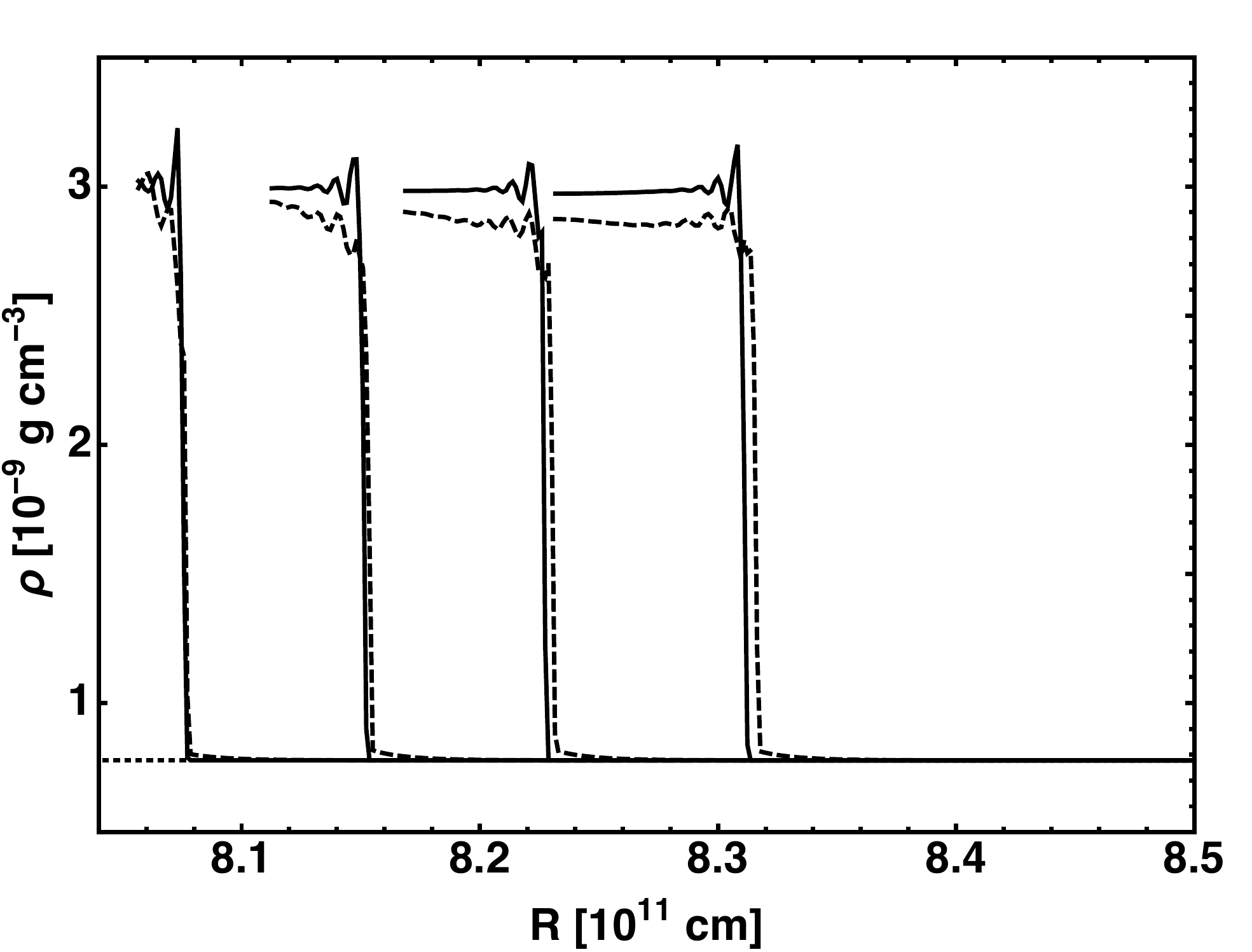}
\includegraphics[width=0.4\textwidth]{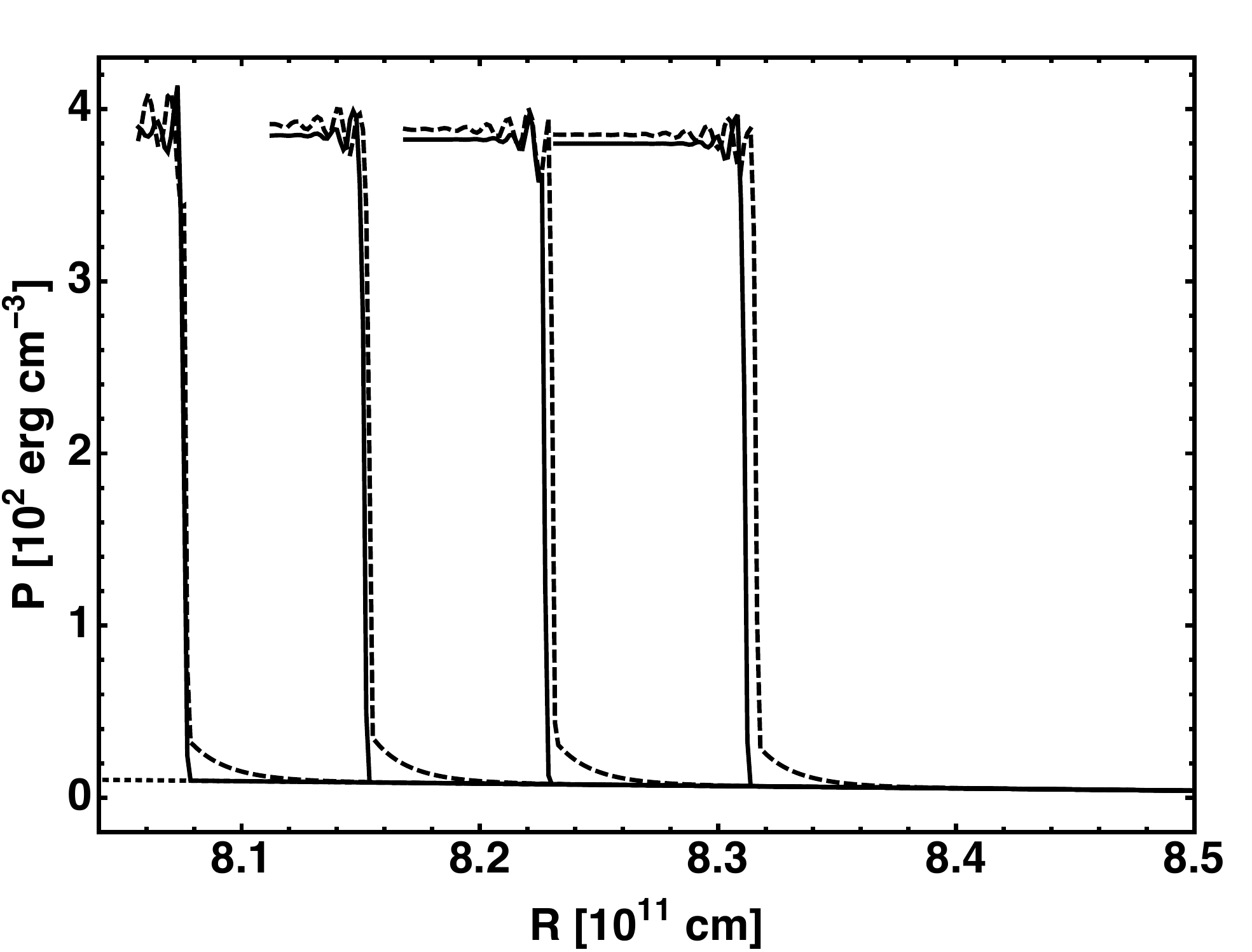}\\
\includegraphics[width=0.4\textwidth]{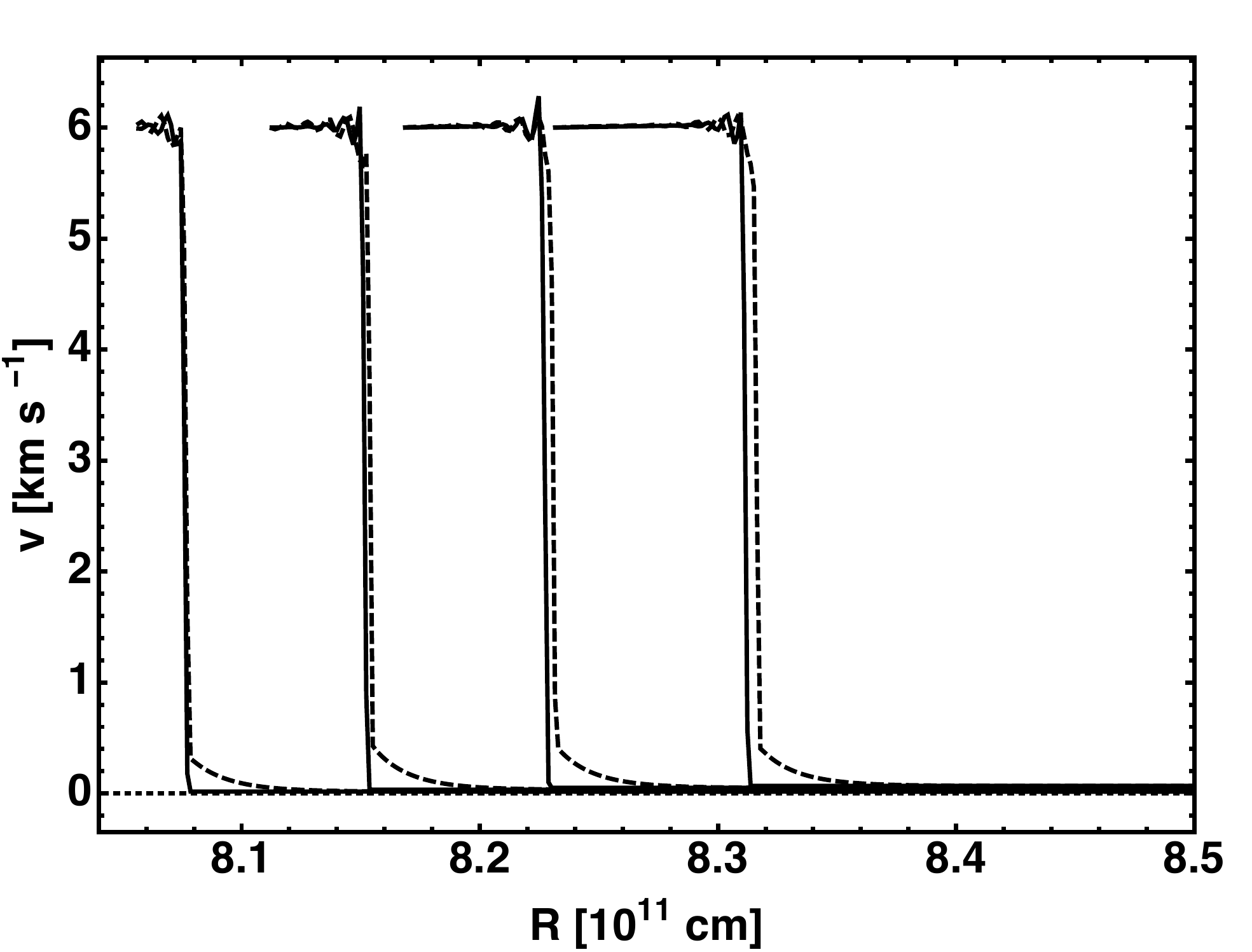}
\includegraphics[width=0.4\textwidth]{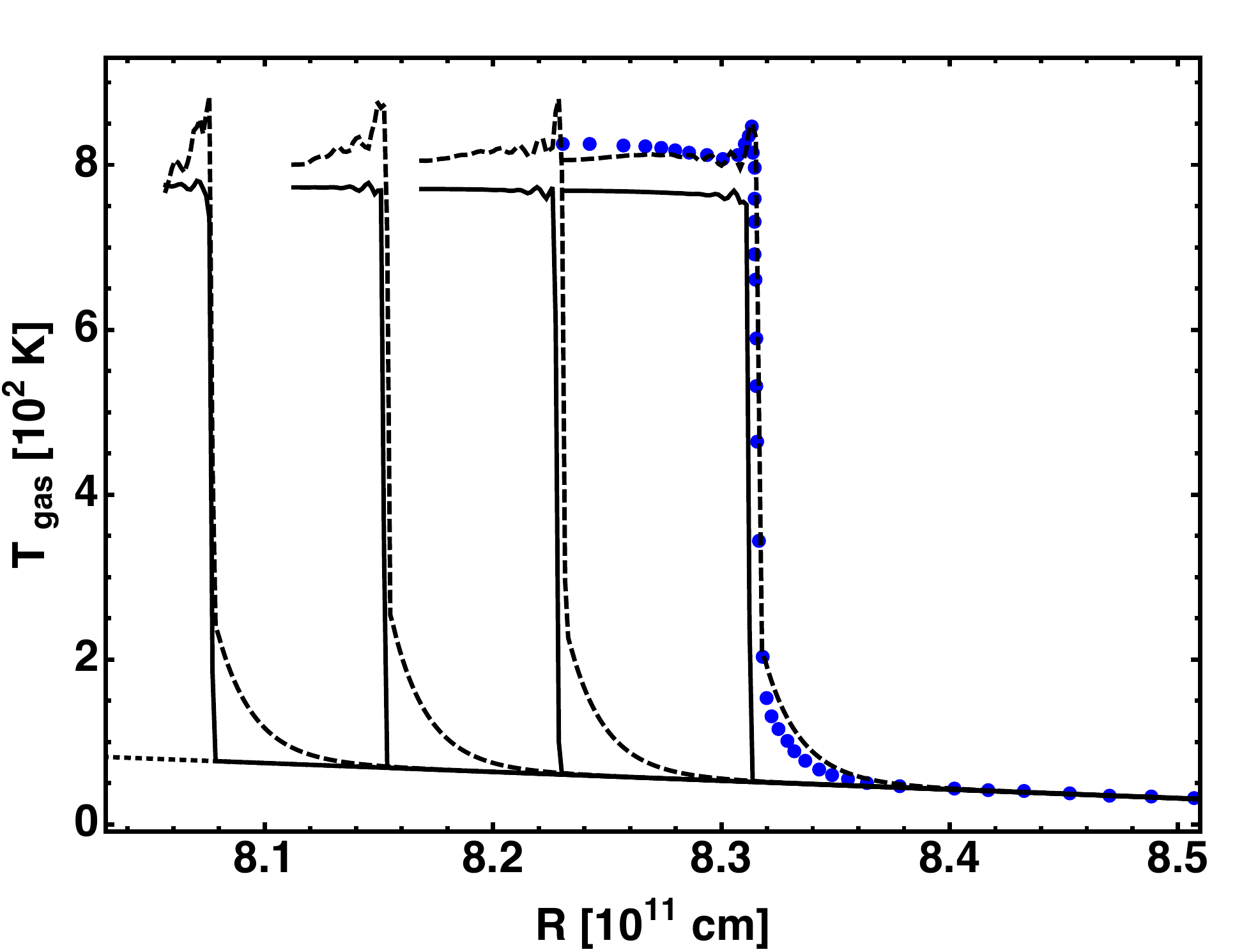}\\
\includegraphics[width=0.4\textwidth]{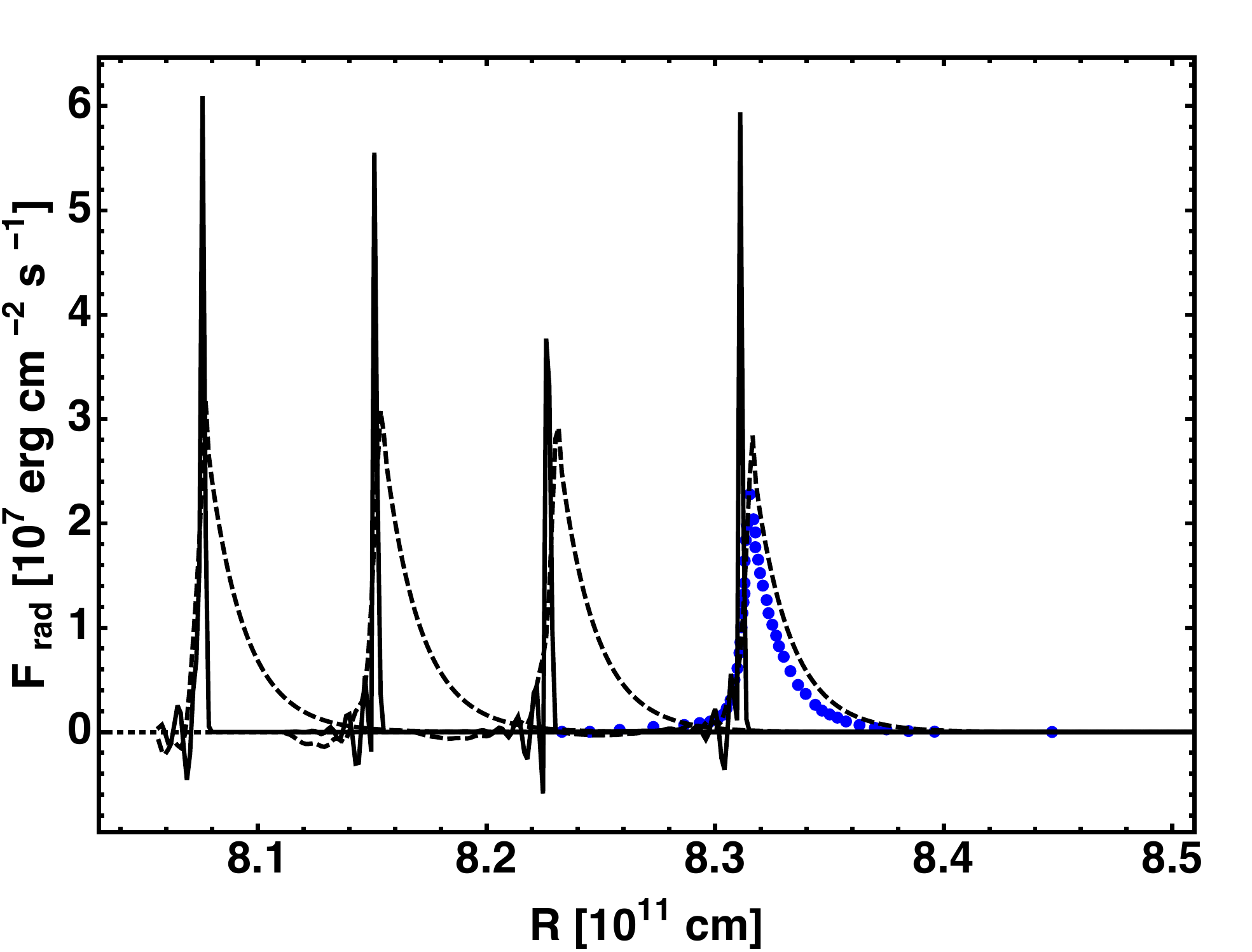}
\includegraphics[width=0.4\textwidth]{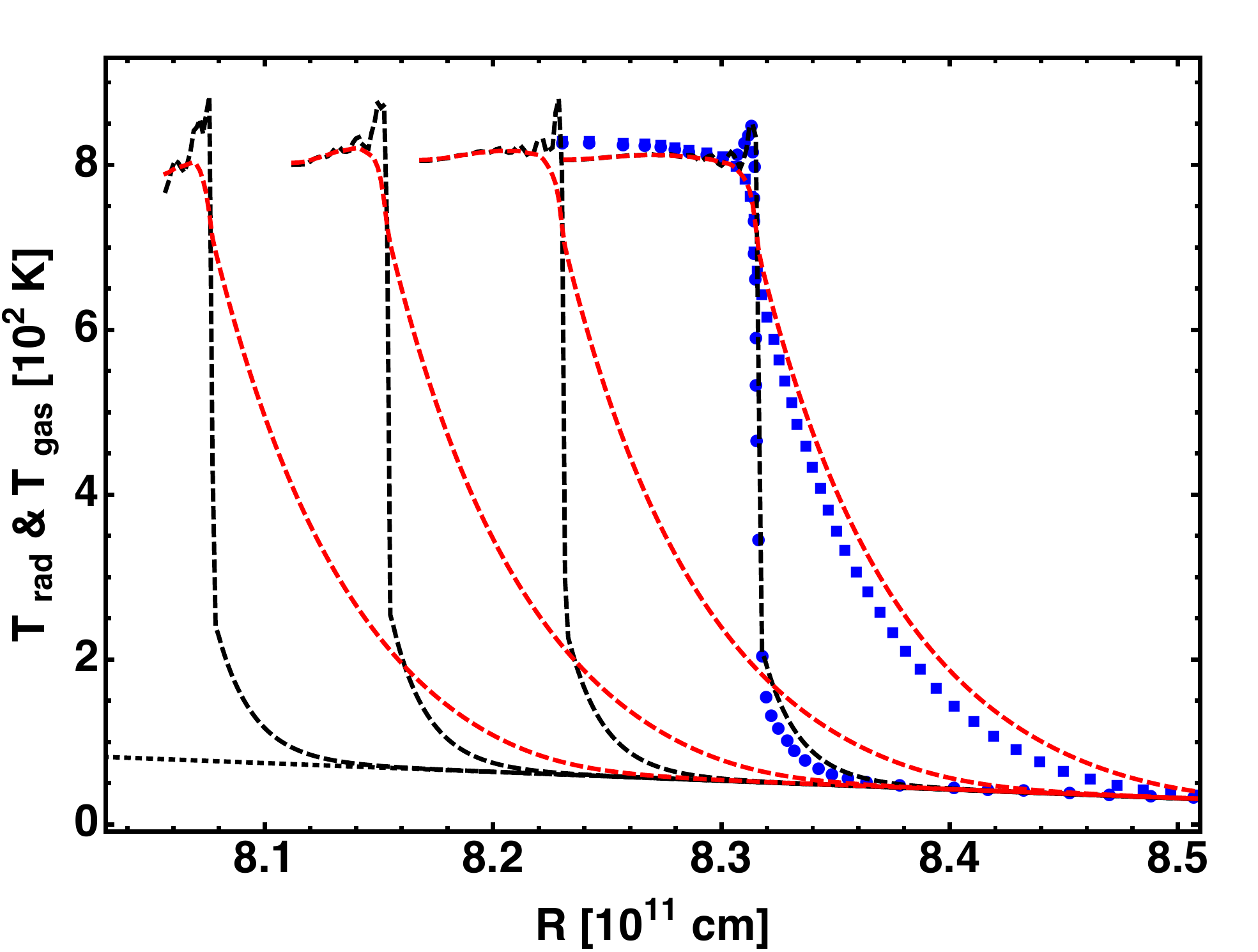}
\caption{
Gas-related properties of the radiation hydrodynamics subcritical shock tube test.
Solid lines denote results using the equilibrium temperature approach,
dashed lines results of the linearization approach.
From top to bottom and left to right, the panels show
gas density,
gas pressure,
gas velocity, 
gas temperature,
radiative flux, and
radiation temperature.
The four snapshots in time (moving from left to right in each panel) correspond to $0.93$, $1.87$, $2.80$, and $3.85 \times 10^4 \mbox{ s}$.
Dotted lines denote the initial setup.
In the bottom-right panel, dashed black lines denote the gas temperature as also shown in the middle-right panel, dashed red lines denote the radiation temperature.
\vONE{
Blue dots denote the numerical data extracted from \citet{1994ApJ...424..275E}, their Fig.~8, using the Web Plot Digitizer tool (\href{https://apps.automeris.io/wpd}{https://apps.automeris.io/wpd}) for $t = 3.8 \times 10^4 \mbox{ s}$.
The abscissa was converted into the original setup configuration of \citet{1994ApJ...424..275E} to ease comparison, i.e., properties are shown in the non-comoving laboratory frame with the piston starting at a radius of $R_\mathrm{min} = 8 \times 10^9 \mbox{ cm}$.
}
}
\label{fig:tests_radiationhydrodynamics_subcriticalshocktube}
\end{center}
\end{figure*}

\paragraph{Context}
Radiative shocks denote a classical test problem of radiation hydrodynamics.
They were studied in semianalytical approaches and are part of a variety of test suites for codes within the astrophysical literature \citep{1963PhFl....6..781H,1967pswh.book.....Z,1980ApJ...236..201W,1984frh..book.....M,1994ApJ...424..275E,1994PhyD...77..320G,1999ShWav...9..391S,1999ShWav...9..403S,2001ApJS..135...95T,2003ApJS..147..197H,2006ApJS..165..188H,2007A&A...464..429G,2007ShWav..16..445L,2008ShWav..18..129L,2010A&A...511A..81K,2011A&A...529A..35C,2013A&A...559A..80K,2014ApJ...797....4K,2015A&A...574A..81R}.
Caution should be exercised for direct comparisons of the different code results and semianalytical solutions due to differences in the thermodynamic properties of the gas used in the different test configurations.
Here, we will rely on the original setup by \citet{1994ApJ...424..275E} with the negligible change of using a grid in Cartesian coordinates rather than the original spherical coordinates in the far-field limit.

\paragraph{Physical setup}
We model a tube of length $7 \times 10^{10} \mbox{ cm}$ filled with gas, which is compressed by a piston moving from one side into the tube.
The gas density is initially set to a uniform value of $7.78 \times 10^{-10} \mbox{ g cm}^{-3}$.
The initial gas temperature linearly decreases from 85~K on the piston boundary toward 75~K at the upstream side.
The radiation hydrodynamics equations are solved with an ideal equation of state with an adiabatic index of $\gamma = 5/3$ and a molar mass of $\mu = 0.5$.
The gas is assumed to be initially at rest, and the piston is moving with a speed of $u_\mathrm{piston} = 6 \mbox{ km s}^{-1}$.

\paragraph{Numerical configuration}
The problem is solved on a one-dimensional grid.
In the original setup of \citet{1994ApJ...424..275E}, the author used a grid in spherical coordinates, extending from $R_\mathrm{min} = 8 \times 10^{11} \mbox{ cm}$ to $R_\mathrm{max} = 8.7 \times 10^{11} \mbox{ cm}$.
Due to the short grid extent far away from the origin of the spherical grid, this setup results into a quasi-Cartesian grid, i.e.~the variation of the interface areas with radius becomes rather small.
Hence, we directly use a Cartesian grid here with an extent from $x_\mathrm{min} = 0$ to $x_\mathrm{max} = 7 \times 10^{10} \mbox{ cm}$.

We model the test problem in the comoving frame of the piston.
The gas velocity is initialized with the negative of the physical piston velocity $u_\mathrm{gas} = -u_\mathrm{piston}$.
The boundary conditions at the side of the piston are set to an impermeable wall utilizing reflective boundary conditions.
The boundary condition at the upstream side $x_\mathrm{max}$ of the tube is set to the initial uniform density, initial value of 75~K in temperature, and negative piston velocity.
We use a uniform resolution with $512$ grid cells.

\paragraph{Results}
The shock structure is depicted in Fig.~\ref{fig:tests_radiationhydrodynamics_subcriticalshocktube} at four instances in time, namely $0.933$, $1.67$, $2.80$, and $3.85 \times 10^4 \mbox{ s}$.
The last snapshot in time corresponds to the last time shown in the figures by \citet{1994ApJ...424..275E}.
The abscissa in our figures was converted into the original setup configuration of \citet{1994ApJ...424..275E} to ease comparison, i.e.~properties are shown in the non-comoving laboratory frame with the piston starting at a radius of $R_\mathrm{min} = 8 \times 10^9 \mbox{ cm}$.

Both radiation transport methods yield the basic shock parameters such as its propagation speed and the physical properties in the shocked and upstream directions.
The equilibrium temperature approach implies that the radiation temperature and gas temperature in the shock tube are the same.
Hence, the bottom-right panel of Fig.~\ref{fig:tests_radiationhydrodynamics_subcriticalshocktube} visualizes the result for the two-temperature linearization approach only.
Moreover, the equilibrium temperature approach does not show the smooth transition of gas temperature from the shocked to the upstream region, as shown Fig.~\ref{fig:tests_radiationhydrodynamics_subcriticalshocktube}.
Moreover, the steep gradient of gas temperature results in a slightly enhanced cooling flux, leading to a cooler temperature and lower pressure in the shocked gas.
As a result of the stronger cooling, the gas is compressed to slightly higher densities than in the linearization approach. 

For these shock tube tests, which assume a gray radiation field, the radiation energy density can be translated into a radiation temperature via
$E_\mathrm{rad} = a ~ T_\mathrm{rad}^4$, cf.~Equation \eqref{eq:RT_temperatureupdateequilibrium}.
The so-called radiative precursor of the radiation temperature in comparison to the gas temperature is presented in Fig.~\ref{fig:tests_radiationhydrodynamics_subcriticalshocktube}, bottom-right panel.

\vONE{
The simulation results of the subcritical shock are in overall very good agreement with the results from the original \citet{1994ApJ...424..275E} study.
Our last \vONE{presented} snapshot seems to be a little bit more advanced in time than the last snapshot from the original article.
The radiative precursor in our simulation is found to be stronger than in the original study.
Keeping in mind the differences between the hydrodynamics schemes of the two studies, and the fact that the exact form of the radiative precursor will depend on the choice of the flux limiter function (we used the one by \citet{1981ApJ...248..321L} for these simulations), the agreement of the simulation results is satisfactory.
}

\subsubsection{Radiative Shock Tube, Supercritical}
\begin{figure*}[!p]
\begin{center}
\includegraphics[width=0.4\textwidth]{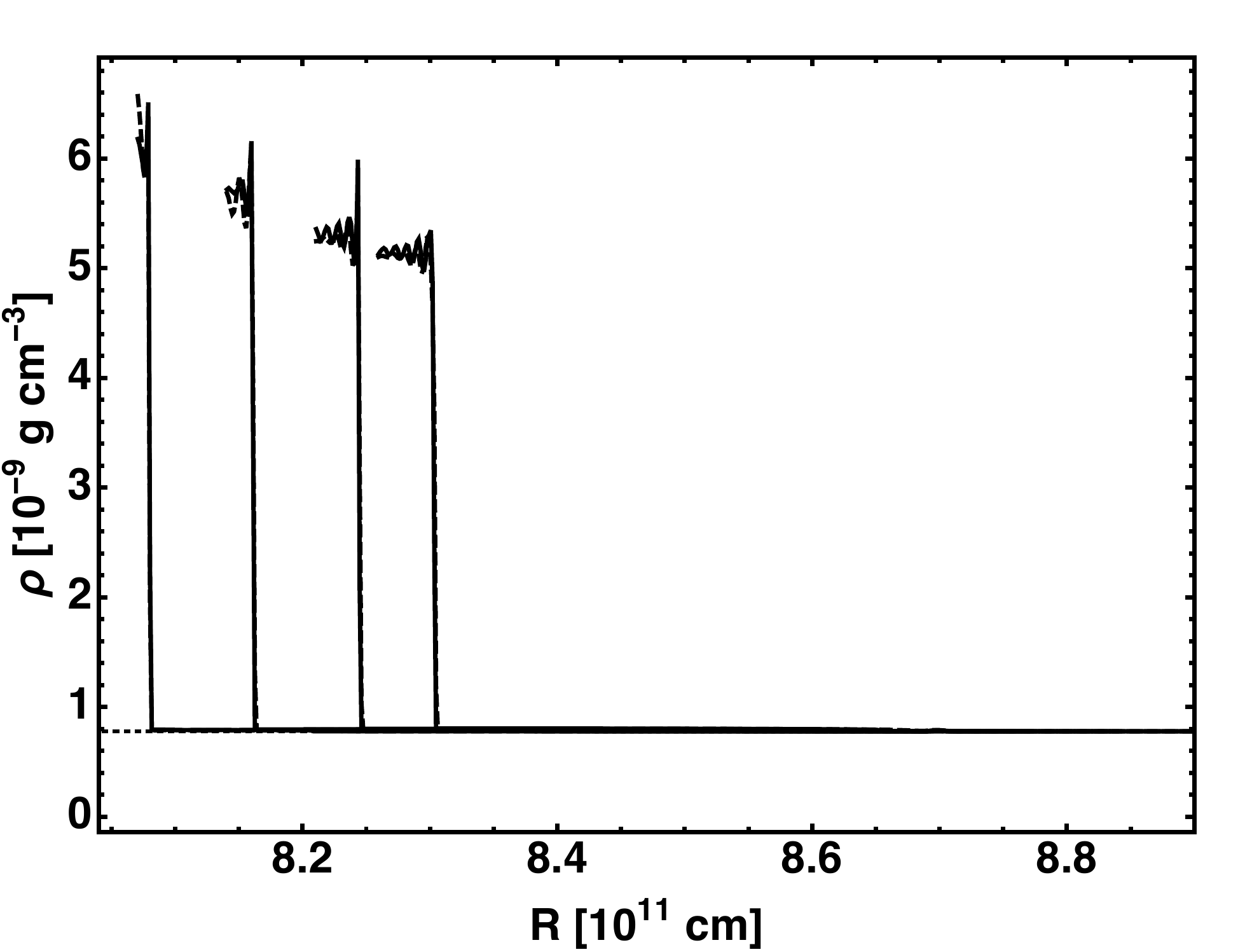}
\includegraphics[width=0.4\textwidth]{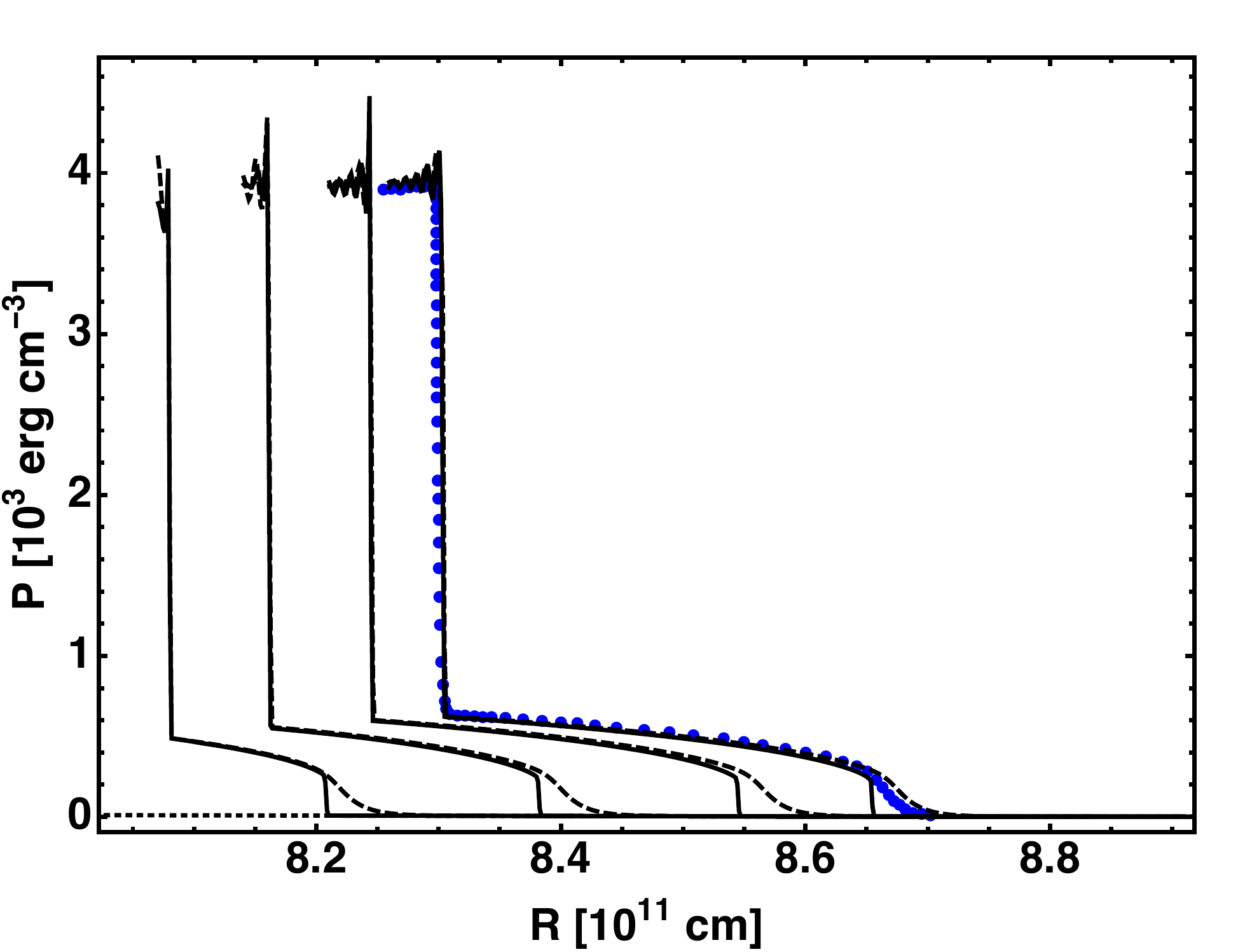}\\
\includegraphics[width=0.4\textwidth]{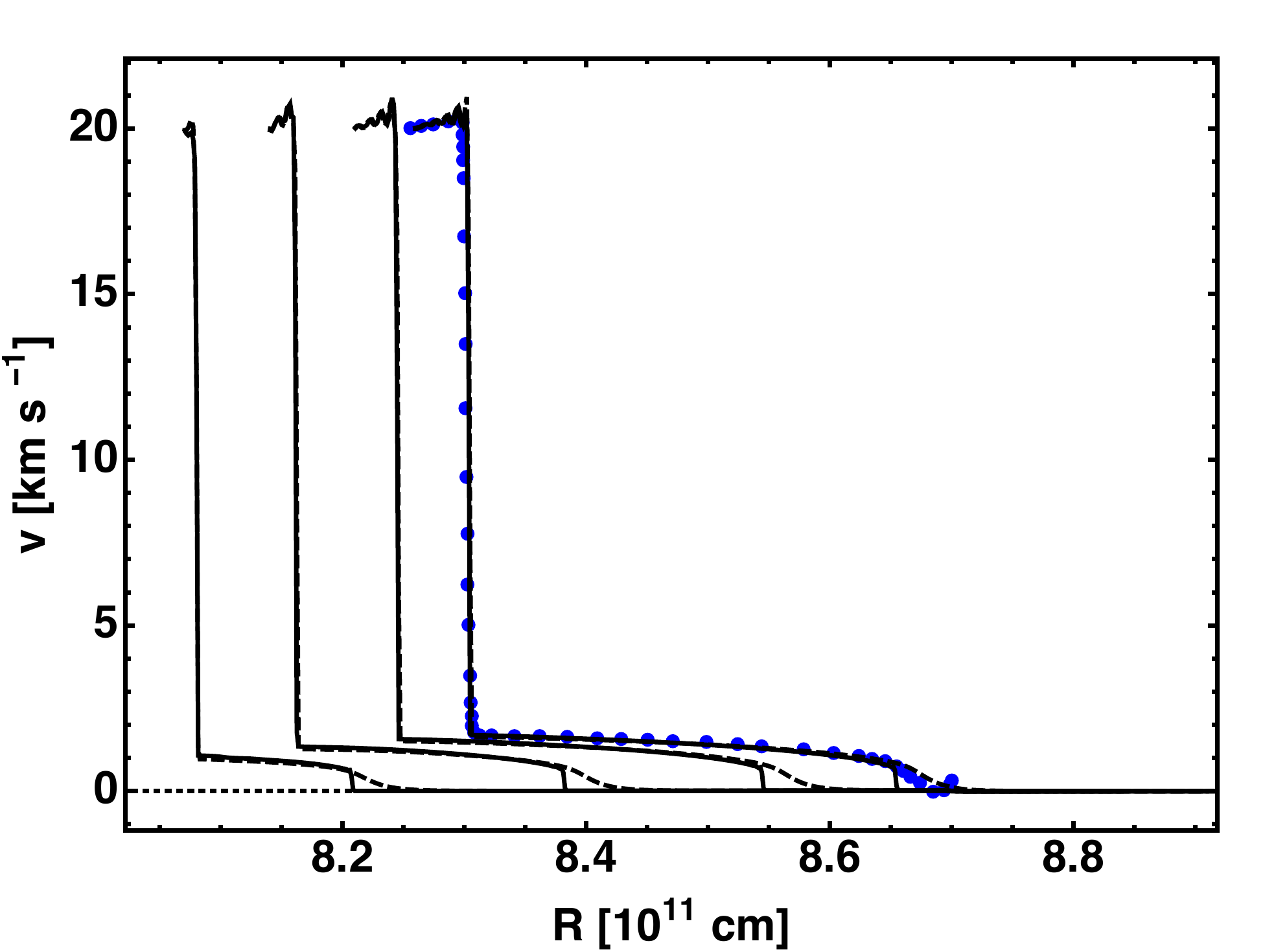}
\includegraphics[width=0.4\textwidth]{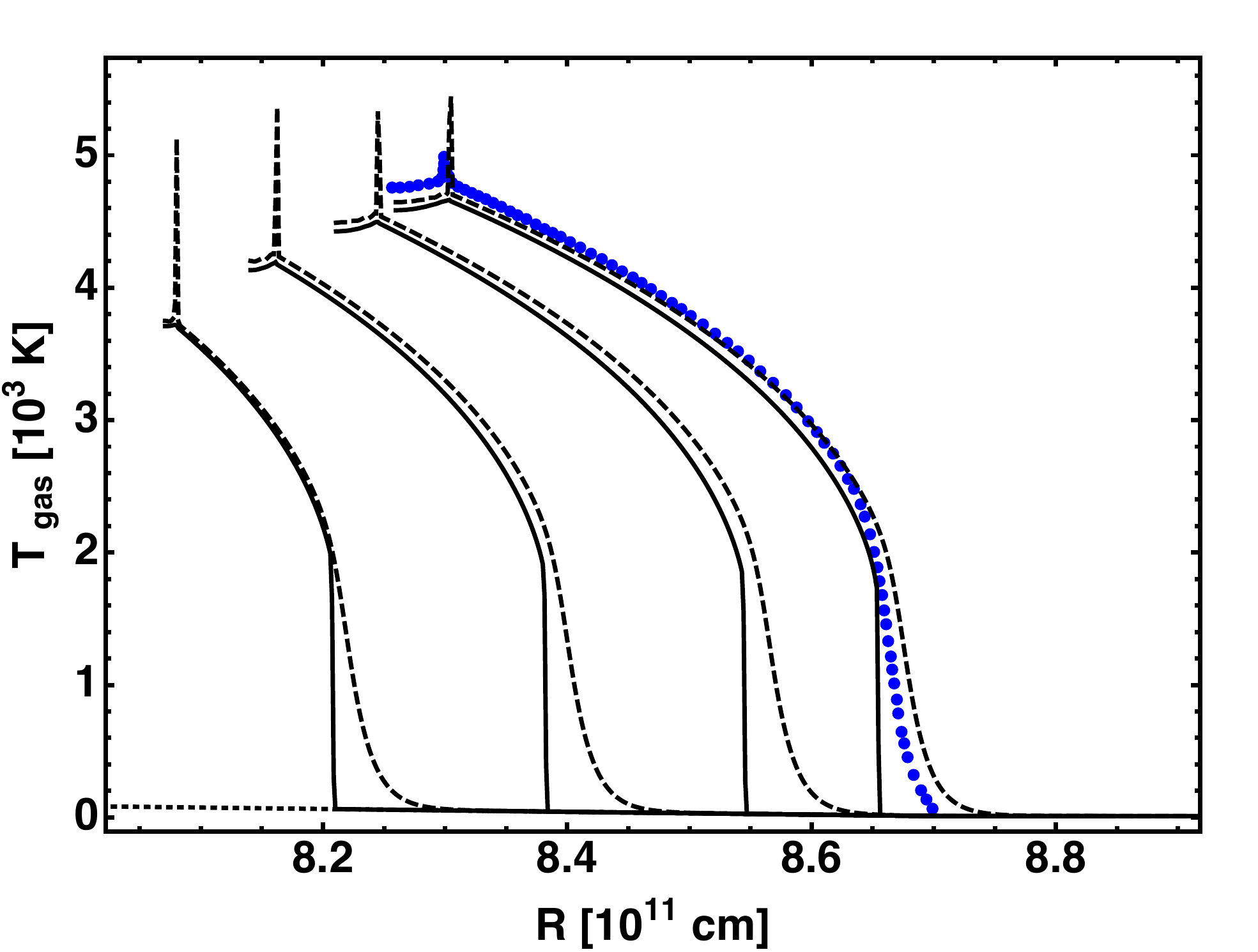}\\
\includegraphics[width=0.4\textwidth]{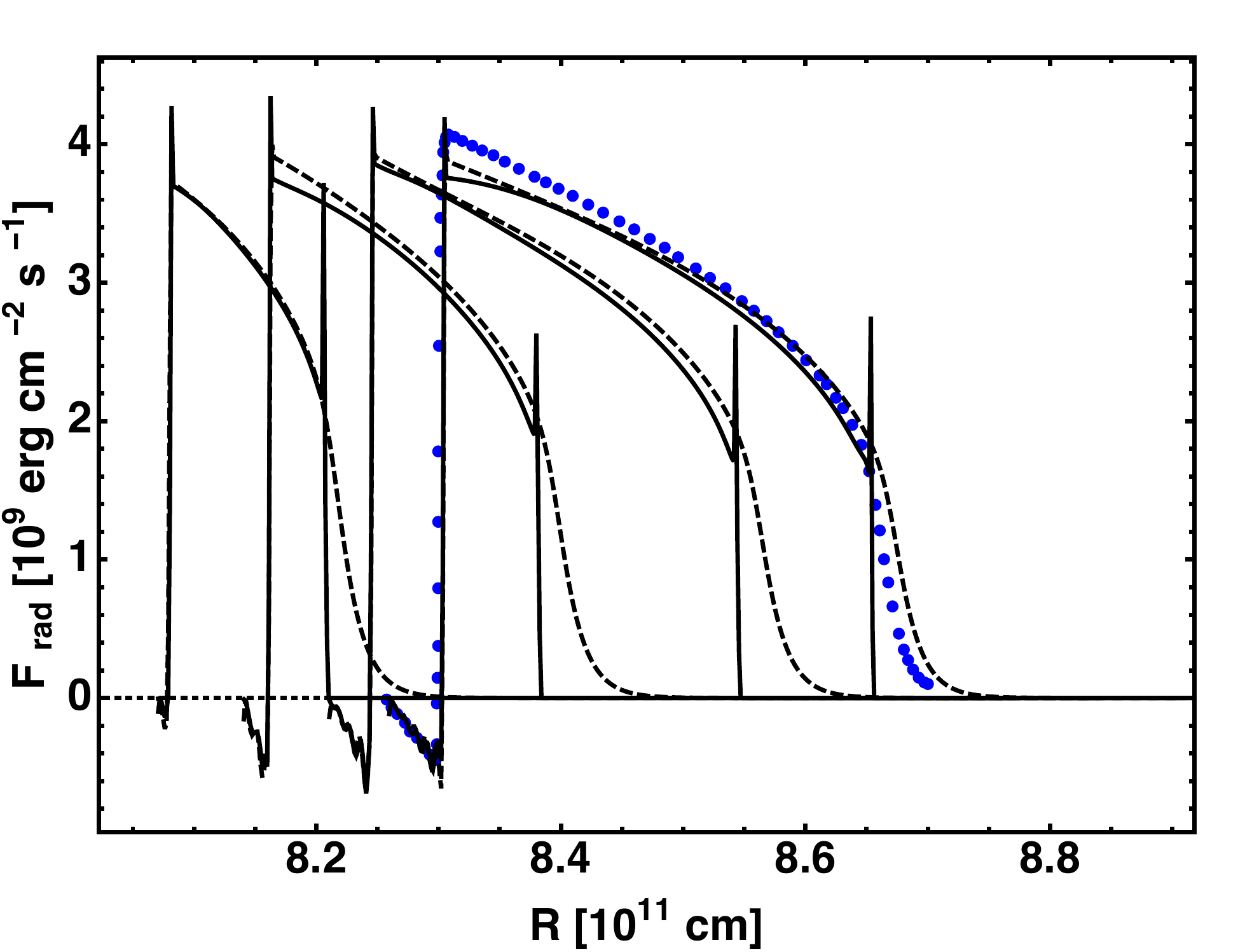}
\includegraphics[width=0.4\textwidth]{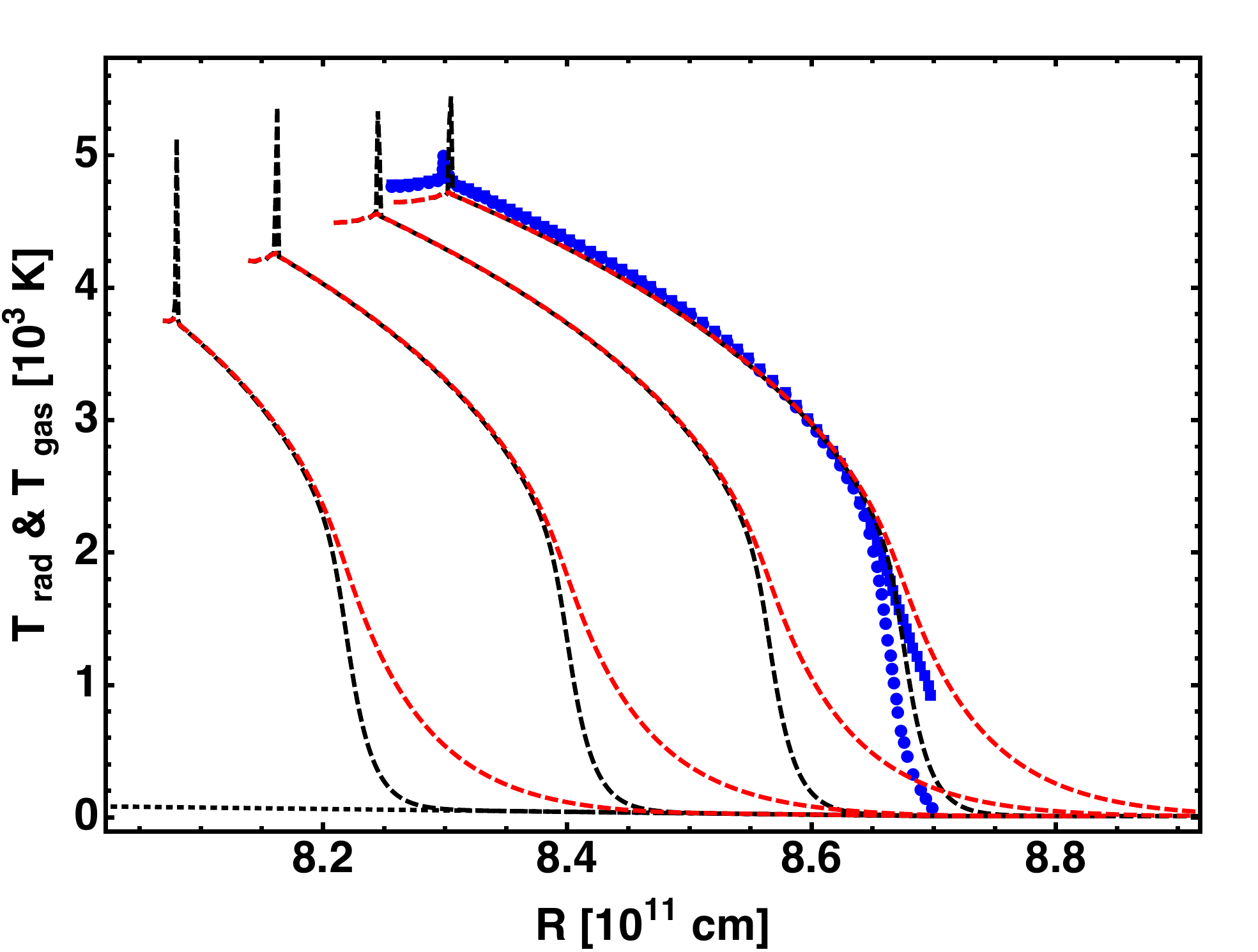}
\caption{
Gas-related properties of the radiation hydrodynamics supercritical shock tube test.
Solid lines denote results using the equilibrium temperature approach,
dashed lines results of the linearization approach.
From top to bottom and left to right, the panels show
gas density,
gas pressure,
gas velocity, 
gas temperature,
radiative flux, and
radiation temperature.
The four snapshots in time (moving from left to right in each panel) correspond to $3.5$, $7.0$, $10.5$, and $13.0 \times 10^3 \mbox{ s}$.
Dotted lines denote the initial setup.
In the bottom-right panel, dashed black lines denote the gas temperature as also shown in the middle-right panel, dashed red lines denote the radiation temperature.
\vONE{
Blue dots denote the numerical data extracted from \citet{1994ApJ...424..275E}, their Figs.~10 and 11, using the Web Plot Digitizer tool (\href{https://apps.automeris.io/wpd}{https://apps.automeris.io/wpd}) for $t = 3.8 \times 10^4 \mbox{ s}$.
The abscissa was converted into the original setup configuration of \citet{1994ApJ...424..275E} to ease comparison, i.e., properties are shown in the non-comoving laboratory frame with the piston starting at a radius of $R_\mathrm{min} = 8 \times 10^9 \mbox{ cm}$.
}
}
\label{fig:tests_radiationhydrodynamics_supercriticalshocktube}
\end{center}
\end{figure*}
The physical setup and numerical configuration of the supercritical radiative shock tube test is identical to the subcritical case of the previous section with the exception that the piston is moving at a higher velocity of $u_\mathrm{piston} = 20 \mbox{ km s}^{-1}$.
\vONE{
A supercritical radiative shock is characterized by the fact that the gain of internal energy of the gas due to compressional heating cannot sufficiently be radiated away.
Hence, the preshock radiation temperature resembles the postshock radiation temperature.
In the case of a subcritical radiative shock, the preshock radiation temperature declines rapidly.
}

\paragraph{Results}
The evolution of the shock structure is presented in Fig.~\ref{fig:tests_radiationhydrodynamics_supercriticalshocktube} for four instances in time, namely $1.17$, $2.33$, $3.50$, and $4.43 \times 10^4 \mbox{ s}$.
The last snapshot in time corresponds to the last time shown in the figures by \citet{1994ApJ...424..275E}.

Again, both radiation transport methods yield the basic shock parameters such as its propagation speed and the physical properties in the shocked and upstream direction.
The equilibrium temperature approach assumes that the radiation temperature is identical to the gas temperature.
Hence, neither a spike in gas temperature at the shock front nor the radiative precursor of the radiation temperature are present.
Hence, the bottom-right panel of Fig.~\ref{fig:tests_radiationhydrodynamics_supercriticalshocktube} shows the result for the two-temperature linearization approach only.
Moreover, the equilibrium temperature approach -- as in the subcritical case -- leads to a steeper gradient in temperature and, hence, to a more efficient cooling of the shock.
As a result, the gas temperature is significantly smaller than in the linearization approach.

Although the two-temperature linearization approach yields a spike in gas temperature, the structure of this spike is not resolved at the grid resolution used.
The maximum value of the gas temperature in the spike increases with higher spatial resolution.
This is also the reason why the maximum values presented herein are higher than in the original \citet{1994ApJ...424..275E} simulation runs, which used 300 grid cells instead of 512.
Fig.~\ref{fig:tests_radiationhydrodynamics_supercriticalshocktube} shows the formation of the radiative precursor when the linearization approach is used.

\vONE{
The simulation results of the supercritical shock are in overall very good agreement with the results from the original \citet{1994ApJ...424..275E} study.
Our last snapshot presented seems to be a little bit more advanced in time than the last snapshot from the original article.
As for the subcritical shock, the radiative precursor in our simulation of a supercritical shock is found to be stronger than in the original study.
Keeping in mind the differences between the hydrodynamics schemes of the two studies, and the fact that the exact form of the radiative precursor will depend on the choice of the flux limiter function (we used the one by \citet{1981ApJ...248..321L} for these simulations), the agreement of the simulation results is satisfactory.
}

\structuring{\clearpage}
\subsection{Ionization}
\subsubsection{Str\"omgren Sphere}
\paragraph{Context}
Str\"omgren spheres relate to the classical solution of the size of a H II region for 
a source of constant EUV photon luminosity injected into 
a medium with a uniform number density of hydrogen gas.
The analytical solution for the radius $R_\mathrm{St}$ of the spherical H II region is derived from the equilibrium condition of photoionization and recombination and is given as \citep[see, e.g.,][]{1986ARA&A..24...49Y} by
\begin{equation}
\label{eq:Stroemgren}
R_\mathrm{St} = \left(
\frac{3}{4 \pi} 
\frac{S_\mathrm{EUV}}{n_\mathrm{H}^2 ~ \alpha_2} 
\right)^{1/3}
\end{equation}
The following test runs compare the numerically computed size of different Str\"omgren spheres with the analytical solution and check for the correct scaling of the H II region size with
the number per unit time of ionizing photons $S_\mathrm{EUV}$ emitted by the star,
the hydrogen number density $n_\mathrm{H}$, and
the coefficient for recombinations into any but the hydrogen ground state $\alpha_2$.

\paragraph{Physical setup}
A source of a fixed luminosity of ionizing photons $S_\mathrm{EUV}$ is placed into a uniform medium of initially neutral atomic hydrogen gas mass density $\rho_\mathrm{H} = n_\mathrm{H}~m_\mathrm{H}$. 
We use three different photon number luminosities $S_\mathrm{EUV} = 10^{49}, 10^{50}, \mbox{ and } 10^{51} \mbox{ s}^{-1}$.
We use various different hydrogen \vONE{gas mass} densities from $\rho_\mathrm{H} = 3 \times 10^{-23} \mbox{ g cm}^{-3}$ up to $\rho_\mathrm{H} = 3 \times 10^{-20} \mbox{ g cm}^{-3}$.
The recombination coefficient is taken to be $\alpha_2 = 2 \times 10^{-13} \mbox{ cm}^{3} \mbox{ s}^{-1}$ or $\alpha_2 = 2 \times 10^{-14} \mbox{ cm}^{3} \mbox{ s}^{-1}$.
In order to mimic the approximation used in the analytical solution that a certain sphere around the luminous source is either fully ionized or fully neutral, we set the cross section for the photoionization to an arbitrarily high value of $\sigma_\mathrm{star} = 10^{10} \mbox{ cm}^2$; this procedure guarantees that the turnover from a fully ionized ($x=1$) to a completely neutral ($x=0$) medium occurs within a single grid cell of the computational domain.
A few additional test runs used a physically reasonable value of $\sigma_\mathrm{star} = 6 \times 10^{-18} \mbox{ cm}^2$; for these tests, the smooth turnover from the fully ionized to the neutral medium can be resolved on the numerical grid, and the resulting size of the H II region is numerically computed as the radius where the ionization degree drops below 50\%.

\paragraph{Numerical configuration}
We use a one-dimensional grid in spherical coordinates from an inner radial boundary at $0.01$~pc up to an outer radial boundary of $40$~pc.
For a comparison with the analytical Str\"omgren result, the inner radial boundary should either be chosen to a value much smaller than the expected size of the H II region (as done in these tests) or the injected photon luminosity at the inner radial boundary has to be corrected analytically to take into account the photoionization of the inner volume between the point source and the inner rim of the computational domain.
The numerical grid consists of $20,000$ grid cells, and the size of each grid cell increases logarithmically toward larger radii.
\vONE{
We chose a logarithmic grid spacing to obtain the same relative spatial resolution $\Delta r / r$ for tests with small and large Str\"omgren radii.
}

Additionally, we have run these tests on a variety of two-dimensional and three-dimensional grids in spherical coordinates with uniform and logarithmic grid spacing in the radial direction as well as uniform in angle and uniform in $\cos(\mbox{angle})$ in the polar direction.

For this one-dimensional problem, the on-the-spot approximation is expected to give a very accurate result.
Nevertheless, we use this as a test for both methods, the direct ray-tracing plus diffuse EUV radiation field as well as the direct ray-tracing making use of the on-the-spot approximation.

\begin{figure}[htbp]
\centering
\includegraphics[width=0.49\textwidth]{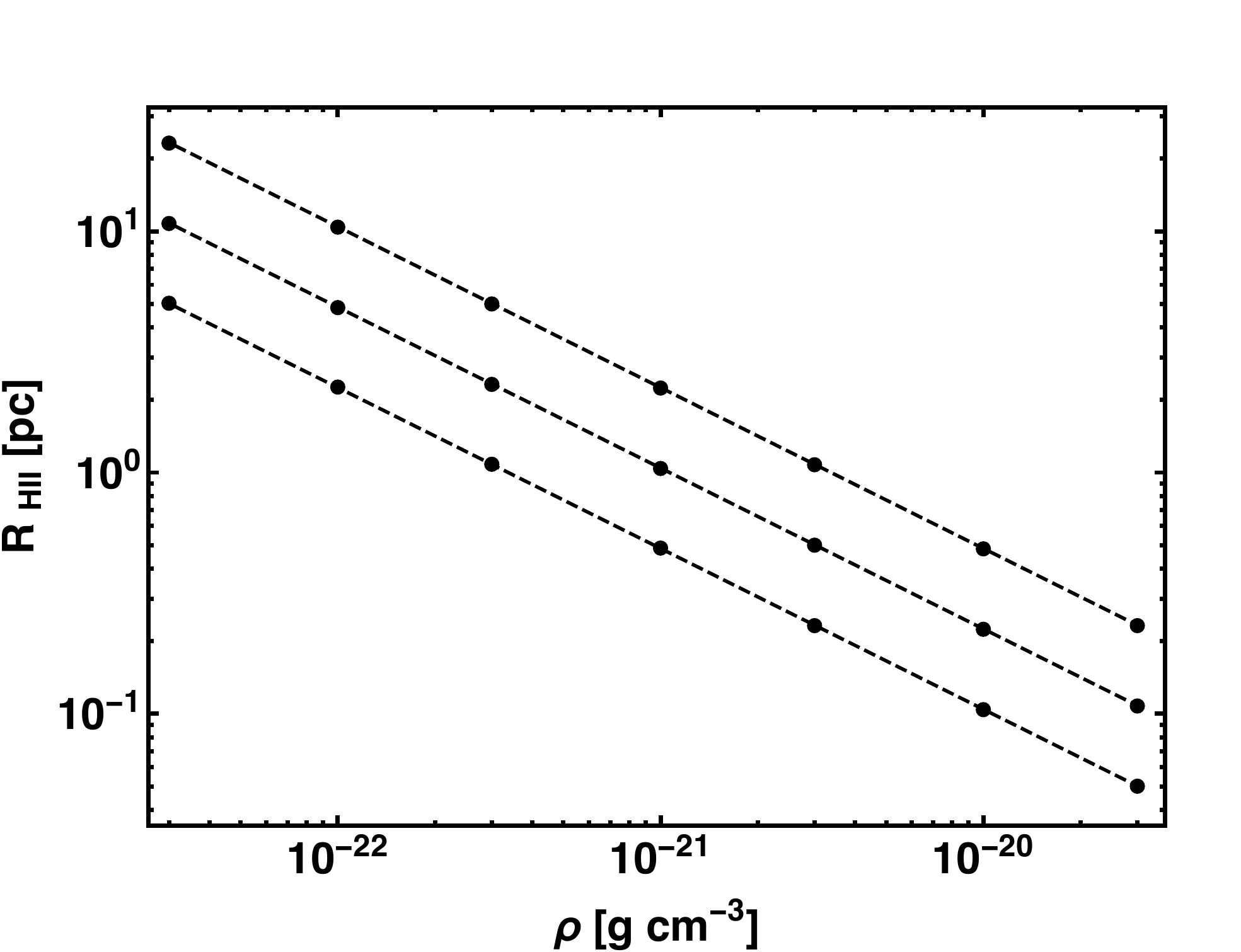}\\
\caption{
Ionization Str\"omgren test.
Size \vONE{$R_\mathrm{H II}$} of a \vONE{spherical H II region}
for hydrogen mass densities from $\rho_{H} = 3 \times 10^{-23} \mbox{ g cm}^{-3}$ up to $3 \times 10^{-20} \mbox{ g cm}^{-3}$  
around a point-like radiation source of 
three different EUV photon number luminosities of $10^{48}$, $10^{49}$, and $10^{50} \mbox{ s}^{-1}$ (from bottom to top line).
Here, \vONE{the recombination coefficient} was set to $\alpha_2 = 2 \times 10^{-13} \mbox{ cm}^{3} \mbox{ s}^{-1}$.
Filled circles represent the numerical results.
Dashed lines represent the analytical solution, \vONE{namely the Str\"omgren sphere radius from Equation \eqref{eq:Stroemgren},} for each of the luminosities, respectively.
}
\label{fig:tests_ionization_stroemgren}
\end{figure}

\paragraph{Results}
Fig.~\ref{fig:tests_ionization_stroemgren} shows the resulting radius of the H II region for the three simulation series of different photon number densities as a function of hydrogen mass density.
Here, the recombination coefficient is taken to be $\alpha_2 = 2 \times 10^{-13} \mbox{ cm}^{3} \mbox{ s}^{-1}$.

All tests result in H II region sizes in agreement with the analytical solution; the absolute value of the difference between the numerical and the analytical result is given by the spatial resolution of the grid.
The dependence of the size of the H II region on the three input parameters is in agreement with the analytical solution.

Results of simulations using the on-the-spot approximation are identical to simulations including the transport of diffuse EUV photons (as expected for such a one-dimensional problem).
Simulations using a cross section for the photoionization of $\sigma_\mathrm{star} = 6 \times 10^{-18} \mbox{ cm}^2$ yield a smooth interface between the fully ionized and the neutral medium, the radius of a turnover value of $x=50\%$ is in agreement with the analytical Str\"omgren estimate.

\structuring{\clearpage}
\subsection{Ionization Hydrodynamics}
\paragraph{Context}
Using the newly developed ionization solver module Sedna, we participated in the first STARBENCH test \citep{2015MNRAS.453.1324B}.
The STARBENCH initiative aims at benchmarking numerical codes used for star formation and stellar feedback. 
The first test problem in \citet{2015MNRAS.453.1324B} investigates the early and late D-type expansion phases of a H II region.
Results obtained with the Sedna + PLUTO code package agree with the other participating code results in terms of the expansion velocity in the early and late expansion phases and the final equilibrium extent of the H II region.
In terms of performance, Sedna + PLUTO allowed for one of the highest resolution results.
Furthermore, due to the use of spherical coordinates, we could apply the code to both the 1D and the 3D test cases, and the expanding H II region maintains the assumed spherical symmetry of the test problem very accurately.

Nevertheless, we repeat here the D-type expansion tests, because we have introduced slight modifications to the ionization solver package.

\vONE{
\paragraph{Physical setup}
The physical setup follows the 1D test of the original benchmark study by \citet{2015MNRAS.453.1324B}.
The ionization hydrodynamics test describes a spherically symmetric D-type expansion of a H II region into a uniform density medium.
The medium is assumed to consist solely of atomic hydrogen, and the uniform gas mass density is set to $\rho_\mathrm{gas} = 5.21 \times 10^{-21} \mbox{ g cm}^{-3}$.
The medium is initially at rest $v_\mathrm{gas} = 0$.
Initially, the gas is fully neutral.
The EUV point-like radiation source is set to $10^{49}$ photons per second.
Using the two-temperature thermodynamics approximation, the gas temperature of the ionized gas is set to
$T_\mathrm{gas}^\mathrm{ion} = 10^4 \mbox{ K}$, 
while the gas temperature of the neutral gas is set to either $T_\mathrm{gas}^\mathrm{neu} = 10^2 \mbox{ K}$ in the early-phase test and to $10^3 \mbox{ K}$ in the late-phase test.
To mimic two-temperature local isothermal conditions, an ideal equation of state is used with an adiabatic index of $\gamma = 1.0001$.
The test is computed using the on-the-spot approximation for the recombination field; the associated recombination coefficient is set to $\alpha_2 = 2.7 \times 10^{-13} \mbox{ cm}^3 \mbox{ s}^{-1}$.
Collisional ionization and radiation forces are ignored.
}

\vONE{
\paragraph{Numerical configuration}
The spherically symmetric test problem is realized on a 1D grid in spherical coordinates with the ionization radiation source placed at the origin of the coordinate system.
In the early-phase test, the simulation domain extends from $0.05 \mbox{ pc}$ up to $2.5 \mbox{ pc}$.
In the late-phase test, the simulation domain extends from $0.25 \mbox{ pc}$ up to $12 \mbox{ pc}$.
In both cases, we use $10^4$ uniformly spaced radial grid cells to cover the computational domain.
}

\begin{figure}[htbp]
\centering
\includegraphics[width=0.49\textwidth]{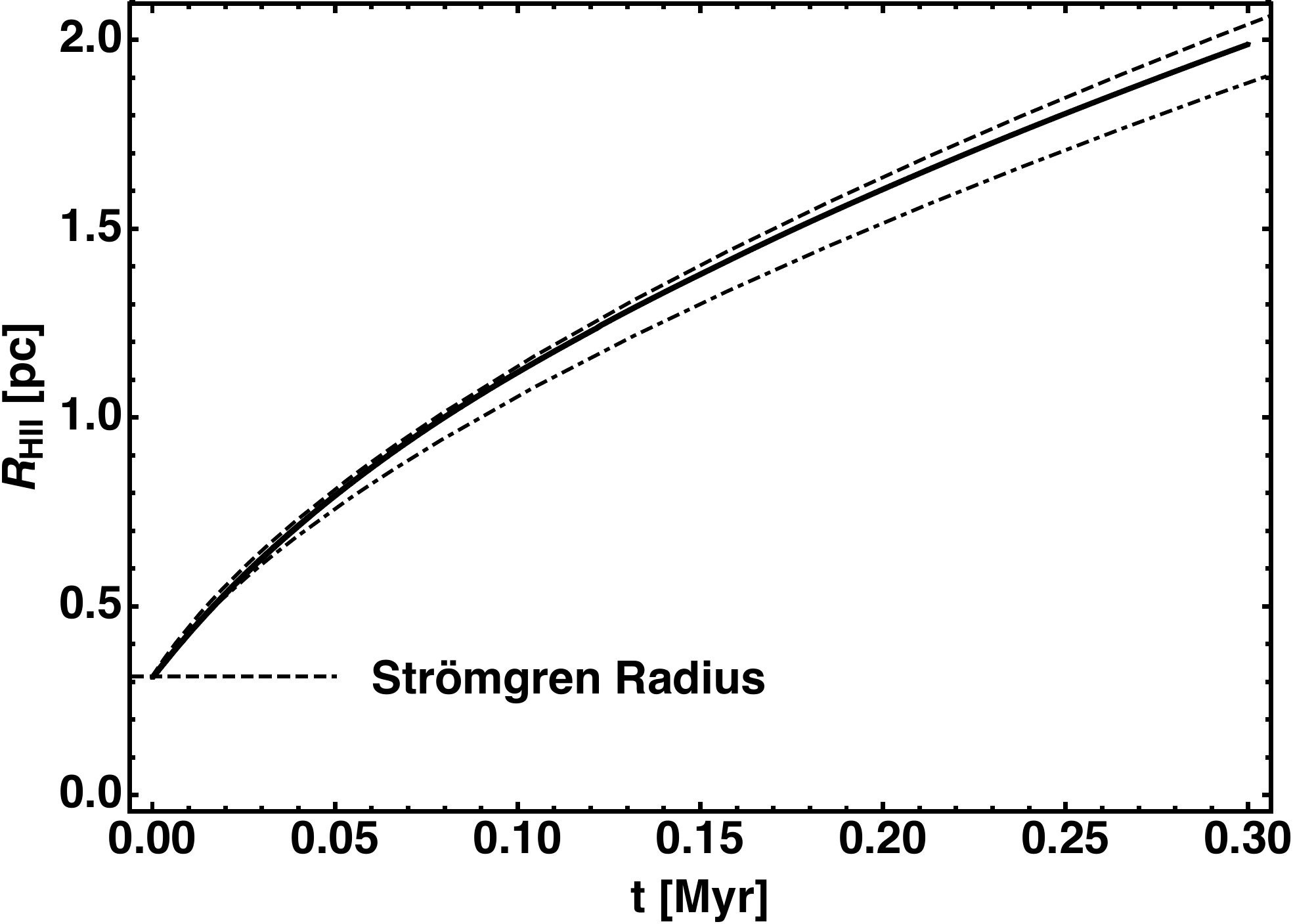}\\
\vspace{5mm}
\includegraphics[width=0.49\textwidth]{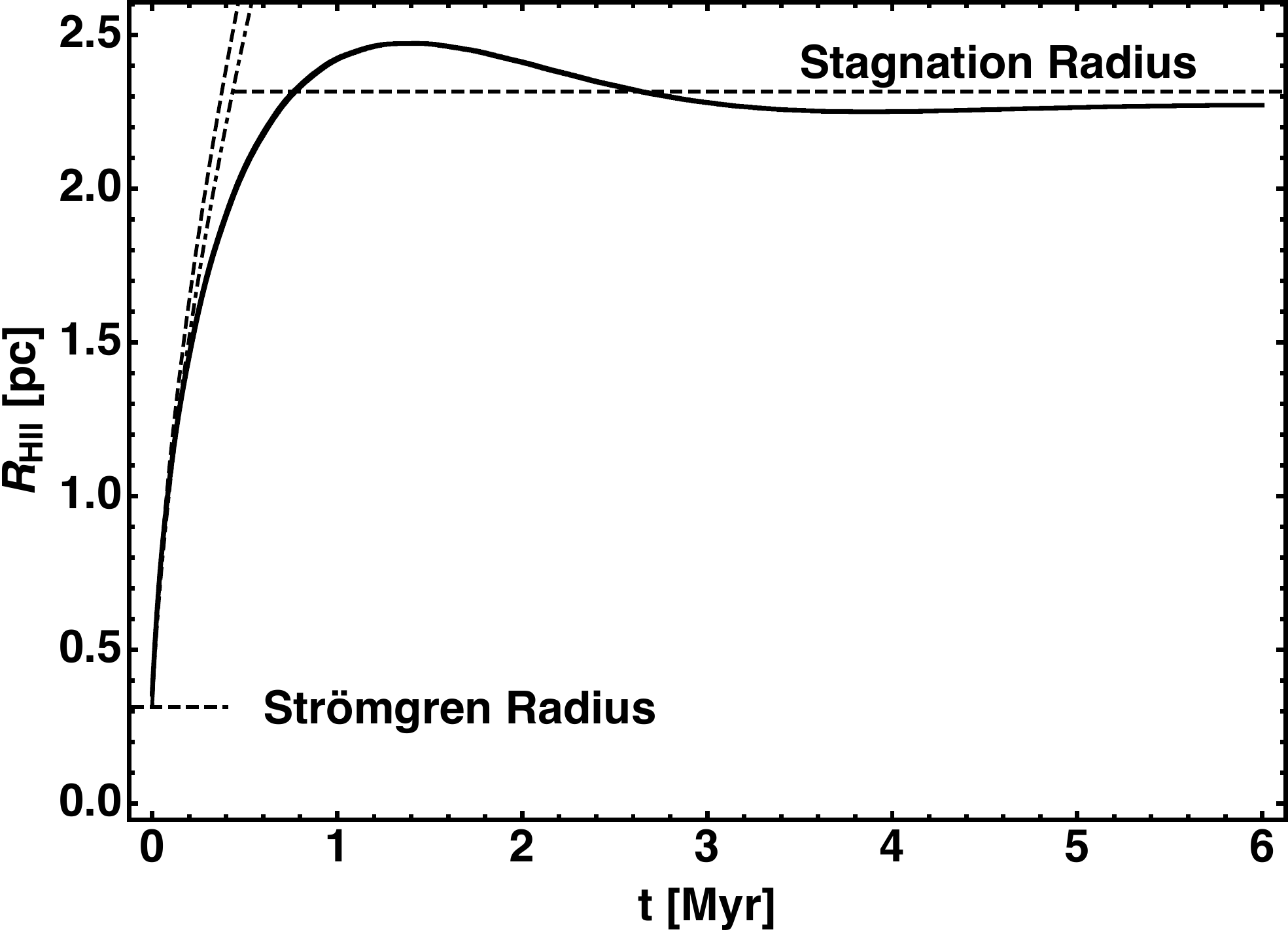}
\caption{
Ionization hydrodynamics test results of the size of an expanding H II region into a uniform density environment as a function of time.
Dotted-dashed lines denote the analytical \citet{1978ppim.book.....S} and dashed lines the analytical \citet{2006ApJ...646..240H} solution.
Horizontal dashed lines denote analytical solutions for the initial Str\"omgren sphere and the final equilibrium stagnation radius.
Solid lines denote the numerical solution.
}
\label{fig:tests_ionization_hydrodynamics}
\end{figure}

\paragraph{Results}
The resulting H II region expansion is shown in Fig.~\ref{fig:tests_ionization_hydrodynamics} for the early-stage setup (upper panel) and the late-phase setup (lower panel).
\vONE{At the onset}, the size of the initial H II region satisfies the Str\"omgren solution $R_\mathrm{H II} (t = 0) = R_\mathrm{St}$ from \vONE{Equation \eqref{eq:Stroemgren}}.
During the early phase (upper panel), the calculated \vONE{H II region} radius lies between the two analytical estimates:
the lower curve for the \citet{1978ppim.book.....S} solution,
\vONE{
\begin{equation}
R_\mathrm{H II}^\mathrm{Spitzer~(1978)}(t) = R_\mathrm{St} \times \left( 1 + \frac{7}{4} ~ \frac{ c_\mathrm{s}^\mathrm{ion}}{ R_\mathrm{St}} ~ t \right)^{4/7}
\end{equation}
}
with the sound speed $c_\mathrm{s}^\mathrm{ion}$ of the ionized gas not taking into account the ram pressure within the expanding H II regions; 
hence, it describes very well the onset of expansion in the simulation.
The upper curve, for the \citet{2006ApJ...646..240H} solution,
\vONE{
\begin{equation}
R_\mathrm{H II}^\mathrm{Hosokawa~\&~Inutsuka~(2006)}(t) = R_\mathrm{St} \times \left( 1 + \frac{7}{4} ~ \sqrt{\frac{4}{3}} ~ \frac{ c_\mathrm{s}^\mathrm{ion}}{ R_\mathrm{St}} ~ t \right)^{4/7}
\end{equation}
}
does take into account the effect of ram pressure, and hence, describes the expansion after the onset. 
During the late phase, the expanding H II region runs into an equilibrium state with the thermal gas pressure of the environment.
As a result, the H II region initially expands over this so-called stagnation radius,
\vONE{
\begin{equation}
R_\mathrm{Stagnation} = R_\mathrm{St} \times \left(\frac{c_\mathrm{s}^\mathrm{ion}}{c_\mathrm{s}^\mathrm{neu}}\right)^{4/3},
\end{equation}
}
with the sound speed $c_\mathrm{s}^\mathrm{neu}$ of the neutral gas
before turning back toward the equilibrium solution.
These results are in very good agreement with our and others' earlier results in \citet{2015MNRAS.453.1324B}.

As described in the original benchmark paper, these tests are done for different environmental neutral gas temperatures in the early- and late-phase setups on purpose, implying
that the two H II expansion simulations will not overlap.
As a further remark on the STARBENCH initiative, we are currently designing a follow-up test, studying the evolution of an unstable ionization front; 
more participants are welcome to join this community effort.

\structuring{\clearpage}
\subsection{Diffuse EUV Radiation from Direct Recombination}
\paragraph{Context}
In addition to the commonly used on-the-spot approximation, the algorithm implemented here allows us to compute the transport of the diffuse EUV radiation field generated by direct recombination of free electrons into the ground state of atomic hydrogen.
The governing equation is given by Equation \eqref{eq:recombination} as
\begin{equation}
\partial_t ~ u_\mathrm{rec}
+ 
\frac{\vec{\nabla} \cdot \vec{F}_\mathrm{rec}}{\langle h \nu \rangle_\mathrm{rec}}
=
+
\alpha_1(T_\mathrm{gas}) ~ n_\mathrm{H}^2 ~ x^2
- 
\chi_\mathrm{rec} ~ u_\mathrm{rec} ~ c
\end{equation}
with $\chi_\mathrm{rec} = n_\mathrm{H} ~ y ~ \sigma_\mathrm{rec}$ in the absence of continuum absorption ($\chi_\mathrm{ext} = 0$) and
$\vec{F}_\mathrm{rec} = - \langle h \nu \rangle_\mathrm{rec} ~ D_\mathrm{rec} ~ \vec{\nabla} u_\mathrm{rec}$ with $D= \lambda_\mathrm{rec} c / \chi_\mathrm{rec}$.
In the following, we label the different terms of the equation above as follows:
the divergence of the radiative flux is called the diffusion term.
The first term on the right-hand side is called the recombination term,
the second term on the right-hand side is called the absorption term.

\paragraph{Analytical solutions}
In the following, we derive analytical solutions to this equation for specific physical regimes and assumptions, which we can afterwards compare to the numerical solutions.
The two source terms on the right-hand side of the equation decouple from each other in terms of their dependence on the ionization degree $x$ of the medium.
The recombination term describes a source of diffuse ionizing photons due to direct recombination of free electrons into hydrogen's ground state and scales quadratically with the ionization degree.
The absorption term is a sink term, which describes the absorption of EUV photons; this terms scales linearly with the neutral degree $y = 1 - x$ of the medium.
Hence, in the following, we will study the equation above either for a fully ionized medium ($x=1$) or a completely neutral medium ($x=0$), in order to eliminate one of the terms on the right-hand side.

First, let us consider a fully \emph{neutral} medium ($x=0$, $y=1$).
The temporal evolution of the ionizing EUV radiation field is governed by the absorption term on the right-hand side and the diffusion term on the left-hand side.
The absorption term scales linearly with the absorption coefficient $\chi_\mathrm{rec}$,
while the diffusion coefficient is inversely proportional to the absorption coefficient;
furthermore, we can assume a uniform medium in terms of density and absorption opacities, so that the absorption coefficient is not affected by the spatial derivative in front of the term on the left-hand side.
Then, we can distinguish the transport physics into its two extreme regimes -- an optically thin and an optically thick medium.
In the \emph{optically thin} regime ($\chi_\mathrm{rec} \ll 1$), the transport of diffuse EUV photons is dominated by the term on the left-hand side, which can further simplified by the fact that the flux in the optically thin regime approaches $F_\mathrm{rec} \approx \langle h \nu \rangle_\mathrm{rec} ~ u_\mathrm{rec} ~ c$.
Hence, in a \emph{neutral optically thin} medium, we can estimate the temporal evolution of the diffuse EUV photons by
\begin{equation}
\partial_t ~ u_\mathrm{rec}
+ 
c ~ \vec{\nabla} ~ u_\mathrm{rec}
=
0
\end{equation}
In the following, we choose the initial distribution of EUV photon number density to be a Gaussian distribution of the form:
\begin{equation}
u_\mathrm{rec}(t=0)
=
u_0 ~ \exp \left(- \left( \frac{\Delta x}{h} \right)^2 \right)
\end{equation}
The temporal evolution of the peak number photon density of the Gaussian, initialized as a sharp pulse, can be estimated as
\begin{equation}
\label{eq:recombination_neutral_opticallythin}
u_\mathrm{rec}^\mathrm{max}(t)
=
u_0~\exp\left( - \sqrt{\frac{c t}{h}} \right)
\end{equation}

In the \emph{optically thick} regime ($\chi_\mathrm{rec} \gg 1$), the diffusion equation is dominated by the absorption term on the right-hand side. 
Hence, in a \emph{neutral optically thick} medium, we can estimate the temporal evolution of the diffuse EUV photons by
\begin{equation}
\partial_t ~ u_\mathrm{rec}
=
\chi_\mathrm{rec} ~ u_\mathrm{rec} ~ c
\end{equation}
This equation leads to a temporal evolution of the maximum photon number density of the form:
\begin{equation}
\label{eq:recombination_neutral_opticallythick}
u_\mathrm{rec}^\mathrm{max}(t)
=
u_0~\exp(-\chi_\mathrm{rec} ~ c t)
\end{equation}

Now, let us consider a fully \emph{ionized} medium ($x=1$, $y=0$).
Here, the temporal evolution of the ionizing EUV radiation field is governed by the recombination term on the right-hand side and the diffusion term on the left-hand side.

In the following, we chose the initial distribution of EUV photon number density to be uniform in space:
\begin{equation}
u_\mathrm{rec}(t=0) = u_0
\end{equation}
For such a constant photon density, and if we treat the outer boundaries as closed (mimicking an infinite region, i.e., zero gradient in photon density), the gradient in the left-hand side term evaluates to zero.
If the absorption opacity is also uniform in space, the absorption term is space independent, and the temporal evolution of the photon field is governed by the equation
\begin{equation}
\partial_t ~ u_\mathrm{rec}
=
\alpha_1 ~ n_\mathrm{H}^2 ~ x^2
\end{equation}
This equation has the solution
\begin{equation}
\label{eq:recombination_ionized_closed}
u_\mathrm{rec}(t)
=
\alpha_1 ~ n_\mathrm{H}^2 ~ t
\end{equation}

On the other hand, to include the effect of the diffusion term, we can set up the same constant photon density but now into a region of finite size $2 \times L_\mathrm{box}$ with open boundaries, i.e.~the photons can freely escape into both side directions.
If we set the photon number density outside the computational domain as zero, the diffusion term can be approximated by $u_\mathrm{rec} ~ c / L_\mathrm{box} = u_\mathrm{rec} / t_\mathrm{c}$ \vONE{with the photon-crossing timescale $t_\mathrm{c}$} and the resulting temporal evolution of the photon field is governed by the equation 
\begin{equation}
\partial_t ~ u_\mathrm{rec}
+
\frac{u_\mathrm{rec}}{t_\mathrm{c}}
=
\alpha_1 ~ n_\mathrm{H}^2 ~ x^2
\end{equation}
For the given initially uniform photon density within an open region, the final solution becomes
\begin{equation}
\label{eq:recombination_ionized_open}
u_\mathrm{rec}(t)
=
\alpha_1 ~ n_\mathrm{H}^2 ~ t_\mathrm{c} ~ \left( 1 - \exp\left( - \frac{t}{t_\mathrm{c}} \right) \right)
\end{equation}
In the following, we compare numerical test results to these analytic solutions/estimates.

\subsubsection{Neutral Medium Tests}
In this section, we check the numerical solver in the regime of a fully neutral medium by comparing the numerical results to the analytically derived estimates of Eqs.~\eqref{eq:recombination_neutral_opticallythin} and \eqref{eq:recombination_neutral_opticallythick} in the optically thin and optically thick regimes, respectively.

\paragraph{Physical setup}
The EUV photon number density is initially given as $u_\mathrm{rec}(t=0) = u_0 ~ \exp \left(- \left( \Delta x / h \right)^2 \right)$ with an initial maximum photon number density of $u_0 = 10^{-10} \mbox{ cm}^{-3}$ and an FWHM value of $h = 0.01$~pc.

For simplicity and ease of reproducibility, we specify all coefficients here as constants: 
the recombination rate into any state is set to $\alpha_{(1)} = 3 \times 10^{-13} \mbox{ cm}^3 \mbox{ s}^{-1}$,
while the recombination rate into the ground state is set to $\alpha_1 = 10^{-13} \mbox{ cm}^3 \mbox{ s}^{-1}$.
The recombination cross section is given as $\sigma_\mathrm{rec} = 5.53 \times 10^{-18} \mbox{ cm}^2$.
Absorption by dust grains is switched off $(\chi_\mathrm{ext} = 0)$.

The uniform hydrogen gas mass density is varied to achieve the different optical regimes. 
\vONE{
The gas is supposed to be atomic hydrogen only.
}
In the optically thin test setup, the \vONE{gas mass} density is set to $\rho_\mathrm{gas} = 10^{-30} \mbox{ g cm}^{-3}$ ($n_\mathrm{H} \approx 6 \times 10^{-7} \mbox{ cm}^{-3}$) and
in the optically thick test setup, the \vONE{gas mass} density is set to $\rho_\mathrm{gas} = 10^{-20} \mbox{ g cm}^{-3}$ ($n_\mathrm{H} \approx 6 \times 10^{+3} \mbox{ cm}^{-3}$).
Afterwards, we scan the density from the optically thin to the optically thick regime to further check the numerical solution in the transition region as well.

\paragraph{Numerical configuration}
The temporal evolution of the system is solved on a one-dimensional grid in Cartesian coordinates.
The grid extents from the left boundary at $-1$~pc up to the right boundary at $+1$~pc.
The grid consists of $1000$ grid cells with uniform grid spacing.
The domain boundaries are treated as open boundaries for the photon flux.

\begin{figure}[htbp]
\centering
\includegraphics[width=0.49\textwidth]{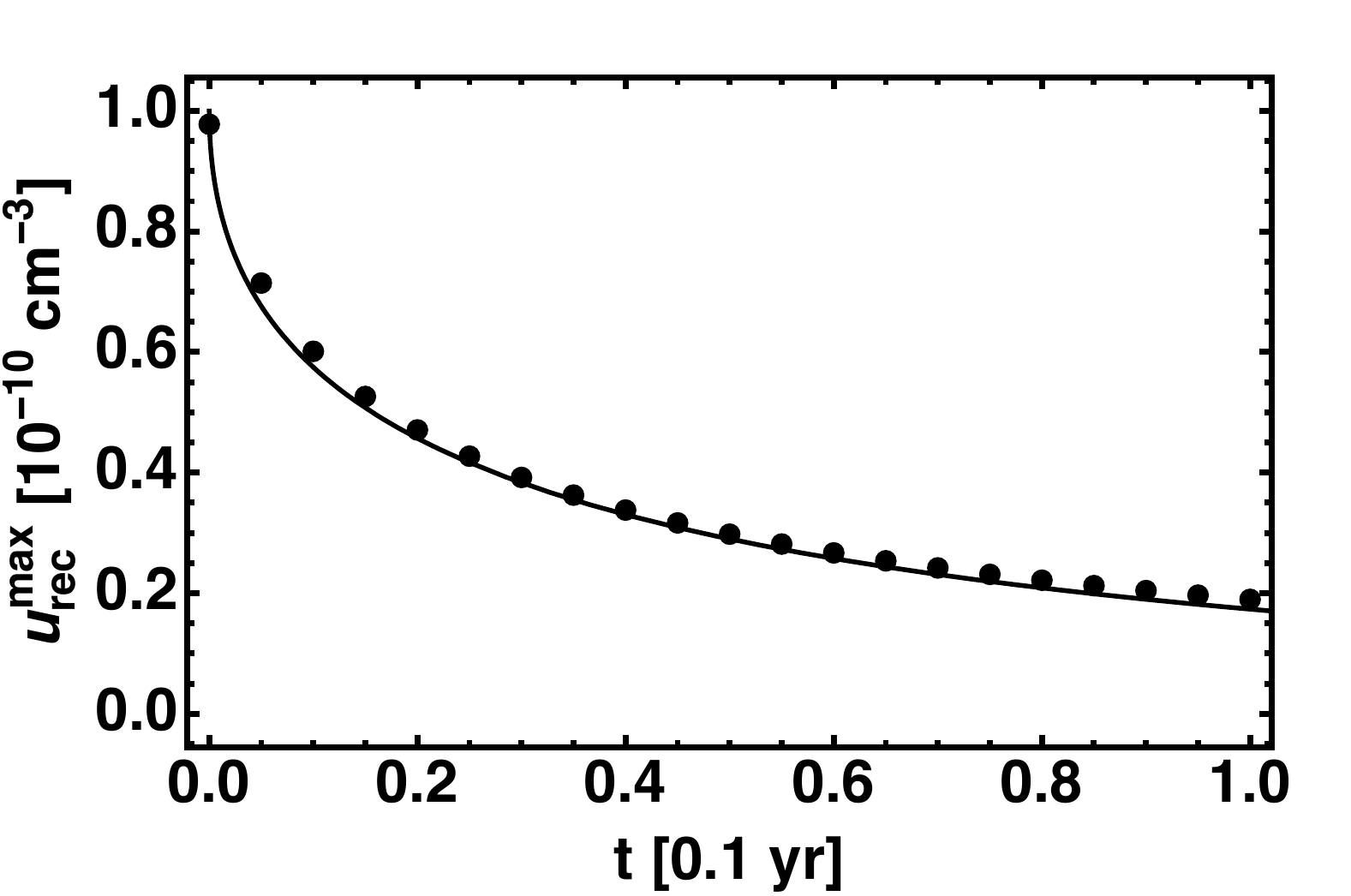}\\
\caption{
Maximum EUV photon number density as function of time for a neutral optically thin medium.
Filled circles denote the numerical result.
The solid line denotes the analytic solution.
}
\label{fig:tests_recombination_neutral_opticallythin}
\end{figure}

\paragraph{Results}
For the optically thin regime, the resulting temporal evolution of the maximum photon number density is compared to the analytic estimate in Fig.~\ref{fig:tests_recombination_neutral_opticallythin}.
The numerical results agree with the analytic solution in the neutral optically thin regime.

\begin{figure}[htbp]
\centering
\includegraphics[width=0.49\textwidth]{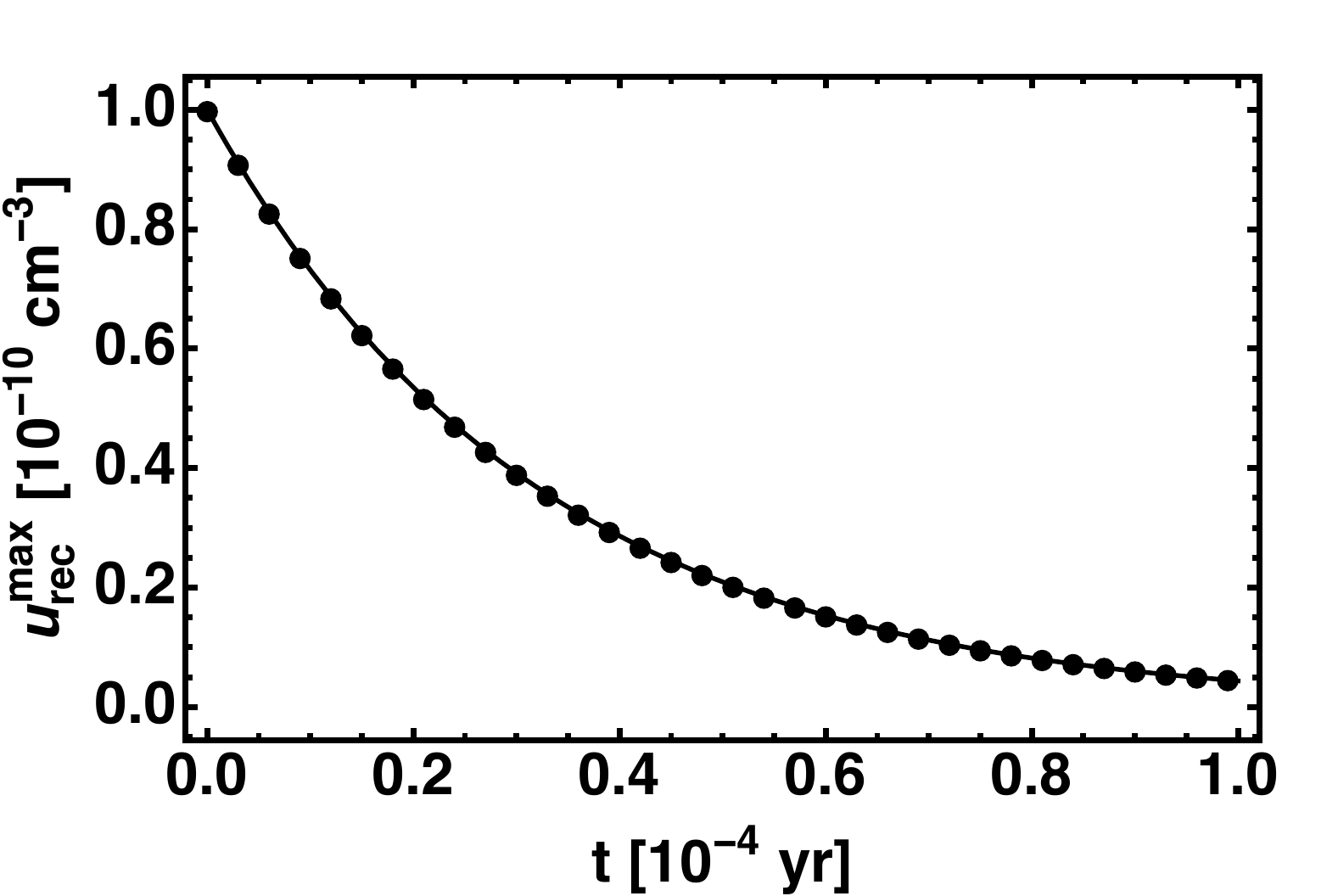}\\
\caption{
Maximum EUV photon number density as function of time for a neutral optically thick medium.
Filled circles denote the numerical result.
The solid line denotes the analytic solution.
}
\label{fig:tests_recombination_neutral_opticallythick}
\end{figure}

For the optically thick regime, the resulting temporal evolution of the maximum photon number density is compared to the analytical estimate in Fig.~\ref{fig:tests_recombination_neutral_opticallythick}.
The numerical results closely agree with the analytical solution in this absorption-dominated regime as well.

\begin{figure}[htbp]
\centering
\includegraphics[width=0.49\textwidth]{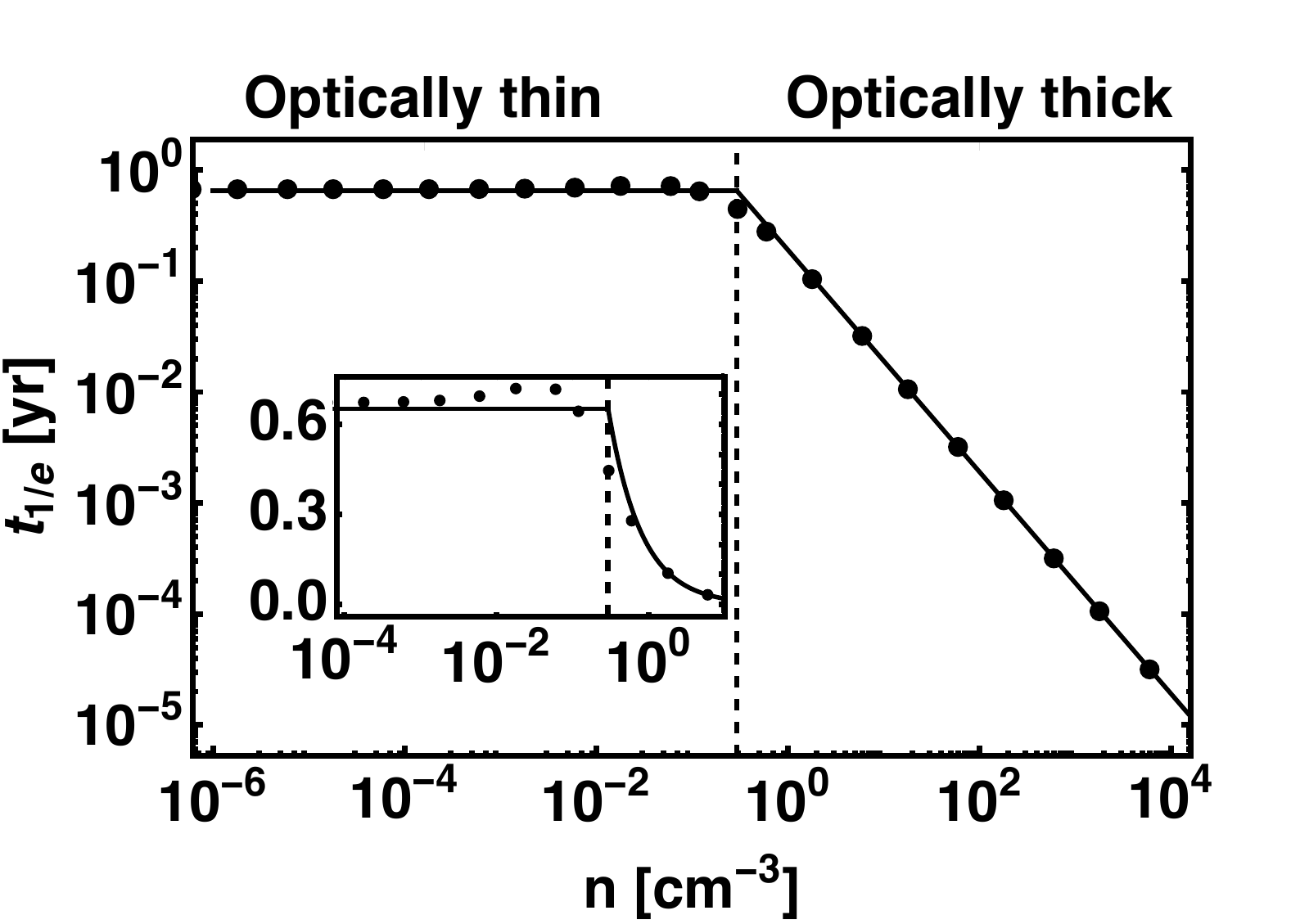}\\
\caption{
Half-life of the initial Gaussian distribution of EUV photon number density as function of hydrogen number density.
Filled circles denote the numerical results for individual simulations.
The two solid lines (horizontally and diagonally declining) denote the analytical estimates for the highly optically thin and thick regimes, respectively.
The inset figure focuses on the results at the transition between optically thin and thick around the $\tau_\mathrm{rec} \approx 1$ transition.
}
\label{fig:tests_recombination_neutral_overview}
\end{figure}

To further check the transition regime from optically thin to optically thick, we run the test case for a variety of different hydrogen number densities from $10^{-6}$ to $10^{+4} \mbox{ cm}^{-3}$, i.e.~spanning 10 orders of magnitude.
In Fig.~\ref{fig:tests_recombination_neutral_overview}, we compare the resulting half-life of the Gaussian photon distribution to the analytical estimates of the two extreme cases (highly optically thick or thin regime).
In both limiting cases (optically thin and optically thick), the resulting half-life of the numerically computed evolution is in agreement with the analytical prediction.
In the transition regime from optically thin to optically thick, we can at least understand the physical behavior of the numerical solution qualitatively:
from low densities in the optically thin limit to high densities in the optically thick limit, the 
\vTWO{
}
velocity decreases from the speed of light to zero.
Due to the fact that the diffusion velocity in the transition regime $\tau_\mathrm{rec} \approx 1$ is lower than in the optically thin limit, the half-life of the Gaussian photon distribution increases in comparison to the analytical solution, which is only valid in the optically thin limit.
By contrast, in the optically thick limit, diffusion is negligible with respect to absorption.
Approaching the transition regime $\tau_\mathrm{rec} \approx 1$, the diffusion becomes nonnegligible, and the combined effect of diffusion and absorption yields a faster decay of the Gaussian photon distribution, or -- in other words -- a half-life shorter than the analytic solution for the optically thick limit.
As a result, the highest deviation from the analytic solutions occurs in the transition regime from optically thin to thick around $\tau_\mathrm{rec} \approx 1$, because the analytic solutions derived for the extreme limits are invalid in the transition regime.
One should also keep in mind that the FLD approximation used in the numerical solver is expected to produce the strongest deviations from the correct radiation transport solution at the transition from the optically thin to the optically thick regimes.

\subsubsection{Ionized Medium Tests}
In this section, we check the numerical solver in the regime of a fully ionized medium by comparing the numerical results to the analytically derived solutions of Eqs.~\eqref{eq:recombination_ionized_closed} and \eqref{eq:recombination_ionized_open} for the infinite and finite domains, respectively.

\paragraph{Physical setup}
The EUV photon number density is initialized to zero, $u_\mathrm{rec}(t=0) = 0$.
The uniform hydrogen gas mass density is set to $\rho_\mathrm{gas} = 10^{-21} \mbox{ g cm}^{-3}$.
The recombination rates and cross sections are identical to the neutral medium tests above.
To keep the medium fully ionized during the entire runtime of the simulation, we irradiate the medium along the x-direction with an incoming direct flux of ionizing EUV photons of $10^{60} \mbox{ s}^{-1}$, which is handled during the ray-tracing step of the overall photoionization solver scheme.

\paragraph{Numerical configuration}
The extent and resolution of the computational domain are identical to the tests of the neutral medium case.
Explicitly, the temporal evolution of the system is solved on a one-dimensional grid in Cartesian coordinates.
The grid extents from the left boundary at $-1$~pc up to the right boundary at $+1$~pc.
The grid consists of $1,000$ grid cells with uniform grid spacing.

The domain boundaries for the photon flux are either treated as closed zero gradient boundaries (to mimic the infinite medium setup) or open free-streaming boundaries (to mimic the finite medium).

\begin{figure}[htbp]
\centering
\includegraphics[width=0.49\textwidth]{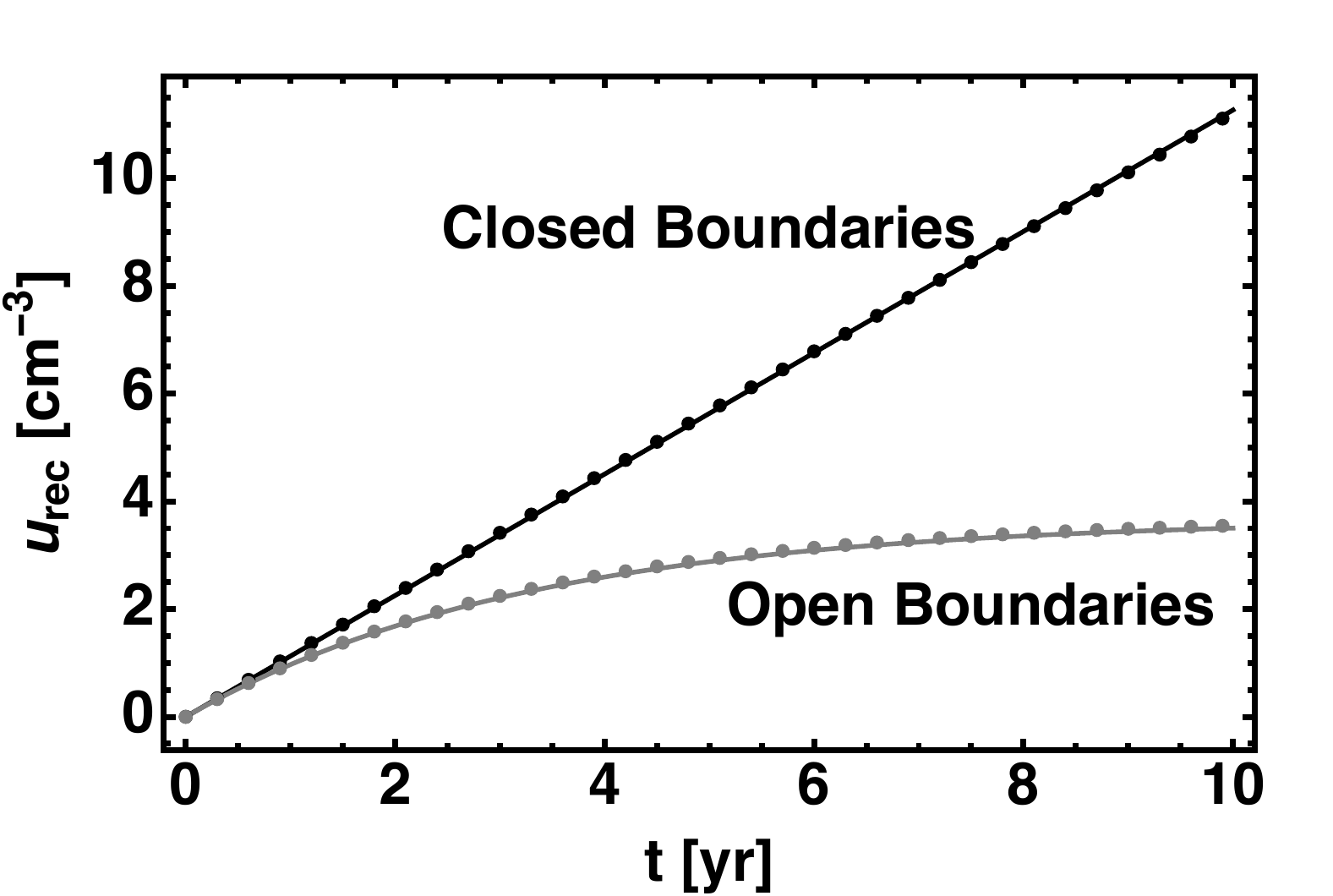}\\
\caption{
Maximum EUV photon number density as function of time for an ionized medium.
Black refers to a medium with closed boundaries.
Gray refers to a medium with open boundaries.
Filled circles denote the numerical results of both simulations.
Solid lines denote the analytically derived solutions.
}
\label{fig:tests_recombination_ionized}
\end{figure}

\paragraph{Results}
The resulting evolution of the maximum value of the diffuse EUV photon number density is shown for both cases -- the open as well as the closed boundaries -- in Fig.~\ref{fig:tests_recombination_ionized}.
In both cases, the numerical results are in agreement with the analytic solutions.
For the case of a closed system (or a medium of infinite size), the maximum photon number density increases linearly with time.
For the case of an open system (or a medium of finite size), the maximum photon number density increases monotonically in time and approaches an equilibrium value determined by the recombination rate and the photon loss across the outer boundaries.

\structuring{\clearpage}
\subsection{Radiation--Ionization Forces}
In the following tests, we check the numerically computed force terms, i.e.~the absorbed radiative momentum for each of the different radiative fluxes.

\paragraph{Physical setup}
A constant source of radiation is placed into a \vONE{finite} gaseous and dusty cloud of uniform mass density.
The various test simulations shown here differ in 
the physics modules included (stellar emission of thermal (nonionizing) radiation, dust emission of thermal radiation, stellar emission of ionizing photons, and recombination radiation from direct recombination into the hydrogen ground state), 
the strength of the different components of the total radiation field, and 
the optical depth of the surrounding medium.

The cloud radius is set to $100.0$~pc,
the dust-to-gas mass ratio is set to 1\% in all tests performed, and -- for simplicity -- the dust evolution routines, which handle evaporation and sublimation, are switched off.
The tests of the thermal continuum radiation (Sect.~\ref{Thermalcontinuumradiation}) utilize a constant opacity of $\kappa = 324.081 \mbox{ cm}^{2} \mbox{ g}^{-1}$ throughout the medium, which yields a total optical depth of $\tau = 1 \times \rho_\mathrm{gas} / (10^{-21} \mbox{ g cm}^{-3})$.
For these tests, we vary the uniform density to compute tests at $\tau = 0.1, 1, \mbox{ and } 100$, respectively.

All other tests in this section make use of the frequency-dependent dust opacities from \citet{1994A&A...291..943O}, which we extend toward the FUV and EUV regimes with two additional frequency bins.
The opacity in these two frequency bins is set to $40,000.0 \mbox{ cm}^{2} \mbox{ g}^{-1}$ per gram dust in the FUV ($6.0 \mbox{ eV} < h \nu < 13.6 \mbox{ eV}$) and $20,000.0 \mbox{ cm}^{2} \mbox{ g}^{-1}$ per gram dust in the EUV bin ($h \nu > 13.6 \mbox{ eV}$), respectively.

In the tests of the stellar and diffuse thermal continuum radiation (Sect.~\ref{Thermalcontinuumirradiationanddiffuseradiation}), we vary the uniform density between 
$\rho_\mathrm{gas} = 10^{-18} \mbox{ g cm}^{-3}$, $3 \times 10^{-18} \mbox{ g cm}^{-3}$, and $10^{-17} \mbox{ g cm}^{-3}$.
In the tests in Sect.~\ref{ThermalcontinuumandionizingEUVirradiation} and Sect.~\ref{IonizingEUVirradiationanddiffuseEUVradiation}, including the photoionization components, the gas mass density is set to $\rho_\mathrm{gas} = 10^{-21} \mbox{ g cm}^{-3}$.

In all tests, the central radiation source has \vONE{an effective photospheric surface temperature corresponding to} a radius of $10 \mbox{ R}_\odot$.
In the tests of Sect.~\ref{ThermalcontinuumandionizingEUVirradiation}, the luminosity is varied from $10^3 \mbox{ L}_\odot$ to $10^6 \mbox{ L}_\odot$, otherwise all tests assume a luminosity of $10^6 \mbox{ L}_\odot$.
We apply the Kurucz stellar atmosphere model (see Sect.~\ref{sect:materialproperties} for details) to compute the emitted spectrum.
Tests, which do not show results for a diffuse ionizing radiation field, use the on-the-spot approximation.
These tests utilize the radiation modules to compute the radiation force terms that enter the hydrodynamic equations; the hydrodynamic evolution itself is not taken into account here, i.e., the density structure is fixed in time.

\paragraph{Numerical configuration}
The absorbed momentum is computed on a one-dimensional grid in spherical coordinates assuming spherical symmetry around the central radiation source.
The grid extends from a left boundary at $0.001$~pc up to the right boundary at $100$~pc.
In all cases, the grid consists of $2048$ grid cells with logarithmically increasing grid spacing toward larger radii.
The outer domain boundary is set to an open boundary; the inner domain boundary is closed for the diffuse fluxes.

In the case of the purely diffuse thermal radiation tests (i.e., no stellar radiation and no ray-tracing), the irradiation luminosity from the central source is added as a source term to the FLD equation and the ray-tracer is switched off.
We checked both FLD approaches, the equilibrium as well as the linearization approach, which yielded identical results, as expected for these time-independent equilibrium test problems.

\paragraph{Results}
The results of the different simulation series are presented in the following subsections, which are ordered by the radiative components included in the numerical model.

\newpage
\subsubsection{Thermal Continuum (Nonionizing) Radiation}
\label{Thermalcontinuumradiation}
\begin{figure}[htbp]
\centering
\includegraphics[width=0.49\textwidth]{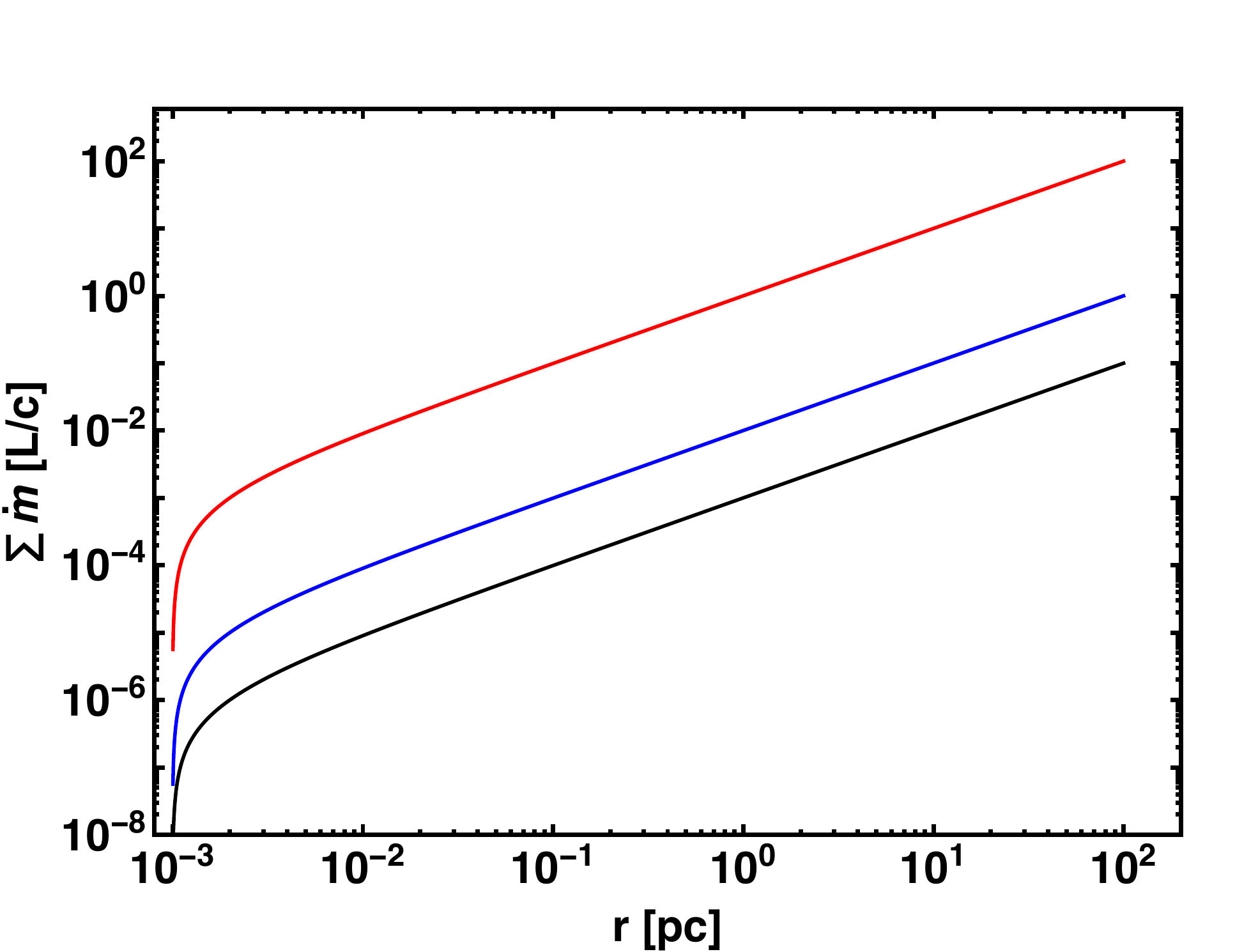}\\
\caption{
Radiative acceleration test of thermal (nonionizing) continuum radiation.
The \vONE{cumulative} absorbed radiative momentum \vONE{per unit time} is shown as a function of cloud radius for three different total optical depths of the cloud:
$\tau = 0.1$ (black),
$\tau = 1$ (blue), and
$\tau = 100$ (red).
}
\label{fig:tests_forces_diffusion}
\end{figure}

For uniform density environments and constant opacity, there is no anisotropy in the optical depth, i.e., there is no preferred path for the photons originating from the central source to diffuse through the medium.
As a result of the absorption and reemission events of the diffuse photon field, the total radiative momentum absorbed \vONE{per unit time} within the cloud scales with the central luminosity and the total optical depth (measured from the central source toward the outer cloud radius $\tau = \kappa ~ \rho ~ r_\mathrm{max}$).
We determine the absorbed radiative momentum in three setups with an optical depth of $\tau = 0.1$, $1$, and $100$; results are shown in Fig.~\ref{fig:tests_forces_diffusion}.
The \vONE{cumulative} absorbed momentum \vONE{per unit time} increases linearly with distance from the central source; this is the expected result for the uniform density and constant opacity used here; see the next section for results of temperature-dependent opacities.
When approaching the surface of the inner sink \vONE{(the inner radial boundary of the computational domain in spherical coordinates)} at $0.001$~pc from the outside, the momentum \vONE{rate} declines due to the fact that photons cannot travel through the sink itself.
As expected, the total radiative momentum absorbed \vONE{per unit time} within the cloud is given by $\dot{m}_\mathrm{tot} = \tau ~ L/c$ \vONE{with the luminosity $L$ of the radiation source at the origin of the computational domain and the speed of light $c$.}

\newpage
\subsubsection{Thermal Continuum Irradiation and Diffuse Radiation}
\label{Thermalcontinuumirradiationanddiffuseradiation}
\begin{figure}[htbp]
\centering
\vspace{-9mm}
\includegraphics[width=0.49\textwidth]{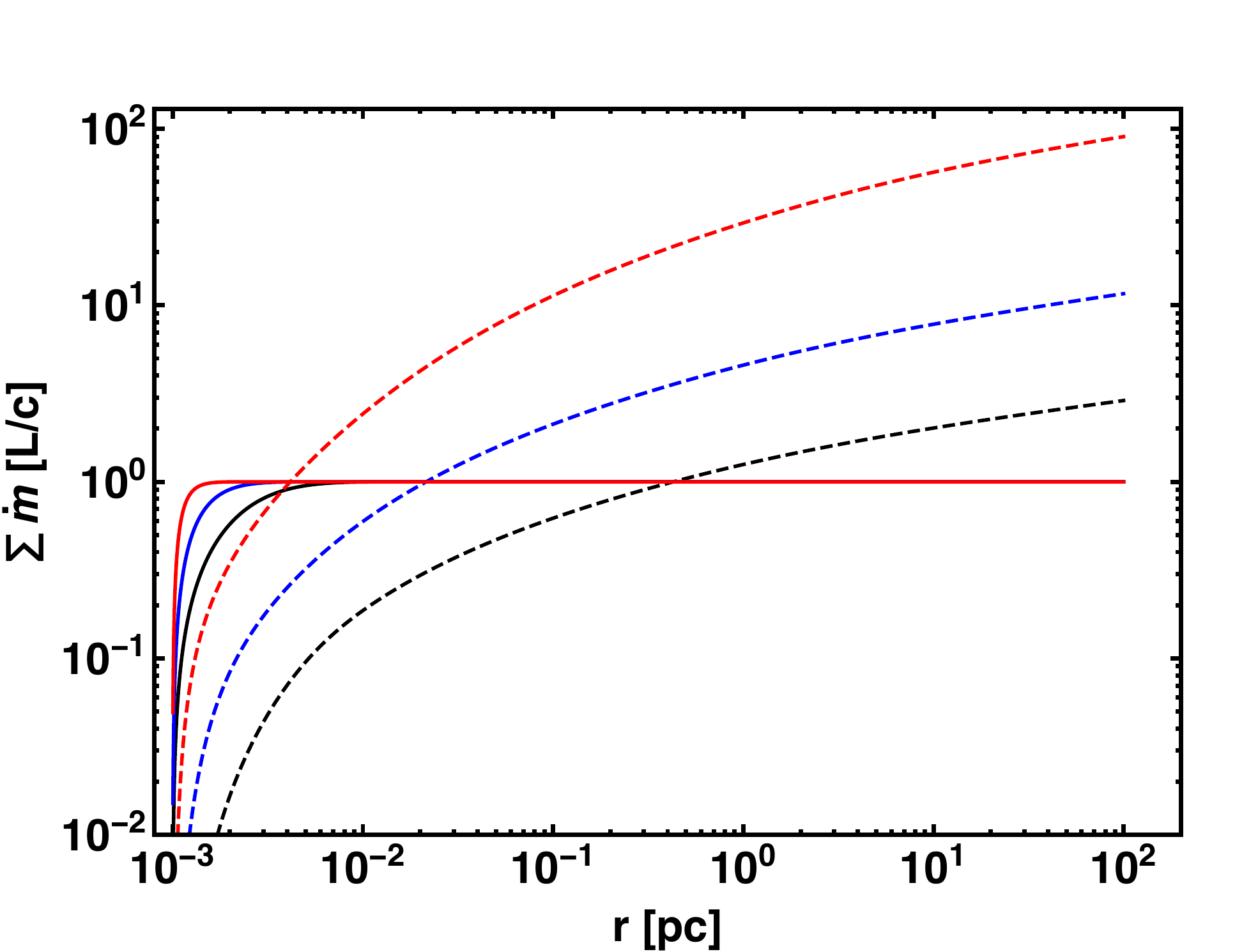}\\
\caption{
Radiative acceleration test of direct irradiation (solid lines) plus thermal (dashed lines) continuum radiation.
The \vONE{cumulative} absorbed radiative momentum \vONE{per unit time} is shown as a function of cloud radius for three different total optical depths of the cloud:
$\tau = 3$ (black),
$\tau = 12$ (blue), and
$\tau = 90$ (red).
}
\label{fig:tests_forces_irradiation+diffusion}
\end{figure}

The main difference from the tests of the previous section is that we now use different radiation transfer solvers for the central radiation source (ray-tracing) and the thermal dust reemission (FLD).
Instead of a constant opacity, we use an extended version of the frequency-dependent dust opacities from \citet{1994A&A...291..943O} here.
We determine the absorbed radiative momentum \vONE{per unit time} for three cases with a total optical depth of $\tau_\mathrm{R} = 3$, $12$, and $90$ for the diffuse component, respectively; $\tau_\mathrm{R}$ denotes the optical depth with respect to the Rosseland mean dust opacity $\tau_\mathrm{R} = \kappa_\mathrm{R, dust} ~ \rho_\mathrm{dust} ~ r_\mathrm{max}$.
The results are shown in Fig.~\ref{fig:tests_forces_irradiation+diffusion}.
The much higher Planck mean opacity for the hot irradiation source compared to the cooler dusty environment leads to an absorption of the direct irradiation component relatively close to the source.
From this first absorption region on, the reemission and absorption within the diffuse component yields an increase in the absorbed momentum \vONE{per unit time} from the diffuse radiation field.
Here, the use of the temperature-dependent Rosseland mean opacities results in a higher opacity in the warmer regions close to the central source than on the larger, cooler cloud scales.
Hence, the \vONE{cumulative} absorbed momentum \vONE{rate} does not scale linearly with the distance to the source, as was the case for the uniform local optical depth tests of the previous section. 
As expected, 
the total radiative momentum absorbed \vONE{per unit time} within the cloud is given by $\dot{m}_\mathrm{tot} = \tau ~ L/c$, whereas 
the \vONE{cumulative} radiative momentum absorbed \vONE{per unit time} from the direct irradiated component within the cloud is given by $\dot{m}_\mathrm{irr} = L/c$,
and the \vONE{cumulative} radiative momentum absorbed \vONE{per unit time} from the diffuse component within the cloud is given by $\dot{m}_\mathrm{diff} = \dot{m}_\mathrm{tot} - \dot{m}_\mathrm{irr} =  (\tau - 1) ~ L/c$.

\newpage
\subsubsection{Thermal Continuum and Ionizing EUV Irradiation}
\label{ThermalcontinuumandionizingEUVirradiation}
\begin{figure}[htbp]
\centering
\includegraphics[width=0.49\textwidth]{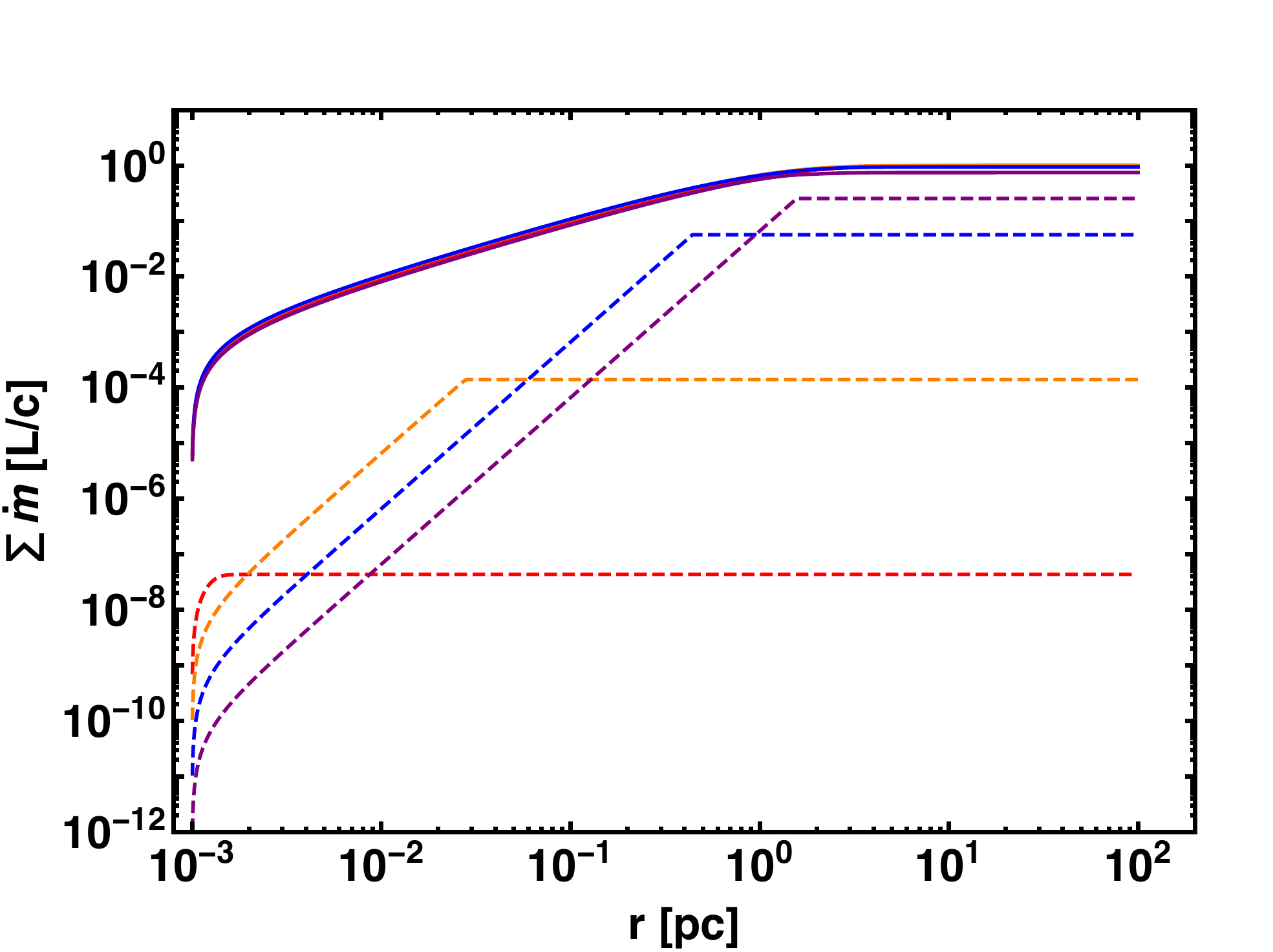}\\
\caption{
Radiative acceleration test of central nonionizing irradiation (solid lines) and ionizing (dashed lines) radiation.
The normalized \vONE{cumulative} absorbed radiative momentum \vONE{per unit time} is shown as a function of cloud radius for four different luminosities of the central radiation source:
$L = 10^3 \mbox{ L}_\odot$ (red),
$L = 10^4 \mbox{ L}_\odot$ (orange),
$L = 10^5 \mbox{ L}_\odot$ (blue), and
$L = 10^6 \mbox{ L}_\odot$ (purple).
The absorbed momentum \vONE{per unit time} is normalized in units of the central luminosity divided by the speed of light, which varies between the simulations.
}
\label{fig:tests_forces_irradiation+ionization}
\end{figure}

In this test, we switch off the diffuse radiation field component and only follow the first absorption of the central source luminosity.
In contrast to the previous tests, we consider thermal continuum (nonionizing) radiation as well as EUV photoionization.
The relative importance of the two components depends on the emitted spectrum of the central source.
We vary the luminosity from $10^3 \mbox{ L}_\odot$ to $10^6 \mbox{ L}_\odot$, which for the constant stellar radius of $10 \mbox{ R}_\odot$ implies an increasing shift of the spectrum toward the ionizing regime.
The amount of available ionizing photons in the EUV regime is determined from the Kurucz stellar atmosphere model; see Sect.~\ref{sect:materialproperties} for details.
Results of these four test runs are shown in Fig.~\ref{fig:tests_forces_irradiation+ionization}.
In all cases, the absorbed momentum \vONE{per unit time} is dominated by the thermal continuum component of the radiation field.
As a result of the lower uniform gas density $\rho_\mathrm{gas} = 10^{-21} \mbox{ g cm}^{-3}$ in contrast to the tests of the previous section, the inner absorption region of the thermal continuum component now extends out to $\approx 2$~pc.
The absorbed momentum \vONE{per unit time} from the photoionization component clearly increases with higher source luminosity.
The total radiative momentum -- i.e.~the sum of both components -- absorbed \vONE{per unit time} within the irradiated cloud is given by $\dot{m}_\mathrm{tot} = L/c$.

\newpage
\subsubsection{Ionizing EUV Irradiation and Diffuse EUV Radiation}
\label{IonizingEUVirradiationanddiffuseEUVradiation}
\begin{figure}[htbp]
\centering
\includegraphics[width=0.49\textwidth]{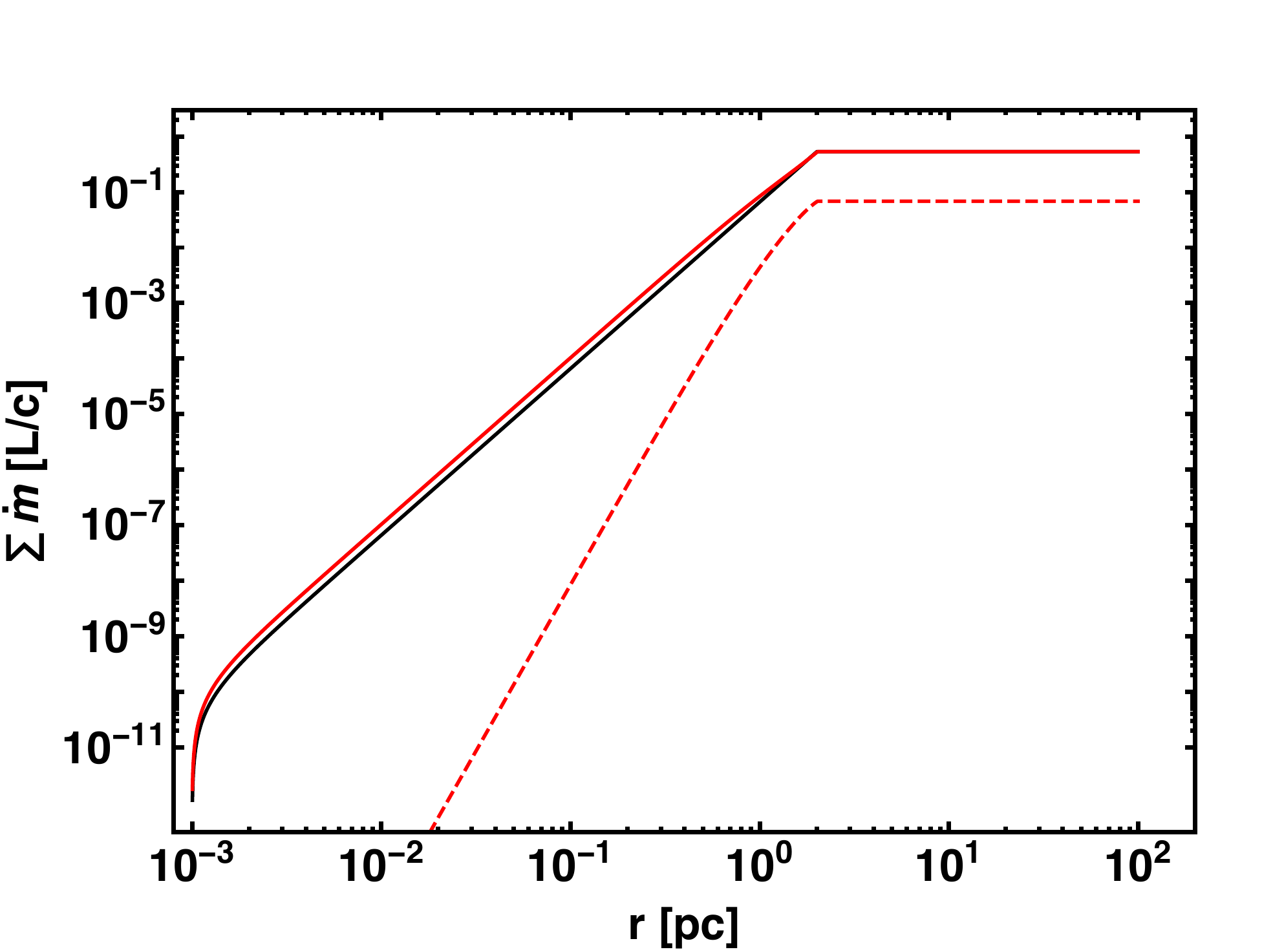}\\
\caption{
Radiative acceleration test of ionizing direct irradiation (solid lines) and ionizing diffuse (dashed line) EUV radiation.
The \vONE{cumulative} absorbed radiative momentum \vONE{per unit time} is shown as a function of cloud radius.
The black solid line denotes the results assuming the on-the-spot approximation; hence, no diffuse EUV component is determined.
The red lines denote the simulation results which include the diffuse EUV radiation field (dashed line).
}
\label{fig:tests_forces_ionization}
\end{figure}

In the tests of this subsection, we switch off the thermal continuum components and only follow the photoionization EUV radiation.
We distinguish between the EUV radiation field from direct irradiation by the central source and the diffuse ionizing EUV radiation originating from direct recombination into hydrogen's ground state.
For comparison, we perform the same simulation run using the on-the-spot approximation, as described in the method Sect~\ref{sect:onthespot}.
Results of both test runs are shown in Fig.~\ref{fig:tests_forces_ionization}.
The radiative momentum absorbed \vONE{per unit time} from the total EUV radiation field is dominated by the direct irradiation component.
Although the use of the on-the-spot recombination coefficients mimics correctly the ionization effects of the diffuse component in an isotropic environment, the approximation leads to a lower momentum absorption \vONE{rate} (the black solid line shown is slightly below the red one).
But frankly, this difference will be of less importance for most ionization simulations; the diffuse EUV component becomes mainly important due to the photoionization properties in multidimensional, anisotropic environments, where it will limit the occurrence of shadows, which are otherwise mistakenly produced when using the on-the-spot approach.
The total radiative momentum absorbed \vONE{per unit time} within the irradiated cloud is lower than $L/c$, because from the source spectrum only photons with an energy higher than $13.6$~eV contribute to the EUV radiation field.

%
%
\structuring{\clearpage}
\newpage
\section{Scientific Applications}
\label{sect:applications}
The introduced software has been used in recent years for a variety of astrophysical studies.
It was successfully applied to the research fields of
super-Earth atmosphere formation \citep{2017MNRAS.471.4662C}, 
\vONE{dynamics of protoatmospheres around low-mass planets with eccentric orbits 
\vTWO{
\citep{2020ApJ...899...54M}, 
}
}
accreting gas giants \citep{2017ApJ...836..221M, 2019ApJ...881..144M}, 
photoevaporation of protoplanetary disks \citep{2018ApJ...857...57N, 2018ApJ...865...75N}, 
low-mass star formation \citep{2018A&A...618A..95B, 2020A&A...638A..86B}, 
magnetized massive core-collapse, magneto-centrifugally driven jets, and magnetic-pressure-driven outflows from high-mass protostars \citep{2018A&A...620A.182K}, 
fragmentation of accretion disk around high-mass protostars and multiplicity in massive star formation \citep{2017MNRAS.464L..90M, 2018MNRAS.473.3615M, 2019A&A...632A..50A, 2020arXiv200813653A},
UV-line driven feedback in massive star formation \citep{2018MNRAS.474..847K, 2018MNRAS.479.4633K, 2019MNRAS.483.4893K}, 
radiation forces and photoionization feedback in massive star formation \citep{2018A&A...616A.101K}, 
stellar wind feedback in massive star formation (Kee \& Kuiper, in prep.),
photoionization feedback in the turbulent interstellar medium \citep{2020MNRAS.493.4643M},
the formation of first stars \citep{2016ApJ...824..119H}, 
the formation of very metal-poor stars 
\vTWO{
\citep{2020MNRAS.497..829F},
}
the formation of the first supermassive black hole progenitors \citep{2017Sci...357.1375H}, and 
accreting intermediate seed black holes 
\vTWO{
\citep{2019MNRAS.483.2031T, 2020MNRAS.496.1909T}.
}

This extensive list of scientific applications covers a broad parameter space in terms of mass densities, optical depths, thermal energies, and radiation energies. 
It furthermore illustrates our software development strategy toward general purpose modules.
A specific feature is thereby given by the ability to utilize grids in spherical coordinates.
Hence, in contrast to a general purpose style, the numerical framework is especially useful for modeling (the environments of) astrophysical objects, which dominate their environment in terms of gravity and/or radiation feedback.
\vONE{
An exception to this rule is the study by \citet{2020MNRAS.493.4643M} in which simulations were carried out on a 3D Cartesian grid with uniform grid spacing in a frame comoving with the bulk velocity of the turbulent and impacted interstellar medium gas.
}

With respect to ongoing code development and maintenance of the software package,
the flux-limited-diffusion solver for the continuum radiation transport has recently been updated to the so-called M1 scheme (V\"olkel \& Kuiper, in prep.).
In \citet{2018ApJ...857...57N, 2018ApJ...865...75N}, \vONE{the authors} further augmented the software by adding a ray-tracing step for the far UV and X-ray emission of a central point source combined with a chemical network to solve for the thermodynamics self-consistently.

%
%
\structuring{\clearpage}
\newpage
\section{Summary}
\label{sect:summary}
We have introduced a newly developed radiation--ionization framework for astrophysical Newtonian fluid dynamics.
The framework includes an update of our continuum radiation transport module Makemake and a newly developed photoionization module called Sedna. 
We have described the equations solved by the overall framework and give details on the derivation of these equations, including their underlying assumptions and approximations.
Numerical specifics to be considered are discussed to help future development of similar tools.

Radiation transport and photoionization are solved within a grid-based approach, and static grids in Cartesian, cylindrical, and spherical coordinates are supported.
One-dimensional, two-dimensional, and three-dimensional geometries are selectable.
A particular focus is given on either systems with a single dominant source of radiation or plane-parallel radiation fields.
Both modules -- continuum radiation and photoionization -- include a ray-tracing algorithm along the first grid coordinate direction and a three-dimensional FLD solver.
In the case of the continuum radiation transport module, the FLD solver is implemented as an equilibrium one-temperature approach (radiation temperature equals dust temperature) as well as in the linearization two-temperature approach.
In the case of the photoionization module, the user can choose between the widely used on-the-spot approximation or additionally solve for the temporal evolution of the diffuse EUV radiation field from direct recombination of free electrons into hydrogen's ground state.

Both modules and both solver steps --  ray-tracing and diffusion -- solve for the appropriate momentum feedback due to the absorption of photons, which are eventually added as additional source terms to the hydrodynamics equations. 

Diffusion-like equations are solved in a fully implicit manner.
The linear systems of equations are solved by modern Krylov subspace iterative algorithms utilizing the open-source Portable Extensible Toolkit for Scientific computing (PETSc) library \citep{Balay:2004ua}.
These modules are parallelized for multiprocessor computing using the MPI standard.

This radiation--ionization framework was combined with the open-source code PLUTO for the MHD modeling of astrophysical fluids.
The functionality, reliability and robustness, as well as the quantitative accuracy of the different modules and their combination is demonstrated in terms of a comprehensive test suite.
The test suite is structured along the module/physics combinations (radiation transport, radiation hydrodynamics, ionization, recombination, ionization hydrodynamics, and radiation--ionization forces).
The test suite includes widely used classical test problems, modern state-of-the-art benchmarks, and newly derived test problems.

Finally, an overview of the current astrophysical applications of the gravito--radiation--ionization hydrodynamics framework (including further subgrid modules for stellar evolution, dust evolution, and protostellar outflow feedback) demonstrates the broad applicability of the solver package from planetary science to star formation and AGN physics, as well as its already successful utilization in a variety of different research fields of astrophysics.

\structuring{\newpage}
\acknowledgments
This code development was conducted within the Emmy Noether research group on ``Accretion Flows and Feedback in Realistic Models of Massive Star Formation'' funded by the German Research Foundation under grant No.~KU 2849/3-1 and KU 2849/3-2.
R.K.~further acknowledges financial support by the German Academy of Science Leopoldina within the Leopoldina Fellowship Programme, grant No.~LPDS 2011-5, for long-term research visits at the Jet Propulsion Laboratory, CA, USA, and the University of Tokyo, Japan.
The authors acknowledge support by the High Performance and Cloud Computing Group at the Zentrum f\"ur Datenverarbeitung of the University of T\"ubingen, the state of Baden-W\"urttemberg through bwHPC, and the German Research Foundation (DFG) through grant No.~INST 37/935-1 FUGG.

\clearpage
\appendix
\section{Overview of Symbols and Constants}

\begin{table}[htbp]
\begin{center}
\begin{tabular}{l l l}
Symbol & Description & Value (in cgs) \\
\hline\hline
$\pi$ 				& Ratio of a circle's circumference to its diameter 	& $\approx 3.1416$ \\
$c$ 					& Speed of light in vacuum 					& $\approx 2.9979 \times 10^{+10} \mbox{ cm} \mbox{ s}^{-1}$ \\
$h$ 					& Planck constant	 						& $\approx 6.6260 \times 10^{-27} \mbox{ erg} \mbox{ s}^{-1}$ \\
$k_\mathrm{B}$ 		& Boltzmann constant 						& $\approx 1.3807 \times 10^{-16} \mbox{ erg} \mbox{ K}^{-1}$ \\
$a_\mathrm{rad}$ 		& Radiation constant 						& $\approx 7.5657 \times 10^{-15} \mbox{ erg} \mbox{ cm}^{-3} \mbox{ K}^{-4}$ \\
$R_\mathrm{gas}$ 		& Universal gas constant						& $\approx 8.3145 \times 10^{+7} \mbox{ g cm}^2 \mbox{ s}^{-2} \mbox{ mol}^{-1} \mbox{ K}^{-1}$ \\
$N_\mathrm{A}$ 		& Avogadro's constant 						& $\approx 6.0221 \times 10^{+23} \mbox{ mol}^{-1}$ \\
$\sigma_\mathrm{SB}$ 	& Stefan--Boltzmann constant 					& $2 ~ \pi^5 ~ R_\mathrm{gas}^4 / (15 ~ h^3 ~ c^2 ~ N_\mathrm{A}^4) \approx 5.6704 \times 10^{-5} \mbox{ erg cm}^{-2} \mbox{ s}^{-1} \mbox{ K}^{-4}$ \\
$u$					& Atomic mass unit							& $\approx 1.6605 \times 10^{-24} \mbox{ g}$ \\
$m_e$				& Electron mass							& $\approx 5.4858 \times 10^{-4} ~ u$ \\
$e$					& Elementary charge							& $\approx 4.8032 \times 10^{-10} \mbox { statC} ~ (\approx 1.6022 \times 10^{-19} \mbox{ C})$ \\
$\alpha_\mathrm{fs}$	& Fine-structure constant						& $2\pi ~ e^2 / (h ~ c) \approx 7.2974 \times 10^{-3}$ \\
$R_\infty$ 			& Rydberg constant							& $\alpha_\mathrm{fs}^2 ~ m_e ~ c / (2 ~ h)$ \\
$A_r$ 				& Recapture constant 						& $2^4 ~ h ~ e^2 / (3^{3/2}  ~ m_e^2 ~ c^3)$ \\
\end{tabular}
\end{center}
\caption{Overview of constants.}
\label{tab:constants}
\end{table}

\begin{longtable}{l l c}
Symbol 							& Description 								& Unit (in cgs) \\
\hline\hline
\multicolumn{3}{l}{Space and Time} \\
\hline
$r$ 								& Spherical radius 							& $\mbox{cm}$ \\
$\theta$ 							& Polar angle 								& $\mbox{rad}$ \\
$\phi$ 							& Azimuthal angle 							& $\mbox{rad}$ \\
$r_\mathrm{min}$ 					& Minimum radius of the computational domain 	& $\mbox{cm}$ \\
$r_\mathrm{max}$ 					& Maximum radius of the computational domain 	& $\mbox{cm}$ \\
$t$ 								& Time 									& $\mbox{s}$ \\
$\Delta t$ 							& Time step 								& $\mbox{s}$ \\
\hline
\multicolumn{3}{l}{Hydrodynamics} \\
\hline
$\rho_\mathrm{gas}$ 				& Gas mass density 							& $\mbox{g} \mbox{ cm}^{-3}$ \\
$\vec{u}_\mathrm{gas}$ 				& Gas velocity 								& $\mbox{cm} \mbox{ s}^{-1}$ \\
$P_\mathrm{gas}$ 					& Gas pressure 								& $\mbox{erg} \mbox{ cm}^{-3}$ \\
$E_\mathrm{tot}$ 					& Total gas energy density 					& $\mbox{erg} \mbox{ cm}^{-3}$ \\
$E_\mathrm{th}$ 					& Thermal gas energy density 					& $\mbox{erg} \mbox{ cm}^{-3}$ \\
$E_\mathrm{kin}$ 					& Kinetic gas energy density			 		& $\mbox{erg} \mbox{ cm}^{-3}$ \\
$\vec{a}_\mathrm{ext}$ 				& External acceleration 						& $\mbox{cm} \mbox{ s}^{-2}$ \\
\hline
\multicolumn{3}{l}{Continuum Radiation} \\
\hline
$I_\mathrm{rad}$ 					& Radiative intensity	 						& $\mbox{erg} \mbox{ cm}^{-2} \mbox{ s}^{-1}$ \\
$\vec{\Omega}$					& Direction of radiative flux					& $1$ \\
$\vec{F}_\mathrm{irr}$ 				& Stellar radiative flux 						& $\mbox{erg} \mbox{ cm}^{-2} \mbox{ s}^{-1}$ \\
$E_\mathrm{rad}$ 					& Thermal radiation energy density 				& $\mbox{erg} \mbox{ cm}^{-3}$ \\
$\vec{F}_\mathrm{rad}$ 				& Thermal radiative flux 						& $\mbox{erg} \mbox{ cm}^{-2} \mbox{ s}^{-1}$ \\
$D_\mathrm{rad}$ 					& Thermal radiation diffusion coefficient 			& $\mbox{cm}^{2} \mbox{ s}^{-1}$ \\
$\lambda_\mathrm{rad}$ 				& Thermal radiation flux limiter 					& $1$ \\
$\chi_\mathrm{scat}$ 				& Scattering coefficient 						& $\mbox{cm}^{-1}$ \\
$\chi_\mathrm{ext}$ 					& Extinction coefficient 						& $\mbox{cm}^{-1}$ \\
$\chi_\mathrm{abs}$ 				& Dust absorption coefficient 					& $\mbox{cm}^{-1}$ \\
$\chi_\mathrm{R}$ 					& Rosseland mean absorption coefficient 			& $\mbox{cm}^{-1}$ \\
$\chi_\mathrm{\nu}$ 					& Frequency-dependent absorption coefficient 		& $\mbox{cm}^{-1}$ \\
$B_\mathrm{rad}$ 					& Blackbody Planck spectrum energy density 		& $\mbox{erg} \mbox{ cm}^{-3}$ \\
$T_\mathrm{dust}$ 					& Dust temperature 							& $\mbox{K}$ \\
$T_\mathrm{gas}$ 					& Gas temperature 							& $\mbox{K}$ \\
$\nu$ 							& Frequency 								& $\mbox{s}^{-1}$ \\
\hline
\multicolumn{3}{l}{Photoionization} \\
\hline
$n_\mathrm{H}$ 					& Total hydrogen number density 				& $\mbox{cm}^{-3}$ \\
$n_\mathrm{H^0}$ 					& Neutral hydrogen number density 				& $\mbox{cm}^{-3}$ \\
$n_\mathrm{H^+}$ 					& Ionized hydrogen number density 				& $\mbox{cm}^{-3}$ \\
$x$ 								& Ionization fraction 							& $1$ \\
$y$ 								& Neutral fraction 							& $1$ \\
$u_\mathrm{EUV}$ 					& Photon number density from direct ray-tracing 	& $\mbox{cm}^{-3}$ \\
$\vec{F}_\mathrm{EUV}$ 				& Ionizing radiative EUV flux from direct ray-tracing 	& $\mbox{erg} \mbox{ cm}^{-2} \mbox{ s}^{-1}$ \\
$S_\mathrm{EUV}$ 					& Number of ionizing photons per unit time		& $\mbox{s}^{-1}$ \\
$u_\mathrm{rec}$ 					& Diffuse recombination photon number density 	& $\mbox{cm}^{-3}$ \\
$\vec{F}_\mathrm{rec}$ 				& Diffuse ionizing EUV flux					& $\mbox{erg} \mbox{ cm}^{-2} \mbox{ s}^{-1}$ \\
$D_\mathrm{rec}$ 					& Recombination diffusion coefficient 				& $\mbox{cm}^2 \mbox{ s}^{-1}$ \\
$\lambda_\mathrm{rec}$ 				& Diffuse recombination flux limiter 				& $1$ \\
$\sigma_\mathrm{EUV}$ 				& Photon cross section for ionizing ray-tracing flux 	& $\mbox{cm}^{2}$ \\
$\sigma_\mathrm{rec}$ 				& Recombination photon cross section 			& $\mbox{cm}^{2}$ \\
$\langle h \nu \rangle_\mathrm{EUV}$	& Mean photon energy of ray-tracing spectrum		& $\mbox{erg}$ \\
$\langle h \nu \rangle_\mathrm{rec}$		& Mean recombination photon energy			& $\mbox{erg}$ \\
$\alpha^{(1)}$ 						& Recombination rate of free electrons into any state & $\mbox{cm}^3 \mbox{ s}^{-1}$ \\
$\alpha^{(2)}$ 						& \multirow{2}{7cm}{Recombination rate of free electrons into any state of atomic hydrogen besides the ground state} & $\mbox{cm}^3 \mbox{ s}^{-1}$ \\
\\
$\alpha_1$ 						& Recombination rate into ground state of hydrogen & $\mbox{cm}^3 \mbox{ s}^{-1}$ \\
$C$ 								& Collisional excitation coefficient 				& $\mbox{cm}^3 \mbox{ s}^{-1}$ \\
$\Phi_1, \Phi_2$					& Recombination coefficient functions				& $1$ \\
 $Z$ 								& Atomic number 							& $1$ \\
\caption{Overview of Symbols.}
\label{tab:symbols}
\end{longtable}

\newpage
\bibliographystyle{aasjournal}
\bibliography{Papers}

\begin{thebibliography}{}
\expandafter\ifx\csname natexlab\endcsname\relax\def\natexlab#1{#1}\fi
\providecommand{\url}[1]{\href{#1}{#1}}
\providecommand{\dodoi}[1]{doi:~\href{http://doi.org/#1}{\nolinkurl{#1}}}
\providecommand{\doeprint}[1]{\href{http://ascl.net/#1}{\nolinkurl{http://ascl.net/#1}}}
\providecommand{\doarXiv}[1]{\href{https://arxiv.org/abs/#1}{\nolinkurl{https://arxiv.org/abs/#1}}}

\bibitem[{Adams {et~al.}(2015)Adams, Colella, Graves, Johnson, Johansen, Keen,
  Ligocki, Martin, McCorquodale, Modiano, Schwartz, Sternberg, \&
  Van~Straalen}]{Chombo2015}
Adams, M., Colella, P., Graves, D.~T., {et~al.} 2015, {Chombo Software Package
  for AMR Applications }, Tech. rep., Lawrence Berkeley National Laboratory,
  Berkeley

\bibitem[{Ahmadi {et~al.}(2019)Ahmadi, Kuiper, \&
  Beuther}]{2019A&A...632A..50A}
Ahmadi, A., Kuiper, R., \& Beuther, H. 2019, A{\&}A, 632, A50

\bibitem[{Auer \& Mihalas(1968)}]{1968ApJ...151..311A}
Auer, L.~H., \& Mihalas, D. 1968, ApJ, 151, 311

\bibitem[{Balay {et~al.}(2004)Balay, Buschelman, Eijkhout, Gropp, Kaushik,
  Knepley, McInnes, Smith, \& Zhang}]{Balay:2004ua}
Balay, S., Buschelman, K., Eijkhout, V., {et~al.} 2004, {PETSc Users Manual
  3.0.0}, Tech. rep., PETSc

\bibitem[{Bhandare {et~al.}(2020)Bhandare, Kuiper, Henning, Fendt, Flock, \&
  Marleau}]{2020A&A...638A..86B}
Bhandare, A., Kuiper, R., Henning, T., {et~al.} 2020, A{\&}A, 638, A86

\bibitem[{Bhandare {et~al.}(2018)Bhandare, Kuiper, Henning, Fendt, Marleau, \&
  K{\"o}lligan}]{2018A&A...618A..95B}
---. 2018, A{\&}A, 618, A95

\bibitem[{Bisbas {et~al.}(2015)Bisbas, Haworth, Williams, Mackey, Tremblin,
  Raga, Arthur, Baczynski, Dale, Frostholm, Geen, Haugb{\o}lle, Hubber, Iliev,
  Kuiper, Rosdahl, Sullivan, Walch, \& W{\"u}nsch}]{2015MNRAS.453.1324B}
Bisbas, T.~G., Haworth, T.~J., Williams, R. J.~R., {et~al.} 2015, MNRAS, 453,
  1324

\bibitem[{Bryan {et~al.}(2014)Bryan, Norman, O'Shea, Abel, Wise, Turk,
  Reynolds, Collins, Wang, Skillman, Smith, Harkness, Bordner, Kim, Kuhlen, Xu,
  Goldbaum, Hummels, Kritsuk, Tasker, Skory, Simpson, Hahn, Oishi, So, Zhao,
  Cen, Li, \& Collaboration}]{2014ApJS..211...19B}
Bryan, G.~L., Norman, M.~L., O'Shea, B.~W., {et~al.} 2014, ApJS, 211, 19

\bibitem[{Cimerman {et~al.}(2017)Cimerman, Kuiper, \&
  Ormel}]{2017MNRAS.471.4662C}
Cimerman, N.~P., Kuiper, R., \& Ormel, C.~W. 2017, MNRAS, 471, 4662

\bibitem[{Collins {et~al.}(2010)Collins, Xu, Norman, Li, \&
  Li}]{2010ApJS..186..308C}
Collins, D.~C., Xu, H., Norman, M.~L., Li, H., \& Li, S. 2010, ApJS, 186, 308

\bibitem[{Commer{\c c}on {et~al.}(2014)Commer{\c c}on, Debout, \&
  Teyssier}]{2014A&A...563A..11C}
Commer{\c c}on, B., Debout, V., \& Teyssier, R. 2014, A{\&}A, 563, A11

\bibitem[{Commer{\c c}on {et~al.}(2011)Commer{\c c}on, Teyssier, Audit,
  Hennebelle, \& Chabrier}]{2011A&A...529A..35C}
Commer{\c c}on, B., Teyssier, R., Audit, E., Hennebelle, P., \& Chabrier, G.
  2011, A{\&}A, 529, A35

\bibitem[{Dubey {et~al.}(2012)Dubey, Daley, ZuHone, Ricker, Weide, \&
  Graziani}]{2012ApJS..201...27D}
Dubey, A., Daley, C., ZuHone, J., {et~al.} 2012, ApJS, 201, 27

\bibitem[{Dubey {et~al.}(2009)Dubey, Reid, Weide, Antypas, Ganapathy, Riley,
  Sheeler, \& Siegal}]{2009arXiv0903.4875D}
Dubey, A., Reid, L.~B., Weide, K., {et~al.} 2009, arXiv, arXiv:0903.4875

\bibitem[{Dullemond(2011)}]{2011ascl.soft08016D}
Dullemond, C.~P. 2011, ASCL, ascl:1108.016

\bibitem[{Dzyurkevich {et~al.}(2016)Dzyurkevich, Commercon, Lesaffre, \&
  Semenov}]{2016arXiv160508032D}
Dzyurkevich, N., Commercon, B., Lesaffre, P., \& Semenov, D. 2016, arXiv,
  arXiv:1605.08032

\bibitem[{Dzyurkevich {et~al.}(2017)Dzyurkevich, Commercon, Lesaffre, \&
  Semenov}]{2017A&A...603A.105D}
---. 2017, A{\&}A, 603, A105

\bibitem[{Ensman(1994)}]{1994ApJ...424..275E}
Ensman, L. 1994, ApJ, 424, 275

\bibitem[{Fromang {et~al.}(2006)Fromang, Hennebelle, \&
  Teyssier}]{2006A&A...457..371F}
Fromang, S., Hennebelle, P., \& Teyssier, R. 2006, A{\&}A, 457, 371

\bibitem[{Fryxell {et~al.}(2000)Fryxell, Olson, Ricker, Timmes, Zingale, Lamb,
  MacNeice, Rosner, Truran, \& Tufo}]{2000ApJS..131..273F}
Fryxell, B., Olson, K., Ricker, P., {et~al.} 2000, ApJS, 131, 273

\bibitem[{Fukushima {et~al.}(2020)Fukushima, Hosokawa, Chiaki, Omukai, Yoshida,
  \& Kuiper}]{2020MNRAS.497..829F}
Fukushima, H., Hosokawa, T., Chiaki, G., {et~al.} 2020, MNRAS, 497, 829

\bibitem[{Gehmeyr \& Mihalas(1994)}]{1994PhyD...77..320G}
Gehmeyr, M., \& Mihalas, D. 1994, Physica D: Nonlinear Phenomena, 77, 320

\bibitem[{Gonz{\'a}lez {et~al.}(2007)Gonz{\'a}lez, Audit, \&
  Huynh}]{2007A&A...464..429G}
Gonz{\'a}lez, M., Audit, E., \& Huynh, P. 2007, A{\&}A, 464, 429

\bibitem[{Gonz{\'a}lez {et~al.}(2015)Gonz{\'a}lez, Vaytet, Commer{\c c}on, \&
  Masson}]{2015A&A...578A..12G}
Gonz{\'a}lez, M., Vaytet, N., Commer{\c c}on, B., \& Masson, J. 2015, A{\&}A,
  578, A12

\bibitem[{Gressel {et~al.}(2013)Gressel, Elstner, \&
  Ziegler}]{2013A&A...560A..93G}
Gressel, O., Elstner, D., \& Ziegler, U. 2013, A{\&}A, 560, A93

\bibitem[{Group \& Division(2012)}]{2012ascl.soft02008A}
Group, A. N.~A., \& Division, L. C.~R. 2012, ASCL, ascl:1202.008

\bibitem[{Hayes \& Norman(2003)}]{2003ApJS..147..197H}
Hayes, J.~C., \& Norman, M.~L. 2003, ApJS, 147, 197

\bibitem[{Hayes {et~al.}(2006)Hayes, Norman, Fiedler, Bordner, Li, Clark,
  ud~Doula, \& Mac~Low}]{2006ApJS..165..188H}
Hayes, J.~C., Norman, M.~L., Fiedler, R.~A., {et~al.} 2006, ApJS, 165, 188

\bibitem[{Heaslet \& Baldwin(1963)}]{1963PhFl....6..781H}
Heaslet, M.~A., \& Baldwin, B.~S. 1963, Physics of Fluids, 6, 781

\bibitem[{Hirano {et~al.}(2017)Hirano, Hosokawa, Yoshida, \&
  Kuiper}]{2017Sci...357.1375H}
Hirano, S., Hosokawa, T., Yoshida, N., \& Kuiper, R. 2017, Sci, 357, 1375

\bibitem[{Hosokawa {et~al.}(2016)Hosokawa, Hirano, Kuiper, Yorke, Omukai, \&
  Yoshida}]{2016ApJ...824..119H}
Hosokawa, T., Hirano, S., Kuiper, R., {et~al.} 2016, ApJ, 824, 119

\bibitem[{Hosokawa \& Inutsuka(2006)}]{2006ApJ...646..240H}
Hosokawa, T., \& Inutsuka, S.-I. 2006, ApJ, 646, 240

\bibitem[{Kee \& Kuiper(2019)}]{2019MNRAS.483.4893K}
Kee, N.~D., \& Kuiper, R. 2019, MNRAS, 483, 4893

\bibitem[{Kee {et~al.}(2018{\natexlab{a}})Kee, Owocki, \&
  Kuiper}]{2018MNRAS.474..847K}
Kee, N.~D., Owocki, S.~P., \& Kuiper, R. 2018{\natexlab{a}}, MNRAS, 474, 847

\bibitem[{Kee {et~al.}(2018{\natexlab{b}})Kee, Owocki, \&
  Kuiper}]{2018MNRAS.479.4633K}
---. 2018{\natexlab{b}}, MNRAS, 479, 4633

\bibitem[{Klassen {et~al.}(2014)Klassen, Kuiper, Pudritz, Peters, Banerjee, \&
  Buntemeyer}]{2014ApJ...797....4K}
Klassen, M., Kuiper, R., Pudritz, R.~E., {et~al.} 2014, ApJ, 797, 4

\bibitem[{Kolb {et~al.}(2013)Kolb, Stute, Kley, \&
  Mignone}]{2013A&A...559A..80K}
Kolb, S.~M., Stute, M., Kley, W., \& Mignone, A. 2013, A{\&}A, 559, A80

\bibitem[{K{\"o}lligan \& Kuiper(2018)}]{2018A&A...620A.182K}
K{\"o}lligan, A., \& Kuiper, R. 2018, A{\&}A, 620, A182

\bibitem[{Kudritzki {et~al.}(1988)Kudritzki, Yorke, \&
  Frisch}]{1988rmgm.book.....K}
Kudritzki, R.-P., Yorke, H.~W., \& Frisch, H. 1988, {Radiation in moving
  gaseous media: Eighteenth Advanced Course of the Swiss Society of
  Astrophysics and Astronomy} (M{\"u}nchen, Universit{\"a}t, Munich, Germany:
  Radiation in moving gaseous media : eighteenth Advanced Course of the Swiss
  Society of Astrophysics and Astronomy)

\bibitem[{Kuiper \& Hosokawa(2018)}]{2018A&A...616A.101K}
Kuiper, R., \& Hosokawa, T. 2018, A{\&}A, 616, A101

\bibitem[{Kuiper {et~al.}(2010{\natexlab{a}})Kuiper, Klahr, Beuther, \&
  Henning}]{2010ApJ...722.1556K}
Kuiper, R., Klahr, H., Beuther, H., \& Henning, T. 2010{\natexlab{a}}, ApJ,
  722, 1556

\bibitem[{Kuiper {et~al.}(2011)Kuiper, Klahr, Beuther, \&
  Henning}]{2011ApJ...732...20K}
---. 2011, ApJ, 732, 20

\bibitem[{Kuiper {et~al.}(2012)Kuiper, Klahr, Beuther, \&
  Henning}]{2012A&A...537A.122K}
---. 2012, A{\&}A, 537, A122

\bibitem[{Kuiper {et~al.}(2010{\natexlab{b}})Kuiper, Klahr, Dullemond, Kley, \&
  Henning}]{2010A&A...511A..81K}
Kuiper, R., Klahr, H., Dullemond, C., Kley, W., \& Henning, T.
  2010{\natexlab{b}}, A{\&}A, 511, A81

\bibitem[{Kuiper \& Klessen(2013)}]{2013A&A...555A...7K}
Kuiper, R., \& Klessen, R.~S. 2013, A{\&}A, 555, A7

\bibitem[{Kuiper {et~al.}(2016)Kuiper, Turner, \& Yorke}]{2016ApJ...832...40K}
Kuiper, R., Turner, N.~J., \& Yorke, H.~W. 2016, ApJ, 832, 40

\bibitem[{Kuiper \& Yorke(2013{\natexlab{a}})}]{2013ApJ...772...61K}
Kuiper, R., \& Yorke, H.~W. 2013{\natexlab{a}}, ApJ, 772, 61

\bibitem[{Kuiper \& Yorke(2013{\natexlab{b}})}]{2013ApJ...763..104K}
---. 2013{\natexlab{b}}, ApJ, 763, 104

\bibitem[{Kuiper {et~al.}(2015)Kuiper, Yorke, \& Turner}]{2015ApJ...800...86K}
Kuiper, R., Yorke, H.~W., \& Turner, N.~J. 2015, ApJ, 800, 86

\bibitem[{Kurucz(1979)}]{1979ApJS...40....1K}
Kurucz, R.~L. 1979, ApJS, 40, 1

\bibitem[{Levermore \& Pomraning(1981)}]{1981ApJ...248..321L}
Levermore, C.~D., \& Pomraning, G.~C. 1981, ApJ, 248, 321

\bibitem[{Lowrie \& Edwards(2008)}]{2008ShWav..18..129L}
Lowrie, R.~B., \& Edwards, J.~D. 2008, Shock Waves, 18, 129

\bibitem[{Lowrie \& Rauenzahn(2007)}]{2007ShWav..16..445L}
Lowrie, R.~B., \& Rauenzahn, R.~M. 2007, Shock Waves, 16, 445

\bibitem[{MacNeice {et~al.}(2000)MacNeice, Olson, Mobarry, de~Fainchtein, \&
  Packer}]{2000CoPhC.126..330M}
MacNeice, P., Olson, K.~M., Mobarry, C., de~Fainchtein, R., \& Packer, C. 2000,
  Computer Physics Communications, 126, 330

\bibitem[{MacNeice {et~al.}(2011)MacNeice, Olson, Mobarry, de~Fainchtein, \&
  Packer}]{2011ascl.soft06009M}
---. 2011, ASCL, ascl:1106.009

\bibitem[{Mai {et~al.}(2020)Mai, Desch, Kuiper, Marleau, \&
  Dullemond}]{2020ApJ...899...54M}
Mai, C., Desch, S.~J., Kuiper, R., Marleau, G.-D., \& Dullemond, C. 2020, ApJ,
  899, 54

\bibitem[{Marleau {et~al.}(2017)Marleau, Klahr, Kuiper, \&
  Mordasini}]{2017ApJ...836..221M}
Marleau, G.-D., Klahr, H., Kuiper, R., \& Mordasini, C. 2017, ApJ, 836, 221

\bibitem[{Marleau {et~al.}(2019)Marleau, Mordasini, \&
  Kuiper}]{2019ApJ...881..144M}
Marleau, G.-D., Mordasini, C., \& Kuiper, R. 2019, ApJ, 881, 144

\bibitem[{Menon {et~al.}(2020)Menon, Federrath, \&
  Kuiper}]{2020MNRAS.493.4643M}
Menon, S.~H., Federrath, C., \& Kuiper, R. 2020, MNRAS, 493, 4643

\bibitem[{Meyer {et~al.}(2018)Meyer, Kuiper, Kley, Johnston, \&
  Vorobyov}]{2018MNRAS.473.3615M}
Meyer, D. M.-A., Kuiper, R., Kley, W., Johnston, K.~G., \& Vorobyov, E. 2018,
  MNRAS, 473, 3615

\bibitem[{Meyer {et~al.}(2017)Meyer, Vorobyov, Kuiper, \&
  Kley}]{2017MNRAS.464L..90M}
Meyer, D. M.-A., Vorobyov, E.~I., Kuiper, R., \& Kley, W. 2017, MNRAS: Letters,
  464, L90

\bibitem[{Mignone {et~al.}(2007)Mignone, Bodo, Massaglia, Matsakos, Tesileanu,
  Zanni, \& Ferrari}]{2007ApJS..170..228M}
Mignone, A., Bodo, G., Massaglia, S., {et~al.} 2007, ApJS, 170, 228

\bibitem[{Mignone {et~al.}(2010)Mignone, Tzeferacos, Zanni, Tesileanu,
  Matsakos, \& Bodo}]{2010ascl.soft10045M}
Mignone, A., Tzeferacos, P., Zanni, C., {et~al.} 2010, ASCL, ascl:1010.045

\bibitem[{Mignone {et~al.}(2012)Mignone, Zanni, Tzeferacos, Van~Straalen,
  Colella, \& Bodo}]{2012ApJS..198....7M}
Mignone, A., Zanni, C., Tzeferacos, P., {et~al.} 2012, ApJS, 198, 7

\bibitem[{Mihalas \& Mihalas(1984)}]{1984frh..book.....M}
Mihalas, D., \& Mihalas, B.~W. 1984, {Foundations of radiation hydrodynamics}
  (New York: Oxford University Press)

\bibitem[{Nakatani {et~al.}(2018{\natexlab{a}})Nakatani, Hosokawa, Nomura, \&
  Kuiper}]{2018ApJ...857...57N}
Nakatani, R., Hosokawa, T., Nomura, H., \& Kuiper, R. 2018{\natexlab{a}}, ApJ,
  857, 57

\bibitem[{Nakatani {et~al.}(2018{\natexlab{b}})Nakatani, Hosokawa, Yoshida,
  Nomura, \& Kuiper}]{2018ApJ...865...75N}
Nakatani, R., Hosokawa, T., Yoshida, N., Nomura, H., \& Kuiper, R.
  2018{\natexlab{b}}, ApJ, 865, 75

\bibitem[{Norman(2000)}]{2000RMxAC...9...66N}
Norman, M.~L. 2000, "Astrophysical Plasmas: Codes, 9, 66

\bibitem[{Norman {et~al.}(2009)Norman, Reynolds, \& So}]{2009AIPC.1171..260N}
Norman, M.~L., Reynolds, D.~R., \& So, G.~C. 2009, in RECENT DIRECTIONS IN
  ASTROPHYSICAL QUANTITATIVE SPECTROSCOPY AND RADIATION HYDRODYNAMICS:
  Proceedings of the International Conference in Honor of Dimitri Mihalas for
  His Lifetime Scientific Contributions on the Occasion of His 70th Birthday.
  AIP Conference Proceedings, Physics Department, U.C. San Diego, La Jolla, CA
  92093; Ctr. for Astrophysics and Space Sciences, U.C. San Diego, La Jolla, CA
  92093 (AIP), 260--272

\bibitem[{Norman {et~al.}(2018)Norman, Smith, \& Bordner}]{2018FrASS...5...34N}
Norman, M.~L., Smith, B.~D., \& Bordner, J. 2018, Front. Astron. Space Sci., 5,
  34

\bibitem[{Oliva \& Kuiper(2020)}]{2020arXiv200813653A}
Oliva, G.~A., \& Kuiper, R. 2020, arXiv, arXiv:2008.13653

\bibitem[{Ormel {et~al.}(2015)Ormel, Shi, \& Kuiper}]{2015MNRAS.447.3512O}
Ormel, C.~W., Shi, J.-M., \& Kuiper, R. 2015, MNRAS, 447, 3512

\bibitem[{O'Shea {et~al.}(2004)O'Shea, Bryan, Bordner, Norman, Abel, Harkness,
  \& Kritsuk}]{2004astro.ph..3044O}
O'Shea, B.~W., Bryan, G., Bordner, J., {et~al.} 2004, arXiv, arXiv:astro

\bibitem[{O'Shea {et~al.}(2010)O'Shea, Bryan, Bordner, Norman, Abel, Harkness,
  \& Kritsuk}]{2010ascl.soft10072O}
---. 2010, ASCL, ascl:1010.072

\bibitem[{Ossenkopf \& Henning(1994)}]{1994A&A...291..943O}
Ossenkopf, V., \& Henning, T. 1994, A{\&}A, 291, 943

\bibitem[{Osterbrock(1989)}]{1989agna.book.....O}
Osterbrock, D.~E. 1989, {Astrophysics of gaseous nebulae and active galactic
  nuclei} (Lick Observatory, Santa Cruz, CA: Research supported by the
  University of California)

\bibitem[{Pascucci {et~al.}(2004)Pascucci, Wolf, Steinacker, Dullemond,
  Henning, Niccolini, Woitke, \& Lopez}]{2004A&A...417..793P}
Pascucci, I., Wolf, S., Steinacker, J., {et~al.} 2004, A{\&}A, 417, 793

\bibitem[{Ramsey \& Dullemond(2015)}]{2015A&A...574A..81R}
Ramsey, J.~P., \& Dullemond, C.~P. 2015, A{\&}A, 574, A81

\bibitem[{Richling \& Yorke(1997)}]{1997A&A...327..317R}
Richling, S., \& Yorke, H.~W. 1997, A{\&}A, 327, 317

\bibitem[{Rosdahl {et~al.}(2013)Rosdahl, Blaizot, Aubert, Stranex, \&
  Teyssier}]{2013MNRAS.436.2188R}
Rosdahl, J., Blaizot, J., Aubert, D., Stranex, T., \& Teyssier, R. 2013, MNRAS,
  436, 2188

\bibitem[{Rosdahl \& Teyssier(2015)}]{2015MNRAS.449.4380R}
Rosdahl, J., \& Teyssier, R. 2015, MNRAS, 449, 4380

\bibitem[{Sincell {et~al.}(1999{\natexlab{a}})Sincell, Gehmeyr, \&
  Mihalas}]{1999ShWav...9..391S}
Sincell, M.~W., Gehmeyr, M., \& Mihalas, D. 1999{\natexlab{a}}, Shock Waves, 9,
  391

\bibitem[{Sincell {et~al.}(1999{\natexlab{b}})Sincell, Gehmeyr, \&
  Mihalas}]{1999ShWav...9..403S}
---. 1999{\natexlab{b}}, Shock Waves, 9, 403

\bibitem[{Spitzer(1968)}]{1968dms..book.....S}
Spitzer, L. 1968, {Diffuse matter in space} (New York: Interscience
  Publication)

\bibitem[{Spitzer(1978)}]{1978ppim.book.....S}
---. 1978, {Physical processes in the interstellar medium}, ed. L.~Spitzer~Jr.
  (Weinheim, Germany: Wiley-VCH Verlag GmbH)

\bibitem[{Teyssier(2002)}]{2002A&A...385..337T}
Teyssier, R. 2002, A{\&}A, 385, 337

\bibitem[{Thompson(1984)}]{1984ApJ...283..165T}
Thompson, R.~I. 1984, ApJ, 283, 165

\bibitem[{Toyouchi {et~al.}(2020)Toyouchi, Hosokawa, Sugimura, \&
  Kuiper}]{2020MNRAS.496.1909T}
Toyouchi, D., Hosokawa, T., Sugimura, K., \& Kuiper, R. 2020, MNRAS, 496, 1909

\bibitem[{Toyouchi {et~al.}(2019)Toyouchi, Hosokawa, Sugimura, Nakatani, \&
  Kuiper}]{2019MNRAS.483.2031T}
Toyouchi, D., Hosokawa, T., Sugimura, K., Nakatani, R., \& Kuiper, R. 2019,
  MNRAS, 483, 2031

\bibitem[{Turner \& Stone(2001)}]{2001ApJS..135...95T}
Turner, N.~J., \& Stone, J.~M. 2001, ApJS, 135, 95

\bibitem[{Winkler \& Newman(1980)}]{1980ApJ...236..201W}
Winkler, K.-H.~A., \& Newman, M.~J. 1980, ApJ, 236, 201

\bibitem[{Wise \& Abel(2011)}]{2011MNRAS.414.3458W}
Wise, J.~H., \& Abel, T. 2011, MNRAS, 414, 3458

\bibitem[{Yorke(1986)}]{1986ARA&A..24...49Y}
Yorke, H.~W. 1986, ARA{\&}A, 24, 49

\bibitem[{Yorke \& Sonnhalter(2002)}]{2002ApJ...569..846Y}
Yorke, H.~W., \& Sonnhalter, C. 2002, ApJ, 569, 846

\bibitem[{Yorke \& Welz(1996)}]{1996A&A...315..555Y}
Yorke, H.~W., \& Welz, A. 1996, A{\&}A, 315, 555

\bibitem[{Zel'Dovich \& Raizer(1967)}]{1967pswh.book.....Z}
Zel'Dovich, Y.~B., \& Raizer, Y.~P. 1967, {Physics of shock waves and
  high-temperature hydrodynamic phenomena} (New York: Academic Press)

\bibitem[{Ziegler(2011)}]{2011ascl.soft01006Z}
Ziegler, U. 2011, ASCL, ascl:1101.006

\end{thebibliography}

\end{document}